\documentclass{JHEP3}

\usepackage{epsfig,multicol,bbm}
\usepackage{graphicx}
\usepackage{dcolumn}
\usepackage{amsmath}
\usepackage{enumerate}

\def\la{\mathrel{\mathpalette\fun <}}

\def\fun#1#2{\lower3.6pt\vbox{\baselineskip0pt\lineskip.9pt
        \ialign{$\mathsurround=0pt#1\hfill##\hfil$\crcr#2\crcr\sim\crcr}}}
        
\def\expec#1{\langle#1\rangle}

\def\vmu{\mbox{\boldmath${\mu}$}}
\def\bfp{\mbox{\bf p}}
\def\bfq{\mbox{\bf q}}

\def\x{{\bf x}}
\def\C{{\bf C}}

\newcommand{\DA}{D\!_A(z)}
\newcommand{\hz}{H(z)}

\newcommand{\Vsur}{V_{\rm survey}}
\newcommand{\Veff}{V_{\rm eff}}

\newcommand{\w}{w}

\newcommand{\ihMpc}{h{\rm\;Mpc^{-1}}}
\newcommand{\kmax}{k_{\rm max}}

\def\la{\mathrel{\mathpalette\fun <}}

\def\fun#1#2{\lower3.6pt\vbox{\baselineskip0pt\lineskip.9pt
        \ialign{$\mathsurround=0pt#1\hfill##\hfil$\crcr#2\crcr\sim\crcr}}}

\title{Neutrino constraints from future nearly all-sky spectroscopic galaxy surveys}

\author{Carmelita Carbone \\ Dipartimento di Astronomia, Universit\a`a di Bologna\\
  Via Ranzani 1, I-40127 Bologna, Italy \\ 
\& INFN, Sezione di Bologna, Viale Berti Pichat 6/2, I-40127
      Bologna, Italy\\
E-mail: \email{carmelita.carbone@unibo.it}}

\author{Licia Verde\\
ICREA \& Instituto de Ciencias del Cosmos (ICC), Universitat de Barcelona  (UB-IEEC)\\
Marti i Franques 1, 08028, Barcelona, Spain \\
E-mail: \email{liciaverde@icc.ub.edu}}

\author{Yun Wang\\
Homer L. Dodge Department of Physics \& Astronomy, University of Oklahoma\\
440 W. Brooks St., Norman, OK 73019, USA \\
E-mail: \email{wang@nhn.ou.edu}}

\author{Andrea Cimatti \\ Dipartimento di Astronomia, Universit\a`a
  di Bologna\\                    
  Via Ranzani 1, I-40127 Bologna, Italy \\ 
E-mail: \email{a.cimatti@unibo.it}}

\abstract{\small We examine whether future, nearly all-sky galaxy redshift surveys, 
in combination with CMB priors,  will be able to detect the signature of the 
cosmic neutrino background and determine the absolute neutrino mass scale.  
We also consider what constraints can be imposed on the effective number of 
neutrino species. 
In particular we consider two spectroscopic strategies in the
near-IR, the so-called ``slitless'' and ``multi-slit'' approaches, whose examples are given
by future space-based galaxy surveys, as EUCLID for the slitless case,
or SPACE, JEDI, and possibly WFIRST in the future, for the multi-slit case.
We find that, in combination with Planck, these galaxy probes will be able to  detect at 
better than 3--sigma level and measure the mass of cosmic neutrinos:  
a) in a cosmology-independent way, if the sum of neutrino masses is above 0.1 eV; 
b) assuming spatial flatness and that dark energy is a cosmological constant, 
otherwise. We find that the sensitivity of such surveys is well suited to  
span the entire range of neutrino masses allowed by neutrino oscillation experiments,
and to yield a clear detection of non-zero neutrino mass.
The detection of the cosmic  relic neutrino background with
cosmological experiments will be a spectacular confirmation of our model 
for the early Universe and a window into one of the oldest relic components 
of our Universe.}


\begin{document}
\section{Introduction}
\label{Intro}
Atmospheric and solar neutrino experiments have demonstrated that
neutrinos have mass,  
implying a lower limit on the total neutrino mass given by  
$M_\nu\equiv \sum m_{\nu}\sim 0.05$ eV
\cite{lesgourgues/pastor:2006}.  
This is a clear indication that the standard model for particle
physics is incomplete and that there must 
be new  physics beyond it.
The neutrino mass splitting required to explain observations of neutrino oscillations
indicates that two hierarchies in the mass spectrum are possible: two
light states and a heavy one (normal hierarchy, NH, with $M_\nu>0.05$ eV),
or two heavy and one light (inverted hierarchy IH, with  $M_\nu>0.1$
eV). A third possibility is that the absolute mass scale is
much larger than the mass splittings  
and therefore the mass  hierarchy does not matter (degenerate neutrino
mass spectrum).

On-going and forthcoming neutrino experiments aim at determining the
parameters of the neutrino 
mixing matrix and the nature of the neutrino mass (Dirac or
Majorana). 
These experiments are sensitive to neutrino flavor and mixing angle, 
and to the absolute mass scale for large neutrino masses. As an
example, beta-decay end-point spectra are sensitive to the neutrino mass, regardless of
whether neutrinos are Dirac or Majaorana particles, and, the current limit on the effective electron
neutrino mass is $< 2.2$ eV, coming from the Mainz and the Troitsk
experiments, while KATRIN is expected to reach a sensitivity of $\sim
0.2$ eV \cite{Lobashev:2003kt,Kraus:2004zw,Thummler:2010tt}.
Near future neutrino oscillation data may resolve the neutrino mass 
hierarchy if one of the still unknown parameters, which relates flavor with mass states, 
is not too small. However, if the mixing angle is too small, oscillation data 
may be unable to solve this issue. 

On the other hand cosmological probes are 
blind to flavor but sensitive to the absolute mass scale even for
small neutrino masses (see Fig.1). 
In fact, a thermal neutrino relic component in the 
Universe impacts both the expansion history and the growth of structure.
Neutrinos with mass $\la 1$ eV become non-relativistic after the epoch of 
recombination probed by the CMB, and this mechanism  allows massive neutrinos
to alter the matter-radiation equality for a fixed $\Omega_mh^2$. 
Neutrino's radiation-like behaviour at early times changes the
expansion rate, shifting the peak positions in the CMB angular power spectrum, 
but this is somewhat degenerate with other cosmological parameters.  
WMAP7 alone constrains $M_\nu<1.3$ eV \cite{Komatsuetal2010}
and, thanks to improved
sensitivity to polarisation and to the angular 
power spectrum damping tail, forecasts for the Planck satellite alone 
give $M_\nu\sim 0.2-0.4$ eV, 
depending on the assumed cosmological model and fiducial neutrino mass 
(e.g., \cite{Perotto, Kitching_nu} and references therein).
Massive neutrinos modify structure
formation on scales $k > k_{\rm  nr}=0.018(m_\nu/1{\rm
  eV})^{1/2}\Omega_m^{1/2}h$/Mpc, where $k_{\rm nr}$ is the
wave-number corresponding to
the Hubble horizon size at the epoch $z_{\rm nr}$, when a given neutrino
species becomes non-relativistic. In
particular, neutrinos free-stream and damp
the galaxy power spectrum on scales $k$ larger than the so called
free-streaming scale $k_{\rm fs}(z)=0.82 H(z)/(1+z)^2 (m_\nu/1{\rm
  eV}) h{\rm Mpc}^{-1}$ \cite{lesgourgues/pastor:2006},
thereby modifying the shape of the matter power spectrum in a 
redshift-dependent manner (see Fig.~\ref{transfer} and
e.g. \cite{HuEisensteinTegmark,0709.0253,1004.4105,1003.2422}). 
Therefore,  much more stringent constraints  can be obtained by combining
CMB data with large-scale structure (LSS) observations. 
Ref.~\cite{Reidnu,0505390} showed that  present data-sets yield a robust upper
limit of $M_\nu<0.3$ eV, almost ruling out the degenerate
mass spectrum; this result was later confirmed by \cite{Lahavmassnu,Concha}.

\begin{figure*}
\includegraphics[width=0.9\textwidth]{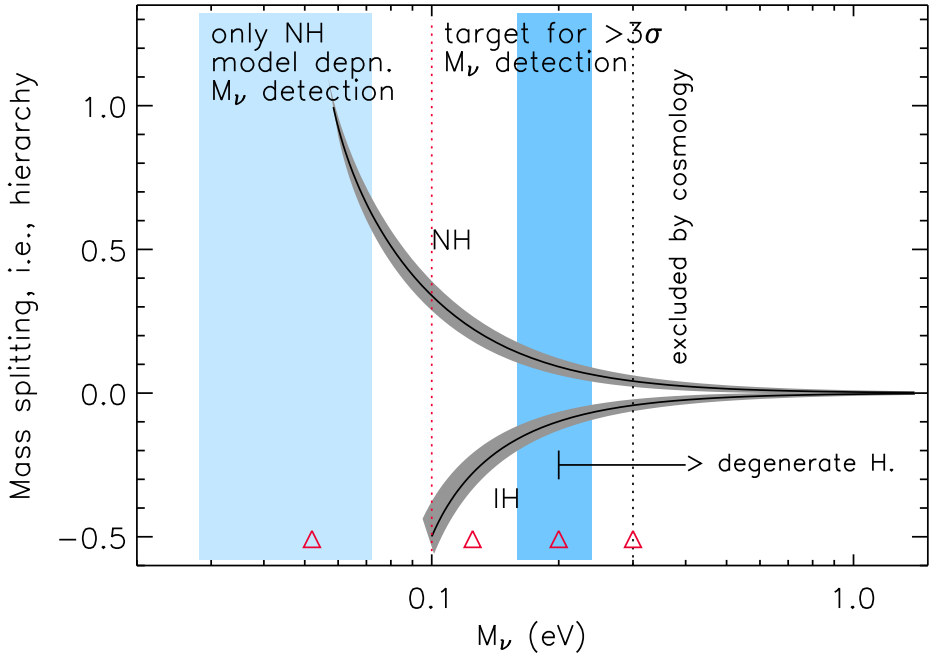}
\caption{Constraints from neutrino oscillations (shaded regions) and
from cosmology. In this parametrisation the sign of the mass splitting specifies
the hierarchy. The red triangles show the fiducial models explored in
this work and the light blue vertical bands our forecasted errors (see
\S 5). For fiducial $M_{\nu}$ values below $0.1$ eV  a LCDM model must
be assumed to obtain  a detection with $> 2$--$\sigma$ statistical
significance. For higher fiducial $M_{\nu}$, we can marginalise over dark energy parameters
and still obtain tight errors on $M_{\nu}$.}
\label{hierarchy}
\end{figure*}

The forecasted sensitivity of future  large-scale structure
experiments, when combined with Planck CMB priors, 
indicate  that cosmology should soon be able to detect signatures of
the cosmic neutrino background  and determine the 
sum of neutrino masses
(e.g. \cite{HannestadWong,Hannestadreview,Kitching_nu,lsstbook,LahavDES} 
and references therein). Since cosmology is only weakly sensitive to
the hierarchy \cite{1003.5918}, a total neutrino mass determination 
from cosmology will be able to determine the hierarchy only if the underlying model 
is normal hierarchy and $M_\nu<0.1$ eV (see e.g. Fig.~\ref{hierarchy}).
A  detection of the cosmic  relic neutrino background (RNG) with
cosmological experiments\footnote{Recall that neutrino experiments
  are not sensitive to relic neutrinos, as current generation of
  experiments do not have sufficient energy resolution to cleanly pin
  down the signature of the RNG. Anyway, the beta-decay end-point
  spectrum is in principle also sensitive to the RNG, and this can be
  foreseen as a plausible perspective for future experiments
only if neutrinos have masses of order eV, thus in the so called
degenerate scheme for neutrino masses, which is still allowed by all present data, though
slightly disfavored by cosmological observations \cite{Cocco:2007za}.}  
would be a spectacular confirmation of our model for the early
Universe and a window into one of the oldest relic components of our Universe
besides the one represented by  
the stochastic gravitational wave background.  This consideration prompts us
to examine whether future galaxy redshift surveys probing LSS   
will be able to detect the signature of the  neutrino background and
to determine the neutrino absolute mass scale.

Beyond neutrino mass, cosmology is also sensitive to the number of
neutrino species. In the standard model for particle physics there
are three neutrinos;  
they decouple early in the cosmic history  and then contribute  to
the relativistic energy density  (i.e. as if they were radiation) with
an effective number 
of neutrino species $N_{\rm eff}=3.046$
(e.g. \cite{lesgourgues/pastor:2006}) until they become
non-relativistic. Cosmology is sensitive to the physical energy  
density of relativistic particles, which include photons and neutrinos:
$\Omega_r=\Omega_{\gamma}+N_{\rm eff}\Omega_{\nu}$, where
$\Omega_{\gamma}$ and $\Omega_{\nu}$ are  
the energy density in photons and in one active neutrino species,
respectively. CMB observations  have constrained exquisitely well
$\Omega_{\gamma}$, thus constraints in  $\Omega_{\rm r}$ can be used 
to study neutrino
properties. 
Deviations from $N_{\rm eff}=3.046$ would indicate non-standard neutrino properties or additional effective relativistic species.  
While the motivation  for considering deviations from the standard model in the form of extra neutrino 
species has now disappeared \cite{mena,miniboone,mangano},  departures from
the standard $N_{\rm eff}$ value could arise from decay of dark-matter
particles \cite{bonometto98,lopez98,hannestad98,kaplinghat01}, early quintessence \cite{bean}, or more exotic models \cite{unparticles}.
 
Relativistic particles affect the CMB and the matter power spectrum
in two ways: {\it a)} through their anisotropic stress
\cite{trotta/melchiorri:2005,Komatsuetal2010}, and {\it b)}
through their relativistic energy density which alters the
epoch of matter radiation equality.  The ratio of CMB
peak heights constrains matter-radiation 
equality yielding a degeneracy between $N_{\rm eff}$ and $\Omega_m h^2$. 
This degeneracy can be lifted by adding either cosmic expansion history data
\cite{deberanrdisneff,Figueroa,hzstern} or adding the large-scale
shape of the matter power 
spectrum: the power spectrum turnover scale is  also related to
matter-radiation equality given by the parameter $\Gamma\sim \Omega_m
h$ (note the different scaling with $h$ compared to the CMB
constraint). LSS surveys can yield
a measurement, at the same time, of  both the cosmic expansion history (via
the Baryon Acoustic Oscillations (BAO) signal), and the large scale turnover
of the power spectrum. 
Present constraints are already competitive  with
nucleosynthesis constraints, and future data will offer the possibility to test
consistency of the standard paradigm for the early Universe. In fact, 
nucleosynthesis constraints rely on physics describing the Universe 
when its energy scale was $T\sim$ MeV, while cosmological constraints rely on physics at $T\sim eV$.
 
In this paper we forecast errors on the total neutrino mass $M_\nu$
and the effective
number of relativistic species $N_{\rm eff}$ by combining Planck priors
with data from future space-based galaxy redshift surveys in the
near-IR. In particular, we consider two main survey strategies: 
\begin{itemize}
\item
The first approach is to use ``multi-slit''                               
spectroscopy aimed at observing a pure magnitude-limited sample of galaxies                    
selected in the near-IR (e.g. in the H-band at 1.6 $\mu$m) with a                                 
limiting magnitude appropriate to cover the desired redshift range.                            
Examples of this approach are given by instruments where the efficient                         
multi-slit capability is provided by micro-shutter arrays (MSA)
(e.g. JEDI\footnote{http://jedi.nhn.ou.edu/} \cite{Wang04,Crotts05,Cheng06}), or by 
digital micromirror devices (DMD) (e.g. SPACE \cite{Cimatti09} and 
possibly WFIRST\footnote{http://wfirst.gsfc.nasa.gov/} in the future). 
With the multi-slit approach, all galaxy types                           
(from passive ellipticals to starbursts) are observed, typically
at $0<z<2-3$, if the observations are done in the near-IR, and
provided that the
targets are randomly selected from the magnitude-limited galaxy
sample.
\item                        
The second approach is based on slitless spectroscopy
(e.g. Euclid\footnote{http://sci.esa.int/euclid} and
JDEM\footnote{http://jdem.gsfc.nasa.gov/} \cite{Glazebrook05,Laureijs09,JDEM})             
which, due to stronger sky background, is                      
sensitive mostly to galaxies with emission lines (i.e. star-forming and                        
AGN systems), and uses mainly H$\alpha$ as a redshift tracer if the observations                    
are done in the near-IR to cover the redshift range $0.5<z<2$.                  
\end{itemize}

Forthcoming surveys will also have a weak gravitational lensing
component,  which  will also be used to constrain neutrino properties
(see e.g. \cite{Kitching_nu}). Here we
concentrate on galaxy clustering as an independent  and complementary
probe.

The rest of the paper is organised as follows. In \S~\ref{Fisher
  matrix approach} we review our method 
and the employed modelling. In \S~\ref{spec_methods}  we report the characteristics of  
the galaxy surveys considered in this work, 
and  in \S~\ref{Fiducial cosmologies} we describe the adopted fiducial 
models and the explored space of cosmological parameters. 
In \S~\ref{Results} we present our results 
on the forecasted errors on the neutrino mass and  
number of neutrino species, and final in \S~\ref{Conclusions}
we draw our conclusions.
 
\section{Fisher matrix approach: $P(k)$--method}
\label{Fisher matrix approach}
In this paper we adopt the Fisher matrix formalism to make predictions
on neutrino masses and relativistic degrees of freedom
from future galaxy redshift surveys. 

The Fisher matrix is defined as the
second derivative of the natural logarithm of the likelihood surface about the maximum. 
In the approximation that the posterior distribution for the parameters is a
multivariate Gaussian\footnote{In practice,
it can happen that the choice of parametrisation                                         
makes the posterior distribution slightly non-Gaussian. However,
for the parametrisation chosen here, the error introduced by assuming Gaussianity in the
posterior distribution can be considered as reasonably small, and therefore the
Fisher matrix approach still holds as an excellent approximation for
parameter forecasts.}
with mean $\vmu\equiv\expec{\x}$ and covariance matrix
$\C\equiv\expec{\x\x^t}-\vmu\vmu^t$, 
its elements are given by \cite{VS96,Tegmark,Jungman,Fisher}
\begin{align}
\label{eq:fish}
F_{ij} = \frac{1}{2}{\rm Tr}
\left[\C^{-1}{\partial\C\over\partial\theta_i}\C^{-1}{\partial\C\over\partial\theta_j}\right]
+{\partial\vmu\over\partial\theta_i}^t\C^{-1}{\partial\vmu\over\partial\theta_j}.
\end{align}
where $\x$ is a N-dimensional vector representing the data set,
whose components $x_i$ are the 
fluctuations in the galaxy density relative to the mean 
in $N$ disjoint cells that cover the three-dimensional 
survey volume in a fine grid. The $\{\theta_i\}$ denote
the cosmological parameters
within the assumed fiducial cosmology.

In order to explore the cosmological parameter
constraints from a given redshift survey, we need to specify
the measurement uncertainties of the galaxy power spectrum.
In general, the statistical error on the measurement of the galaxy
power spectrum $P_{\rm g}(k)$
at a given wave-number bin is \cite{FKP}
\begin{equation}
\left[\frac{\Delta P_{\rm g}}{P_{\rm g}}\right]^2=
\frac{2(2\pi)^2 }{\Vsur k^2\Delta k\Delta \mu}
\left[1+\frac1{n_{\rm g}P_{\rm g}}\right]^2,
\label{eqn:pkerror}
\end{equation}
where $n_{\rm g}$ is the mean number density of galaxies, 
$\Vsur$ is the comoving survey volume of the galaxy survey, and $\mu$
is the cosine of the angle between $\bf{k}$ and the line-of-sight
direction $\mu = \vec{k}\cdot \hat{r}/k$.

\begin{figure*}
\begin{tabular}{l l}
\includegraphics[width=7.3cm]{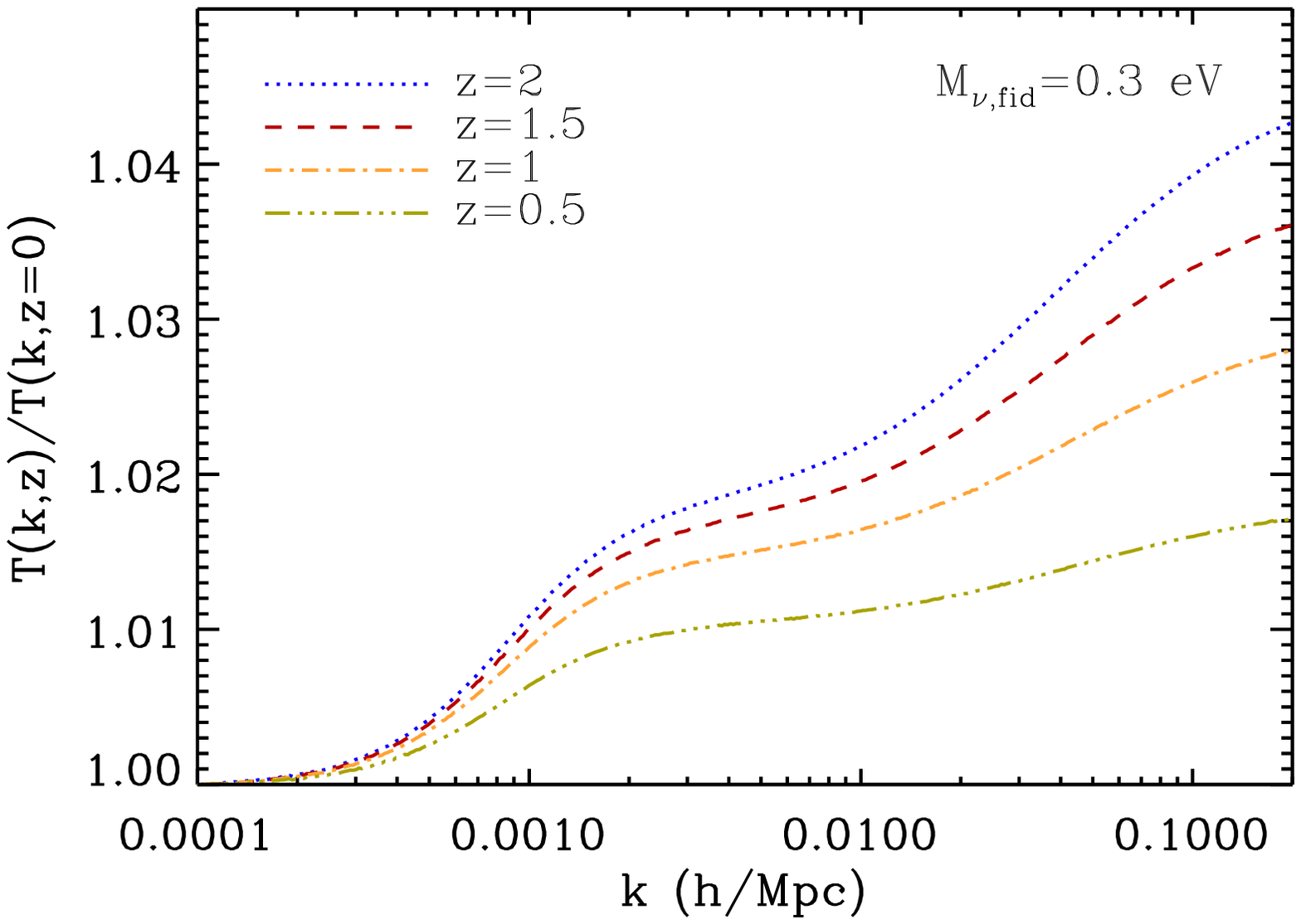}&
\includegraphics[width=7.3cm]{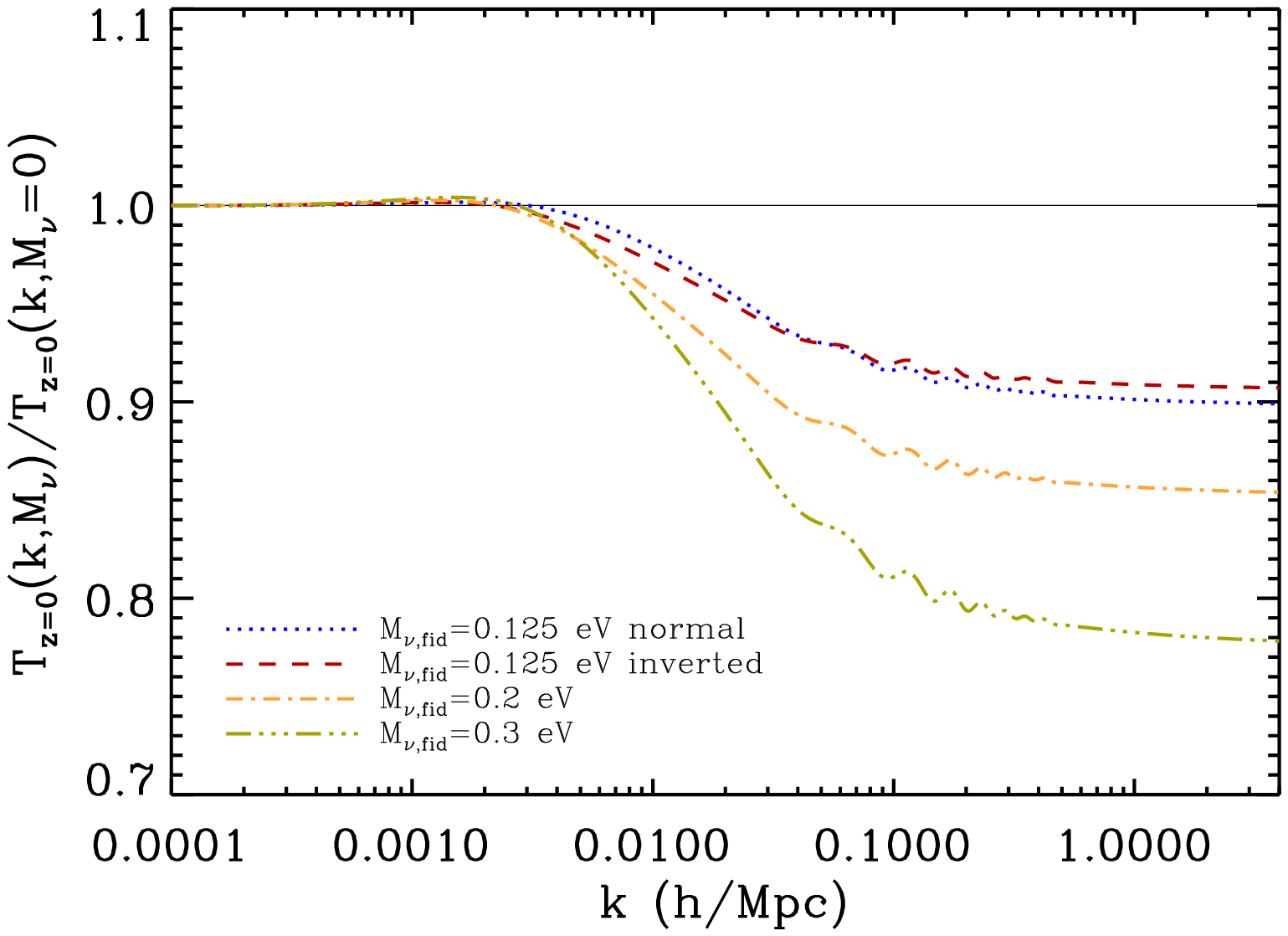}\\
\includegraphics[width=7.3cm]{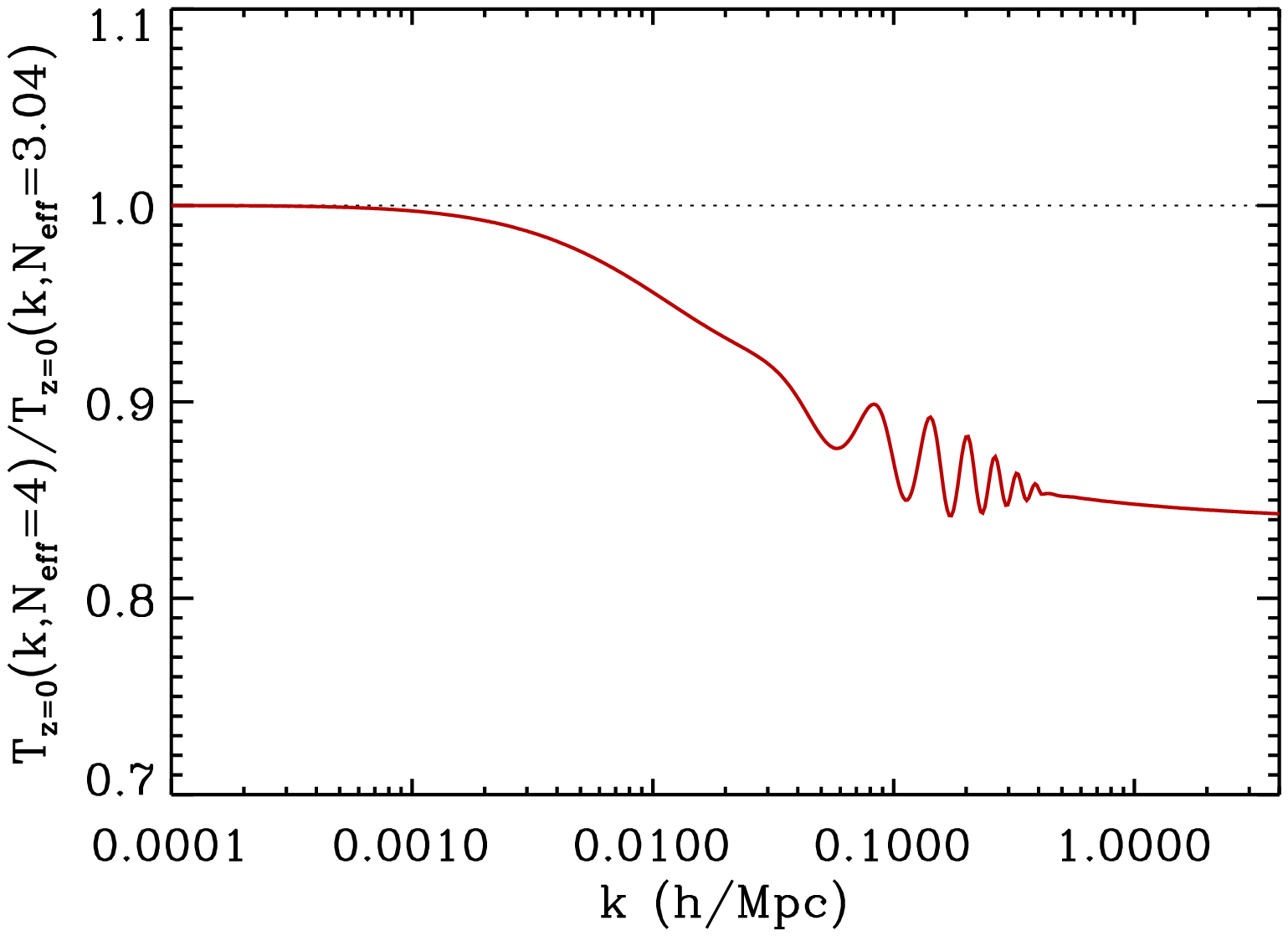}&
\includegraphics[width=7.3cm]{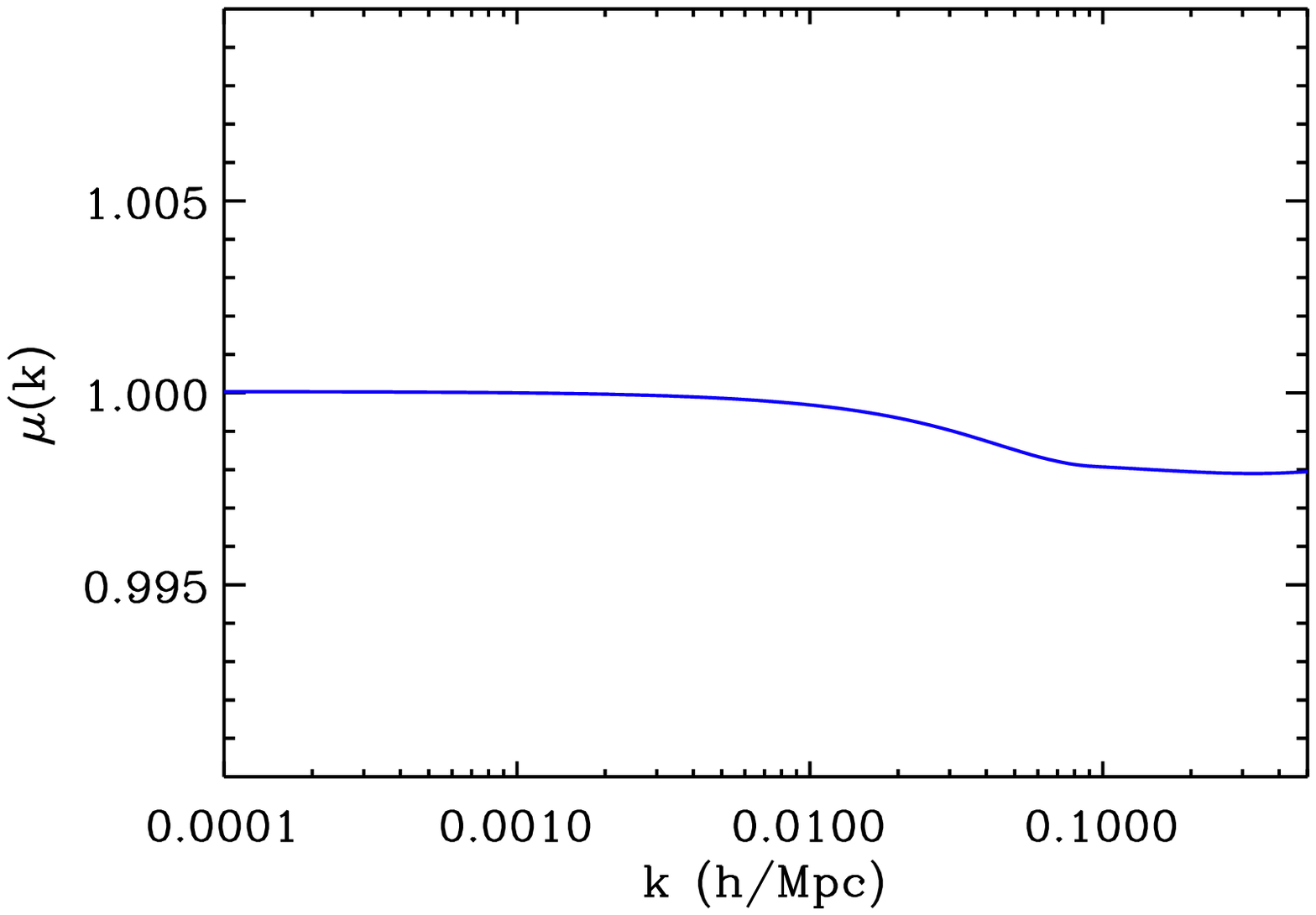}\\
\end{tabular}
\caption{Top Left: Ratio $T(k,z)/T(k,z=0)$ of the linear transfer functions
computed with CAMB at redshifts $z=0,0.5,1,1.5,2$, for a
  fiducial cosmology with a total neutrino
mass $M_\nu=0.3$ eV and a degenerate mass spectrum.
Top Right: Ratio $T_{z=0}(k,M_\nu)/T_{z=0}(k,M_\nu=0)$ between the linear
  transfer functions computed with CAMB for the different $M_\nu$--cosmologies
  described in Sec.~4 and the linear transfer function obtained
  assuming massless neutrinos.
Bottom Left: Ratio $T_{z=0}(k,N_{\rm eff}=4)/T_{z=0}(k,N_{\rm eff}=3.03)$ between the
  linear transfer functions computed with CAMB assuming massless
  neutrinos and an effective number of relativistic species $N_{\rm
    eff}=4$ and $N_{\rm eff}=3.04$, respectively.
Bottom Right: the function 
$\mu(k,f_\nu,\Omega_{de})\equiv f_g(M_\nu \neq 0)/f_g(M_\nu = 0)$, 
where $f_\nu=\Omega_\nu/\Omega_m$. Note that $\mu(k,f_\nu,\Omega_{de})$ represents
the scale dependent correction to $f_g(z)$, evaluated at $M_\nu=0.05$.}
\label{transfer}
\end{figure*}

In general, the \emph{observed} galaxy power spectrum is different
from the \emph{true} spectrum, and it can be reconstructed approximately
assuming a reference cosmology (which we consider to be our fiducial
cosmology) as (e.g. \cite{SE03})
\begin{align}
P_{\rm obs}(k_{{\rm ref}\perp},k_{{\rm ref}\parallel},z)
=\frac {\DA _{\rm ref} ^2 \hz}{\DA ^2 \hz _{\rm ref}} P_{\rm g}(k_{{\rm ref}\perp},k_{{\rm ref}\parallel},z)
+P_{\rm shot}\,,
\label{eq:Pobs}
\end{align}
where
\begin{align}
P_{\rm g}(k_{{\rm ref}\perp},k_{{\rm
    ref}\parallel},z)=b(z)^2\left[1+\beta(z,k) 
\frac{k_{{\rm ref}\parallel}^2}{k_{{\rm ref}\perp}^2+k_{{\rm ref}\parallel}^2}\right]^2\times
P_{{\rm matter}}(k,z)\,.
\label{eq:Pg}
\end{align}
In Eq.~(\ref{eq:Pobs}), $H(z)$ and $D_A(z)$ are the Hubble parameter and the angular
diameter distance, respectively, and the prefactor 
$(\DA _{\rm ref} ^2 \hz)/(\DA ^2 \hz _{\rm ref})$ encapsulates the
geometrical distortions due to the  Alcock-Paczynski
effect \cite{SE03,9605017}.  Their values in the reference cosmology are
distinguished by the subscript `ref', while those in the true cosmology have no
subscript. $k_\perp$ and $k_\parallel$ are the wave-numbers across and along
the line of sight in the true cosmology, and they are related to the 
wave-numbers calculated assuming the reference
cosmology by
$k_{{\rm ref}\perp} = k_\perp D_A(z)/D_A(z)_{\rm ref}$ and
$k_{{\rm ref}\parallel} = k_\parallel H(z)_{\rm ref}/H(z)$. 
$P_{shot}$ is the unknown white shot noise that 
remains even after the conventional shot noise of inverse number density has been 
subtracted \cite{SE03}, and which could arise from galaxy clustering bias even
on large scales due to local bias \cite{Seljak00}.
In Eq.~(\ref{eq:Pg}), $b(z)$ is the \emph{linear bias} factor between galaxy and
matter density distributions, and $\beta(z,k)=f_g(z,k)/b(z)$ is the linear 
redshift-space distortion parameter \cite{Kaiser1987}, which in the presence 
of massive neutrinos depends on both redshift and
wave-numbers, since in this case the linear growth rate $f_g(z,k)$ 
is \emph{scale dependent} even at the linear level. We estimate $f_g(z,k)$ using the
fitting formula of Ref.~\cite{0709.0253} (see the bottom-right
panel of Fig.~\ref{transfer}). For the
linear matter power spectrum 
$P_{{\rm matter}}(k,z)$, we can
encapsulate the effect of
massive neutrino free-streaming into a \emph{redshift dependent}
total matter linear transfer function $T(k,z)$ \cite{0512374,0606533,9710252}, so that 
$P_{{\rm matter}}(k,z)$ in Eq.~(\ref{eq:Pobs}) takes the form
\begin{eqnarray}
P_{{\rm matter}}(k,z)=\frac{8\pi^2c^4k_0\Delta^2_{\cal R}(k_0)}{25
  H_0^4\Omega_{m}^2} T^2(k,z) \left [\frac{G(z)}{G(z=0)}\right]^2
\left(\frac{k}{k_0}\right)^{n_s}e^{-k^2\mu^2\sigma_r^2},
\label{eq:Pm}
\end{eqnarray}
where $G(z)$ is the usual \emph{scale independent} linear
growth-factor in the absence of massive neutrino free-streaming,
i.e. for $k\rightarrow 0$  (see Eq.~(25)
in Ref.~\cite{9710252}), whose fiducial value in each redshift bin is computed 
through numerical integration of the differential equations governing the growth
of linear perturbations in the presence of dark-energy
\cite{astro-ph/0305286}. 
The \emph{redshift-dependent}
linear transfer function $T(k,z)$ depends on matter, baryon and massive neutrino
densities (neglecting dark-energy at early times), 
and is computed in each redshift bin using
CAMB\footnote{http://camb.info/} \cite{CAMB}.
As an example of its redshift dependence, 
in the top-left panel of Fig.~\ref{transfer} we consider the linear transfer
function and show the ratio $T(k,z)/T(k,z=0)$
computed with CAMB at redshifts $z=0.5,1,1.5,2$ for a total neutrino
mass $M_\nu=0.3$ eV.
On the other hand, in the top-right panel of Fig.~\ref{transfer}, as an example of the
neutrino free-streaming effect, we fix the redshift at
$z=0$ and compute, for different neutrino masses, the ratio of the linear transfer function
 to the linear transfer function in absence
of massive neutrinos. The power suppression due to neutrino
free-streaming is evident and increases with the neutrino mass as well
as the free-streaming scale. This
suppression is also slightly dependent on the assumed mass hierarchy, as the
blue-dotted and red-dashed lines clearly show.

In Eq.~(\ref{eq:Pm}) we have added the damping factor
$e^{-k^2\mu^2\sigma_r^2}$,
due to redshift uncertainties, where $\sigma_r=(\partial r/\partial
z)\sigma_z$, $r(z)$ being the comoving
distance \cite{0904.2218,SE03},
and we have assumed the power spectrum of primordial curvature
perturbations, $P_{\cal R}(k)$, to be
\begin{equation}
\Delta^2_{\cal R}(k) \equiv \frac{k^3P_{\cal R}(k)}{2\pi^2}
= \Delta^2_{\cal R}(k_0)\left(\frac{k}{k_0}\right)^{n_s},
\label{eq:pR}
\end{equation}
where $k_0=0.002$/Mpc, $\Delta^2_{\cal R}(k_0)|_{\rm fid}=2.45\times
10^{-9}$ is the dimensionless amplitude of the primordial curvature perturbations
evaluated at a pivot scale $k_0$, and $n_s$ is the scalar spectral
index \cite{arXiv:1001.4635}.

With the aim to make forecasts on the ability of future redshift
galaxy surveys to constrain neutrino
features, adopting different spectroscopic
approaches (as discussed in \S \ref{spec_methods}),
in this work we consider separately the effect on $P_{\rm obs}$
of the total neutrino mass $M_\nu$
and the number of relativistic degrees of freedom $N_{\rm eff}$, exploiting
information from both the galaxy power spectrum shape and BAO distance
indicators. In \S \ref{growth&FoG} we will analyse also the impact 
on neutrino mass constraints due to the inclusion
of both growth--information and a Gaussian damping due to
random peculiar velocities.

In each redshift shell, with size $\Delta z=0.1$ and centred
at redshift $z_i$, we choose the following set of parameters to
describe $P_{\rm obs}(k_{{\rm ref}\perp},k_{{\rm ref}\parallel},z)$: 
\begin{equation}
\left\{H(z_i), D_A(z_i), \bar{G}(z_i), \beta(z_i,k), 
P_{shot}^i, \omega_m, \omega_b, \zeta, n_s, h\right\}, 
\end{equation}
where  $\zeta=N_{\rm eff}$ or
$\zeta=\omega_\nu\equiv\Omega_\nu h^2$ (depending on the assumed
fiducial cosmology, see \S \ref{Fiducial cosmologies}), 
$\omega_m=\Omega_m h^2$, $\omega_b=\Omega_b h^2$,
where $h$ is given
by $H_0=100 h$ km s${}^{-1}$ Mpc${}^{-1}$ , $H_0$ being the Hubble
constant. $\Omega_m$, $\Omega_\nu=M_\nu/(h^2 93.8){\rm eV}$, and $\Omega_b$
are respectively the total matter,
massive neutrino, and baryon 
present-day energy densities, 
in units of the critical energy density of the Universe.
Finally,  since $G(z)$, $b(z)$, and the power spectrum
normalisation $P_0$ are completely degenerate, we have introduced the
quantity $\bar{G}(z_i)=(P_0)^{0.5} b(z_i) G(z_i)/G(z_0)$ \cite{0710.3885}.

In the limit where the survey volume is much larger than the scale of 
any features in $P_{\rm obs}(k)$, it has been shown \cite{Tegmark97}
that it is possible to redefine $x_n$ to be not the density fluctuation
in the $n^{th}$ spatial volume element, but the
average power measured with the FKP method \cite{FKP} in a thin shell 
of radius $k_n$ in Fourier space. Under these assumptions
the redshift survey Fisher matrix can be approximated as \cite{Tegmark,Tegmark97}
\begin{eqnarray}
F_{ij}^{\rm LSS}&=&\int_{\vec{k}_{\rm min}} ^ {\vec{k}_{\rm max}} \frac{\partial
  \ln P_{\rm obs}(\vec{k})}{\partial p_i} \frac{\partial \ln P_{\rm
    obs}(\vec{k})}{\partial p_j} \Veff(\vec{k})
\frac{d\vec{k}}{2(2 \pi)^3}\\ \nonumber
&=&\int_{-1}^{1} \int_{k_{\rm min}}^{\kmax}\frac{\partial \ln
  P_{\rm obs}(k,\mu)}{\partial p_i} \frac{\partial \ln P_{\rm obs}(k,\mu)}{\partial p_j} 
\Veff(k,\mu) \frac{2\pi k^2 dk d\mu}{2(2\pi)^3}
\label{Fisher}                 
\end{eqnarray}
where the derivatives are evaluated at the parameter values $p_i$
of the fiducial model,
and $\Veff$ is the effective volume of the survey:
\begin{eqnarray}
\Veff(k,\mu) =
\left [ \frac{{n_{\rm g}}P_{\rm g}(k,\mu)}{{n_{\rm g}}P_{\rm g}(k,\mu)+1} \right ]^2 \Vsur,
\label{V_eff}                                                                                        
\end{eqnarray}
where we have assumed that the comoving number density
$n_{\rm g}$ is constant in position.
Due to azimuthal symmetry around the line of sight,
the three-dimensional galaxy redshift power spectrum
$P_{\rm obs}(\vec{k})$ depends only on $k$ and $\mu$, i.e. is reduced
to two dimensions by symmetry \cite{SE03}.

To minimise nonlinear effects, we restrict wave-numbers to the 
quasi-linear regime, so that $\kmax$ is given
by requiring that the variance of matter fluctuations in a sphere of
radius $R$ is $\sigma^2(R)=0.25$ for $R=\pi/(2\kmax)$. This gives 
$\kmax\simeq 0.1 \ihMpc$ at $z=0$ and $\kmax\simeq 0.2 \ihMpc$
at $z=1$, well within the quasi-linear regime. In addition, we
impose a uniform upper limit of $\kmax \leq 0.2\ihMpc$ (i.e. $\kmax=0.2
\ihMpc$ at $z>1$), to ensure that we are only considering the
conservative linear regime, essentially unaffected by nonlinear effects.
In each bin we adopt $k_{\rm min}= 10^{-4} h$/Mpc, and we have
verified that changing the survey maximum scale $k_{\rm min}$ with
the shell volume has almost no effect on the results.

For the moment, we do not include information from the amplitude $\bar{G}(z_i)$ and
the redshift space distortions $\beta(z_i,k)$, so we marginalise over
these parameters\footnote{In this case,
we make derivatives of $P_{\rm obs}(k,z)$                                  
with respect to $\beta(z_i,k)$. These derivatives are scale dependent,                                
independently on the scale-dependence of $\beta$. Then we integrate                                
over $k$, as written Eq.~(2.8); in this way we are left with the                                
$\beta$ redshift-dependence alone. Finally, we marginalise over it.}
and also over $P_{shot}^i$. Then we
project $\bfp=\{H(z_i), D_A(z_i), \omega_m, \omega_b, \zeta, n_s, h\}$
into the final sets $\bfq$ of cosmological parameters described in 
\S \ref{Fiducial cosmologies} \cite{Wang06,Wang08a}. 
In this way we adopt the so-called  ``full $P(k)$--method, marginalised
over growth--information'' \cite{1006.3517}, and,
to change from one set of parameters to another, we use \cite{Wang06}
\begin{equation}
F_{\alpha \beta}^{\rm LSS}= \sum_{ij} \frac{\partial p_i}{\partial q_{\alpha}}\,
F_{ij}^{\rm LSS}\, \frac{\partial p_j}{\partial q_{\beta}},
\label{eq:Fisherconv}
\end{equation}
where $F_{\alpha \beta}^{\rm LSS}$ is the survey 
Fisher matrix for the set of parameters $\bfq$, and 
$F_{ij}^{\rm LSS}$ is the survey Fisher matrix for the set of equivalent 
parameters $\bfp$.

We derive neutrino constraints with and without cosmic microwave
background (CMB) priors; to this end we use the specifications of the 
Planck\footnote{www.rssd.esa.int/index.php?project=planck} satellite. As
explained in Appendix \ref{sec:Planck}, in order to describe CMB 
temperature and polarisation power spectra, we
choose the parameter set $\vec{\theta}= \{\omega_m, \omega_b, \zeta,
100\theta_S,\ln (10^{10}\Delta^2_{\cal R}(k_0)), n_S, \tau\}$, 
where $\theta_S$ is the
angular size of the sound horizon at last scattering, and
$\tau$ is the optical depth due to reionisation. After marginalisation
over the optical depth, we propagate the Planck CMB Fisher matrix $F_{ij}^{\rm CMB}$
into the final sets of parameters $\bfq$,
by using the appropriate Jacobian for the involved parameter
transformation.

The 1--$\sigma$ error on $q_\alpha$ marginalised over the
other parameters is
$\sigma(q_\alpha)=\sqrt{({F}^{-1})_{\alpha\alpha}}$, 
where ${F}^{-1}$ is the inverse of the Fisher matrix.
We then consider constraints in a two-parameter
subspace, marginalising over the remaining parameters, in order 
to study the covariance between $N_{\rm eff}$ or $M_\nu$ and the other
cosmological parameters, respectively.

Furthermore, to quantify the level of degeneracy between
the different parameters, we estimate the so-called
correlation coefficients, given by
\begin{equation}
r\equiv \frac{({F}^{-1})_{p_\alpha p_\beta}}
{\sqrt{({F}^{-1})_{p_\alpha p_\alpha}({F}^{-1})_{p_\beta p_\beta}}},
\label{correlation}
\end{equation}
where $p_\alpha$ denotes one of the model parameters.  When the
coefficient $|r|=1$, the two parameters are totally degenerate, while
$r=0$ means they are uncorrelated.

We evaluate $\sigma(q_\alpha)$ and $r$ both from survey data $F^{\rm
  LSS}_{\alpha\beta}$, and from the combined CMB+LSS data 
$F_{\alpha\beta} =F^{\rm LSS}_{\alpha\beta}+F^{\rm CMB}_{\alpha\beta}$.

\section{Surveys: two spectroscopic strategies}
\label{spec_methods}
In this work we forecast neutrino constraints
using two different spectroscopic approaches. In particular,
as mentioned in \S \ref{Intro}, we consider 
the cases relative to two space mission concepts under study:
\begin{itemize}
\item
a EUCLID-like survey of H$\alpha$ emission line galaxies, based
on slitless spectroscopy of the sky. We adopt the empirical redshift
distribution of H$\alpha$ emission line galaxies derived by
\cite{Geach10} from observed H$\alpha$ luminosity functions,
and the bias function derived by \cite{Orsi10}
using a galaxy formation simulation. In particular, we choose
a flux limit of 4$\times 10^{-16}$erg$\,$s$^{-1}$cm$^{-1}$, a survey
area of 20,000 deg$^2$, a redshift success rate $e=0.5$, a redshift accuracy
of $\sigma_z/(1+z)\le 0.001$, and a redshift range $0.5\leq z \leq
2.1$.
The total number of galaxies with redshift errors $\sigma_z/(1+z)\le
0.001$ from a slitless survey is well approximated by
\cite{1006.3517}
\begin{equation}
\frac{N_{gal}}{10^6}=276.74\,\frac{[area]}{20000}\,
 \frac{e}{0.5} \,\left(\bar{f}\right)^{-0.9
 \left(\bar{f}\right)^{0.14}},
\end{equation}
where $\bar{f}\equiv f/[10^{-16}$erg$\,$s$^{-1}$cm$^{-2}$]\footnote{We note
that this case is similar also to JDEM and ADEPT.}.
For this type of space-based slitless redshift survey we add
in our forecasts also information from the ongoing Sloan Digital Sky Survey III (SDSS-III) Baryon
Oscillation Spectroscopic Survey (BOSS)\footnote{http://www.sdss3.org/surveys/boss.php} 
of luminous red galaxies (LRG). For this galaxy survey we assume that the
LRG redshifts are measured over $0.1<z<0.5$ \footnote{For BOSS the actual redshift range is
$0.1<z<0.7$, but we do not take into account the shell $0.5<z<0.7$
in order to avoid overlapping in $z$ between the space-- and
ground--based redshift surveys discussed in this work.} with
standard deviation $\sigma_z /(1+z)=0.001$, for a galaxy population
with a fixed number density of $n=3\times 10^{-4} h^3$Mpc$^{-3}$, and a
fixed linear bias of $b=1.7$ \cite{Reid et al 2010}, over a survey area of 10,000 deg$^2$.
\item
a H-band magnitude limited survey of randomly
sampled galaxies enabled by multi-slit spectroscopy (e.g., SPACE
\cite{Cimatti09}, JEDI \cite{Wang04,Crotts05,Cheng06}, and possibly WFIRST
in the future). To predict galaxy densities for such
surveys we use the empirical galaxy redshift distribution compiled by
Zamorani et al. from existing data \cite{Laureijs09}, and we use
predictions of galaxy bias from galaxy formation simulations
\cite{Orsi10}. We consider multi-slit surveys with a limiting magnitude
of $H_{AB}$=22, a redshift success rate of 90\%, a sampling rate of
35\%, a survey area of 20,000 deg$^2$, a redshift accuracy
of $\sigma_z/(1+z)\le 0.001$, and a redshift range $0.1\leq z \leq
2.1$.
The total number of galaxies with redshift errors $\sigma_z/(1+z)\le
0.001$ from a multi-slit survey is well approximated by
\cite{1006.3517}
\begin{equation}
\frac{N_{gal}}{10^6}=\left[192.21+197.03\,(H_{AB}-22)^{1.3}\right]
\,\frac{[area]}{20000}\,\frac{e}{0.9\times 0.35}.
\end{equation}
\end{itemize}
Note that BOSS data are not
added to the multi-slit galaxy redshift survey, since the latter has redshift
ranges that extend to $z\sim 0.1$ \cite{Cimatti09,Laureijs09}.
The case is different for H$\alpha$ flux selected galaxies observed
from space, since a wavelength range between 1 and 2 $\mu$m
naturally imposes a redshift range $0.52<z<2.05$ in
which H$\alpha$ will be visible \cite{Laureijs09}.

Furthermore, the bias functions for H$\alpha$ flux and H-band magnitude selected
galaxies increase with redshift, with the former being less strongly
biased than the latter \cite{Orsi10}. In fact, 
the H-band traces massive structures (similar to selecting              
galaxies in the K-band), which makes them strongly biased. Star forming 
galaxies (which are selected by H$\alpha$ flux), on the other hand, appear 
to avoid the cores of clusters and populate the filaments of the dark-matter 
structure, making them less biased than H-band galaxies \cite{Orsi10}. 

In conclusion, multi-slit surveys allow accurate redshift measurement for a larger number
of galaxies (and these galaxies are more biased tracers of large-scale structure
than star-forming galaxies), and over a greater redshift range (extending
to $z\sim 0.1$) than slitless surveys. This can improve the
constraints on neutrino masses, and, moreover, 
the ability to split a galaxy catalogue into red and blue galaxies could
provide an important diagnostic test of potential systematic
errors when measuring neutrino masses \cite{1006.2825}.
However, multi-slit surveys have substantially stronger requirements in
instrumentation and mission implementation \cite{Cimatti09}. 

\section{Fiducial cosmologies: $M_\nu$ and $N_{\rm eff}$}
\label{Fiducial cosmologies}
The Fisher matrix approach propagates errors of galaxy
power spectrum measurements Eq.~(\ref{eqn:pkerror}) into errors of the cosmological
parameters which characterise the underlying fiducial cosmology.
According to the latest observations (e.g. \cite{Komatsuetal2010} and
refs. therein), we assume the fiducial cosmological model
adopted in the Euclid Assessment Study Report \cite{Laureijs09}
with the exception that we normalise to the amplitude of the
primordial curvature perturbations $\Delta^2_{\cal  R}(k_0)$ instead
of $\sigma_8$:
$\Omega_m=0.25$, $h=0.7$,  
$\Delta^2_{\cal  R}(k_0)=2.45\times 10^{-9}$,
$\Omega_b=0.0445$, $n_s=1${}\footnote{CMB constraints give $n_s\sim
  0.96$, anyway we have verified that adopting the latter value of
  $n_s$ changes neutrino mass constraints by $\sim$5\%, and the dark-energy equation of state parameter errors by $\sim$10\%, at the maximum.}.
We consider neither primordial gravitational waves nor a scale
dependent component of the scalar spectral index, and 
assume the matter energy density $\Omega_{m}$ to include the neutrino
contribution when neutrinos are non-relativistic
\begin{equation}
\Omega_{m} = \Omega_{c} + \Omega_{b} + \Omega_{\nu}~,
\label{Omega_m}
\end{equation}

Moreover, we also assume dark-energy to be a cosmic
fluid described by a \emph{redshift dependent} equation of state 
\begin{equation}
\label{eq:wz0}
\w_{de}(z)=\frac{p_{de}(z)}{\rho_{de}(z)}
\end{equation}
where $p_{de}$ and $\rho_{de}$ represent respectively the pressure and
energy density of the dark-energy fluid.

This in turn yields a redshift dependent dark-energy density
\begin{equation}
\label{eq:rhox}
\rho_{de}(z)=\rho_{de}(0) \exp \left[ 3 \int_0^z
  \frac{1+\w(z')}{1+z'}dz'\right],
\end{equation}
for which we take a fiducial value $\Omega_{de}=0.75$.
Finally, to compute our forecasts on dark-energy parameters, we adopt
the widely used linear dark-energy equation of state
\cite{Chev01,Linder03}
\begin{equation}
  w_{de}(a)=w_0+(1-a)w_a,
\label{eq:w0wa}
\end{equation} 
where $a\equiv 1/(1+z)$ is the scale factor normalised to unity at
present, and we assume as fiducial values $w_0=-0.95$ and $w_a=0$, which
lie well within the current $95$\% C.L. limits. In what follows, the
dark-energy figure of merit (FoM) will be computed in terms of the
conventional FoM for ($w_0$,$w_a$) as proposed by DETF
\cite{albrecht09} to compare dark-energy surveys.

We will constraint the following final set of eight parameters
\begin{equation}
\bfq=\left\{\Omega_m,\Omega_{de},h,\Delta^2_{\cal
      R}(k_0),\Omega_b,w_0,w_a,n_s\right\},
\label{q_set}
\end{equation}
which constitutes what we call our ``base'' parameters\footnote{Let us 
 notice the adopted full $P(k)$--method marginalised over
 growth--information 
does not give any constraint on $\Delta^2_{\cal
    R}(k_0)$, 
since the normalisation of the galaxy power spectrum is                                  
  marginalised over. Therefore, the $\Delta^2_{\cal R}(k_0)$--errors 
shown in this work are forecasts from Planck alone.}, to which we add the $\zeta$ parameter as we
explain below\footnote{Finally, in \S \ref{growth&FoG} including
  growth--information we will constrain the matter spectrum normalisation $\sigma_8$
  together with the parameters in Eq.~(\ref{q_set})}.

In what follows, we will consider six different fiducial models
matching our ``base'' fiducial $\Lambda$CDM cosmology, in which we adopt
the same fiducial values for the eight ``base'' parameters.
Then we specify the assumed cosmological models:
\begin{itemize} 
\item 
The first model assumes a {\bf $N_{\rm eff}$--cosmology}, 
where neutrinos are effectively massless  but the
the number of relativistic species $\zeta\equiv N_{\rm eff}$ can deviate 
from the standard value $N_{\rm eff}=3.04$. In this case the fiducial value 
$N_{\rm eff}|_{\rm fid}=3.04$ is chosen, fixing
$M_{\nu}={\rm const}=0$\footnote{A different choice  for the fiducial 
$M_{\nu}$, would not affect the forecasted errors on $N_{\rm eff}$.} 
\cite{mena}.
$N_{\rm eff}$ is given by the energy density associated to the total radiation
\begin{equation}
\Omega_{r} = \Omega_{\gamma} \left(1+0.2271 N_{\rm eff}\right)~,
\label{Or}
\end{equation}
where $\Omega_{\gamma}=2.469\times 10^{-5} h^{-2}$ is the
present-day
photon energy density parameter for $T_{\rm cmb}=2.725$~K
\cite{Komatsuetal08}. 

\item 
The remaining models assume a {\bf $M_\nu$--cosmology}, where $N_{\rm eff}$ is fixed at the
  fiducial value and $\zeta\equiv M_\nu$ is allowed to vary accordingly to the
  assumed fiducial
  neutrino mass spectrum \cite{1003.5918}. In this case, we
choose the following five fiducial values for the total neutrino mass
consistent with current data \cite{0910.0008,Komatsuetal2010}:
\begin{eqnarray}
M_\nu|_{\rm fid}=\Bigg\{
\begin{array}{cc} 
0.3, 0.2 \;{\rm eV} & \mbox{for degenerate spectrum}\\
0.125 \;{\rm eV} & \mbox{for inverted hierarchy}\\
0.125, 0.05 \;{\rm eV} & \mbox{for normal hierarchy}
\end{array}
\end{eqnarray}

\end{itemize}

The motivation for considering different fiducial models is that the dependence of 
the power spectrum on $M_{\nu}$ is nonlinear and thus the size of the forecasted
error bar on $M_{\nu}$ depends on the fiducial value chosen.

At the CMB level, if neutrinos are still relativistic
at the decoupling epoch, $z \simeq 1090$, i.e. if the
mass of the heaviest neutrino specie is $m_\nu<0.58$~eV,
massive neutrinos do not affect the CMB power spectra, except through
the gravitational lensing effect \cite{Komatsuetal08,1009.3204,0511735}, and, as a
consequence, the dark-energy equation of state $w_{de}$ is not degenerate with the
neutrino mass. However, the limit on the the sum of neutrino
masses degrades significantly when the dark-energy
equation of state is a function of redshift as we assume in the present work,
since dark-energy and massive neutrinos both affect the growth rate of
structures \cite{Saito1}. However, as we will show in the next section,
the combination of CMB and LSS probes reduces or even breaks these degeneracies. 
The same does not happen for the number of relativistic species
$N_{\rm eff}$. Moreover, the degeneracies between $M_\nu$ and the
other cosmological parameters can increase as the number of free
parameters of the model increases, which could potentially bias the
results for large $k$ values \cite{1006.2825}. As evident from
Eqs.~(\ref{eq:Pobs})-(\ref{eq:Pg}),  the model adopted in this paper falls within
the two bias parameter models discussed in Ref.~\cite{1006.2825}, which
seem to mimic accurately the broad features of
galaxy bias and redshift-space distortions from
SDSS, leading to consistent constraints in the presence of massive neutrinos.
\begin{table*}
\caption{Parameter 1-$\sigma$ errors for slitless spectroscopy}
\setlength{\tabcolsep}{1pt}
\begin{tabular}{l c c c c c c c}
\hline
\hline
{}&{}&{}&\footnotesize{{\small slitless+BOSS}}&{}&{}&{}
\\
\hline
\footnotesize{{\small {\rm fiducial}$\to$ }}&
\footnotesize{{\small $M_\nu$=0.3 eV}}${}^a$& 
\footnotesize{{\small $M_\nu$=0.2 eV}}${}^a$ & 
\footnotesize{{\small $M_\nu$=0.125 eV}}${}^b$ &
\footnotesize{{\small $M_\nu$=0.125 eV}}${}^c$ & 
\footnotesize{{\small $M_\nu$=0.05 eV}}${}^b$ &
\footnotesize{{\small $N_{\rm eff}$=3.04}}${}^d$
\\
\hline
$\Omega_m$ & $0.0140$ & $0.0137$ & $0.0139$ & $0.0138$ & $0.0137$ & $0.0124$
\\
$\Omega_{de}$ & $0.0260$ & $0.0265$ & $0.0258$ & $0.0257$ & $0.0253$ &$0.0256$
\\
$\Omega_b$ & $0.0032$ & $0.0031$ & $0.0032$ & $0.0032$ & $0.0032$& $0.0034$
\\
$h$  & $0.0116$ & $0.0112$ & $0.0113$ & $0.0114$ & $0.0113$ & $0.0137$
\\
$M_\nu$ & $0.1459$ & $0.1461$ & $0.1795$ & $0.1435$ & $0.1428$ & $--$
\\
$N_{\rm eff}$ & $--$ & $--$ & $--$ & $--$ & $--$ & $0.5435$
\\
$ns$ & $0.0233$ & $0.0225$ & $0.0314$ & $0.0228$ & $0.0220$ & $0.0326$
\\
$w_0$ & $0.0815$ & $0.0837$ & $0.0812$ & $0.0808$ & $0.0801$ & $0.0807$
\\
$w_a$ & $0.3461$ & $0.3573$ & $0.3450$ & $0.3440$ & $0.3386$ & $0.3324$
\\
FoM & $51.75$ & $48.44$ & $52.44$ & $52.68$ & $54.25$ & $55.23$\\
\hline
\hline
{}&{}&{}&\footnotesize{{\small slitless+BOSS+Planck}}&{}&{}&{}
\\
\hline
$\Omega_m$ & $0.0034$ & $0.0035$ & $0.0033$ & $0.0035$ & $0.0035$ &$0.0031$
\\
$\Omega_{de}$ & $0.0064$ & $0.0064$ & $0.0062$ & $0.0064$ & $0.0065$ &$0.0064$
\\
$\Omega_b$ & $0.0006$ & $0.0056$ & $0.0006$ & $0.0005$ & $0.0005$& $0.0005$
\\
$h$  & $0.0043$ & $0.0043$ & $0.0043$ & $0.0043$ & $0.0042$ & $0.0046$
\\
$M_\nu$ & $0.0347$ & $0.0433$ & $0.0311$ & $0.0441$ & $0.0526$ & $--$
\\
$N_{\rm eff}$ & $--$ & $--$ & $--$ & $--$ & $--$ & $0.0865$
\\
$ns$ & $0.0022$ & $0.0022$ & $0.0022$ & $0.0022$ & $0.0022$ & $0.0041$
\\
$w_0$ & $0.0732$ & $0.0721$ & $0.0716$ & $0.0715$ & $0.0709$ & $0.0705$
\\
$w_a$ & $0.1760$ & $0.1742$ & $0.1713$ & $0.1725$ & $0.1722$ & $0.1664$
\\
$\Delta^2_{\cal R}(k_0)$ & $0.0250$ & $0.0226$ & $0.0226$ &$0.0227$ & $0.0244$ & $0.0227$
\\
FoM & $245.07$ & $242.80$ & $259.32$ & $247.96$ & $240.31$ & $294.15$
\\
\hline
\end{tabular}
\begin{flushleft}
${}^a$\footnotesize{for degenerate spectrum: $m_1\approx m_2\approx m_3$};
${}^b$\footnotesize{for normal hierarchy: $m_3\neq 0$, $m_1\approx m_2\approx 0$}\\
${}^c$\footnotesize{for inverted hierarchy: $m_1\approx m_2$, $m_3\approx 0$};
${}^d$\footnotesize{fiducial cosmology with massless neutrinos}
\end{flushleft}
\label{slitless_errors}
\end{table*}
\begin{table*}
\caption{Parameter 1-$\sigma$ errors for multi-slit spectroscopy}
\setlength{\tabcolsep}{3pt}
\begin{tabular}{l c c c c c c c}
\hline
\hline
{}&{}&{}&\footnotesize{{\small multi-slit}}&{}&{}&{}
\\
\hline
\footnotesize{{\small {\rm fiducial}$\to$ }}&
\footnotesize{{\small $M_\nu$=0.3 eV}}${}^a$&
\footnotesize{{\small $M_\nu$=0.2 eV}}${}^a$ &
\footnotesize{{\small $M_\nu$=0.125 eV}}${}^b$ &
\footnotesize{{\small $M_\nu$=0.125 eV}}${}^c$ &
\footnotesize{{\small $M_\nu$=0.05 eV}}${}^b$ &
\footnotesize{{\small $N_{\rm eff}$=3.04}}${}^d$
\\
\hline
$\Omega_m$ & $0.0090$ & $0.0091$ & $0.0092$ & $0.0090$ & $0.0091$ &$0.0086$
\\
$\Omega_{de}$ & $0.0179$ & $0.0178$ & $0.0177$ & $0.0177$ & $0.0176$ &$0.0185$
\\
$\Omega_b$ & $0.0020$ & $0.0020$ & $0.0020$ & $0.0020$ & $0.0020$& $0.0023$
\\
$h$  & $0.0079$ & $0.0078$ & $0.0079$ & $0.0079$ & $0.0079$ & $0.0098$
\\
$M_\nu$ & $0.1229$ & $0.1113$ & $0.1321$ & $0.1110$ & $0.1126$ & $--$
\\
$N_{\rm eff}$ & $--$ & $--$ & $--$ & $--$ & $--$ & $0.4059$
\\
$ns$ & $0.0190$ & $0.0160$ & $0.0221$ & $0.0158$ & $0.0151$ & $0.0242$
\\
$w_0$ & $0.0617$ & $0.0614$ & $0.0619$ & $0.0613$ & $0.0612$ & $0.0629$
\\
$w_a$ & $0.2399$ & $0.2399$ & $0.2411$ & $0.2400$ & $0.2389$ & $0.2427$
\\
FoM & $94.16$ & $94.78$ & $94.49$ & $95.12$ & $96.02$ & $94.09$\\
\hline
\hline
{}&{}&{}&\footnotesize{{\small multi-slit+Planck}}&{}&{}&{}
\\
\hline
$\Omega_m$ & $0.0026$ & $0.0027$ & $0.0026$ & $0.0027$ & $0.0028$ & $0.0025$
\\
$\Omega_{de}$ & $0.0050$ & $0.0051$ & $0.0050$ & $0.0051$ & $0.0053$ &$0.0054$
\\
$\Omega_b$ & $0.0004$ & $0.0004$ & $0.0004$ & $0.0004$ & $0.0004$& $0.0004$
\\
$h$  & $0.0033$ & $0.0033$ & $0.0033$ & $0.0033$ & $0.0033$ & $0.0038$
\\
$M_\nu$ & $0.0296$ & $0.0376$ & $0.0268$ & $0.0388$ & $0.0463$ & $--$
\\
$N_{\rm eff}$ & $--$ & $--$ & $--$ & $--$ & $--$ & $0.0817$
\\
$ns$ & $0.0021$ & $0.0021$ & $0.0021$ & $0.0021$ & $0.0021$ & $0.0039$
\\
$w_0$ & $0.0554$ & $0.0552$ & $0.0552$ & $0.0552$ & $0.0551$ & $0.0551$
\\
$w_a$ & $0.1307$ & $0.1311$ & $0.1294$ & $0.1309$ & $0.1320$ & $0.1274$
\\
$\Delta^2_{\cal R}(k_0)$ & $0.0247$ & $0.0242$ & $0.0224$ &$0.0224$ & $0.0237$ & $0.0226$
\\
FoM & $391.39$ & $376.55$ & $399.17$ & $377.41$ & $359.33$ & $441.78$
\\
\hline
\end{tabular}
\begin{flushleft}
${}^a$\footnotesize{for degenerate spectrum: $m_1\approx m_2\approx
  m_3$};
${}^b$\footnotesize{for normal hierarchy: $m_3\neq 0$, $m_1\approx
  m_2\approx 0$}\\
${}^c$\footnotesize{for inverted hierarchy: $m_1\approx m_2$,
  $m_3\approx 0$};
${}^d$\footnotesize{fiducial cosmology with massless neutrinos}
\end{flushleft}
\label{dmd_errors}
\end{table*}
\begin{table*}
\caption{Neutrino correlation coefficients for slitless spectroscopy}
\setlength{\tabcolsep}{1pt}
\begin{tabular}{l c c c c c c c}
\hline
\hline
{}&{}&{}&\footnotesize{{\small slitless+BOSS}}&{}&{}&{}
\\
\hline
\footnotesize{{\small {\rm fiducial}$\to$ }}&
\footnotesize{{\small $M_\nu$=0.3 eV}}${}^a$&
\footnotesize{{\small $M_\nu$=0.2 eV}}${}^a$ &
\footnotesize{{\small $M_\nu$=0.125 eV}}${}^b$ &
\footnotesize{{\small $M_\nu$=0.125 eV}}${}^c$ &
\footnotesize{{\small $M_\nu$=0.05 eV}}${}^b$ &
\footnotesize{{\small $N_{\rm eff}$=3.04}}${}^d$
\\
\hline
$\Omega_m$ & $0.601$ & $0.573$ & $0.554$ & $0.549$ & $0.544$ &$-0.316$
\\
$\Omega_{de}$ & $-0.208$ & $-0.188$ & $-0.188$ & $-0.182$ & $-0.134$ &$0.221$
\\
$\Omega_b$ & $0.323$ & $0.331$ & $0.232$ & $0.303$ & $0.249$ & $-0.439$
\\
$h$  & $0.252$ & $0.259$ & $0.165$ & $0.230$ & $0.129$ & $0.602$
\\
$M_\nu$ & $1$ & $1$ & $1$ & $1$ & $1$ & $--$
\\
$N_{\rm eff}$ & $--$ & $--$ & $--$ & $--$ & $--$ & $1$
\\
$ns$ & $0.567$ & $0.340$ & $0.717$ & $0.338$ & $0.364$ & $0.763$
\\
$w_0$ & $0.325$ & $0.310$ & $0.311$ & $0.298$ & $0.310$ & $-0.305$
\\
$w_a$ & $-0.463$ & $-0.437$ & $-0.424$ & $-0.417$ & $-0.413$ & $0.307$
\\
\hline
\hline
{}&{}&{}&\footnotesize{{\small slitless+BOSS+Planck}}&{}&{}&{}
\\
\hline
$\Omega_m$ & $0.344$ & $0.404$ & $0.331$ & $0.416$ & $0.485$ &$0.140$
\\
$\Omega_{de}$ & $-0.312$ & $-0.371$ & $-0.290$ & $-0.361$ & $-0.445$ &$0.389$
\\
$\Omega_b$ & $-0.141$ & $-0.131$ & $-0.136$ & $-0.117$ & $-0.169$& $-0.062$
\\
$h$  & $0.019$ & $0.007$ & $0.023$ & $0.006$ & $0.019$ & $0.414$
\\
$M_\nu$ & $1$ & $1$ & $1$ & $1$ & $1$ & $--$
\\
$N_{\rm eff}$ & $--$ & $--$ & $--$ & $--$ & $--$ & $1$
\\
$ns$ & $-0.089$ & $-0.071$ & $-0.034$ & $0.048$ & $0.020$ & $0.844$
\\
$w_0$ & $0.011$ & $0.025$ & $0.008$ & $0.029$ & $0.031$ & $0.001$
\\
$w_a$ & $-0.166$ & $-0.202$ & $-0.157$ & $-0.204$ & $-0.266$ & $0.074$
\\
$\Delta^2_{\cal R}(k_0)$ & $0.106$ & $0.229$ & $0.073$ &$0.105$ & $0.288$ & $0.123$
\\
\hline
\end{tabular}
\begin{flushleft}
${}^a$\footnotesize{for degenerate spectrum: $m_1\approx m_2\approx
  m_3$};
${}^b$\footnotesize{for normal hierarchy: $m_3\neq 0$, $m_1\approx
  m_2\approx 0$}\\
${}^c$\footnotesize{for inverted hierarchy: $m_1\approx m_2$,
  $m_3\approx 0$};
${}^d$\footnotesize{fiducial cosmology with massless neutrinos}
\end{flushleft}
\label{slitless_corr}
\end{table*}
\begin{table*}
\caption{Neutrino correlation coefficients for multi-slit spectroscopy}
\setlength{\tabcolsep}{3pt}
\begin{tabular}{l c c c c c c c}
\hline
\hline
{}&{}&{}&\footnotesize{{\small multi-slit}}&{}&{}&{}
\\
\hline
\footnotesize{{\small {\rm fiducial}$\to$ }}&
\footnotesize{{\small $M_\nu$=0.3 eV}}${}^a$&
\footnotesize{{\small $M_\nu$=0.2 eV}}${}^a$ &
\footnotesize{{\small $M_\nu$=0.125 eV}}${}^b$ &
\footnotesize{{\small $M_\nu$=0.125 eV}}${}^c$ &
\footnotesize{{\small $M_\nu$=0.05 eV}}${}^b$ &
\footnotesize{{\small $N_{\rm eff}$=3.04}}${}^d$
\\
\hline
$\Omega_m$ & $0.607$ & $0.578$ & $0.577$ & $0.561$ & $0.544$ &$-0.447$
\\
$\Omega_{de}$ & $-0.151$ & $-0.144$ & $-0.136$ & $-0.136$ & $-0.134$ &$0.324$
\\
$\Omega_b$ & $0.284$ & $0.292$ & $0.217$ & $0.271$ & $0.249$& $-0.551$
\\
$h$  & $0.175$ & $0.179$ & $0.110$ & $0.158$ & $0.129$ & $0.604$
\\
$M_\nu$ & $1$ & $1$ & $1$ & $1$ & $1$ & $--$
\\
$N_{\rm eff}$ & $--$ & $--$ & $--$ & $--$ & $--$ & $1$
\\
$ns$ & $0.640$ & $0.442$ & $0.758$ & $0.434$ & $0.364$ & $0.814$
\\
$w_0$ & $0.341$ & $0.326$ & $0.342$ & $0.315$ & $0.310$ & $-0.368$
\\
$w_a$ & $-0.468$ & $-0.444$ & $-0.444$ & $-0.425$ & $-0.413$ & $0.407$
\\
\hline
\hline
{}&{}&{}&\footnotesize{{\small multi-slit+Planck}}&{}&{}&{}
\\
\hline
$\Omega_m$ & $0.355$ & $0.422$ & $0.337$ & $0.437$ & $0.485$ &$0.130$
\\
$\Omega_{de}$ & $-0.319$ & $-0.384$ & $-0.292$ & $-0.375$ & $-0.445$ &$0.472$
\\
$\Omega_b$ & $-0.201$ & $-0.184$ & $-0.195$ & $-0.167$ & $-0.169$& $-0.049$
\\
$h$  & $0.052$ & $0.034$ & $0.058$ & $0.033$ & $0.019$ & $0.480$
\\
$M_\nu$ & $1$ & $1$ & $1$ & $1$ & $1$ & $--$
\\
$N_{\rm eff}$ & $--$ & $--$ & $--$ & $--$ & $--$ & $1$
\\
$ns$ & $0.015$ & $0.013$ & $0.046$ & $0.029$ & $0.020$ & $0.842$
\\
$w_0$ & $-0.011$ & $0.011$ & $-0.015$ & $0.016$ & $0.031$ & $0.005$
\\
$w_a$ & $-0.168$ & $-0.217$ & $-0.156$ & $-0.222$ & $-0.266$ & $0.080$
\\
$\Delta^2_{\cal R}(k_0)$ & $0.060$ & $0.177$ & $0.029$ &$0.068$ & $0.288$ & $0.153$
\\
\hline
\end{tabular}
\begin{flushleft}
${}^a$\footnotesize{for degenerate spectrum: $m_1\approx m_2\approx
  m_3$};
${}^b$\footnotesize{for normal hierarchy: $m_3\neq 0$, $m_1\approx
  m_2\approx 0$}\\
${}^c$\footnotesize{for inverted hierarchy: $m_1\approx m_2$,
  $m_3\approx 0$};
${}^d$\footnotesize{fiducial cosmology with massless neutrinos}
\end{flushleft}
\label{dmd_corr}
\end{table*}

\section{Results}
\label{Results}
In this Section we present the predicted 1--$\sigma$ marginalised
errors and correlations for the cosmological parameters considered in
this work, focusing on the
total neutrino mass $M_\nu$ and the number of relativistic species
$N_{\rm eff}$, and comparing the results between the two spectroscopic
strategies described in \S \ref{spec_methods}. 
We show forecasts from LSS alone and in combination with Planck priors.

In Tables~\ref{slitless_errors}-\ref{dmd_errors} we show the
marginalised errors for the six fiducial
cosmologies considered here, computed adopting slitless and multi-slit
spectroscopy, respectively; in Tables~\ref{slitless_corr}-\ref{dmd_corr}  
we report the corresponding correlation coefficients. Both LSS alone and
LSS+CMB results are reported.

\begin{figure*}
\begin{tabular}{l l}
\includegraphics[width=7.3cm]{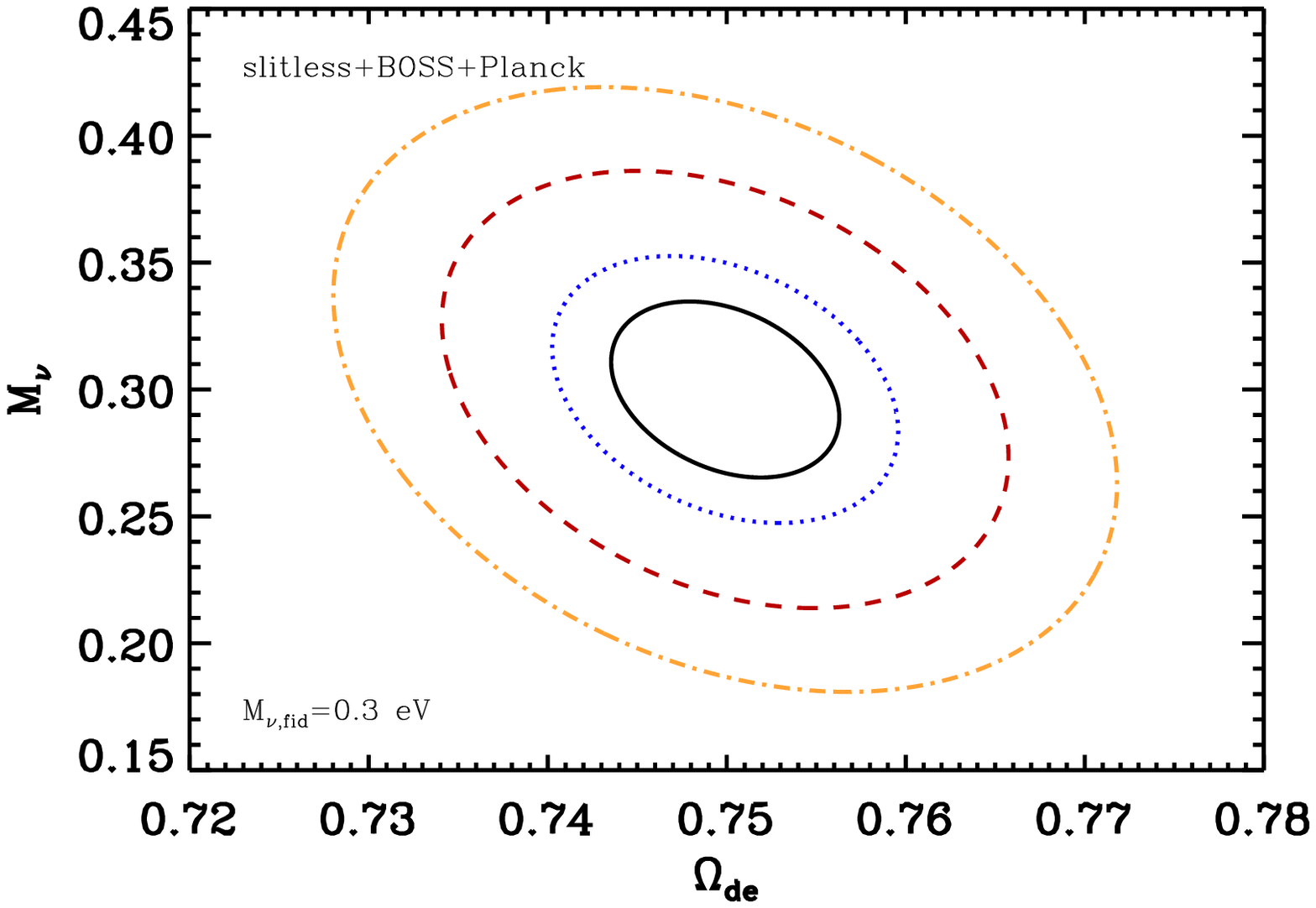}&
\includegraphics[width=7.3cm]{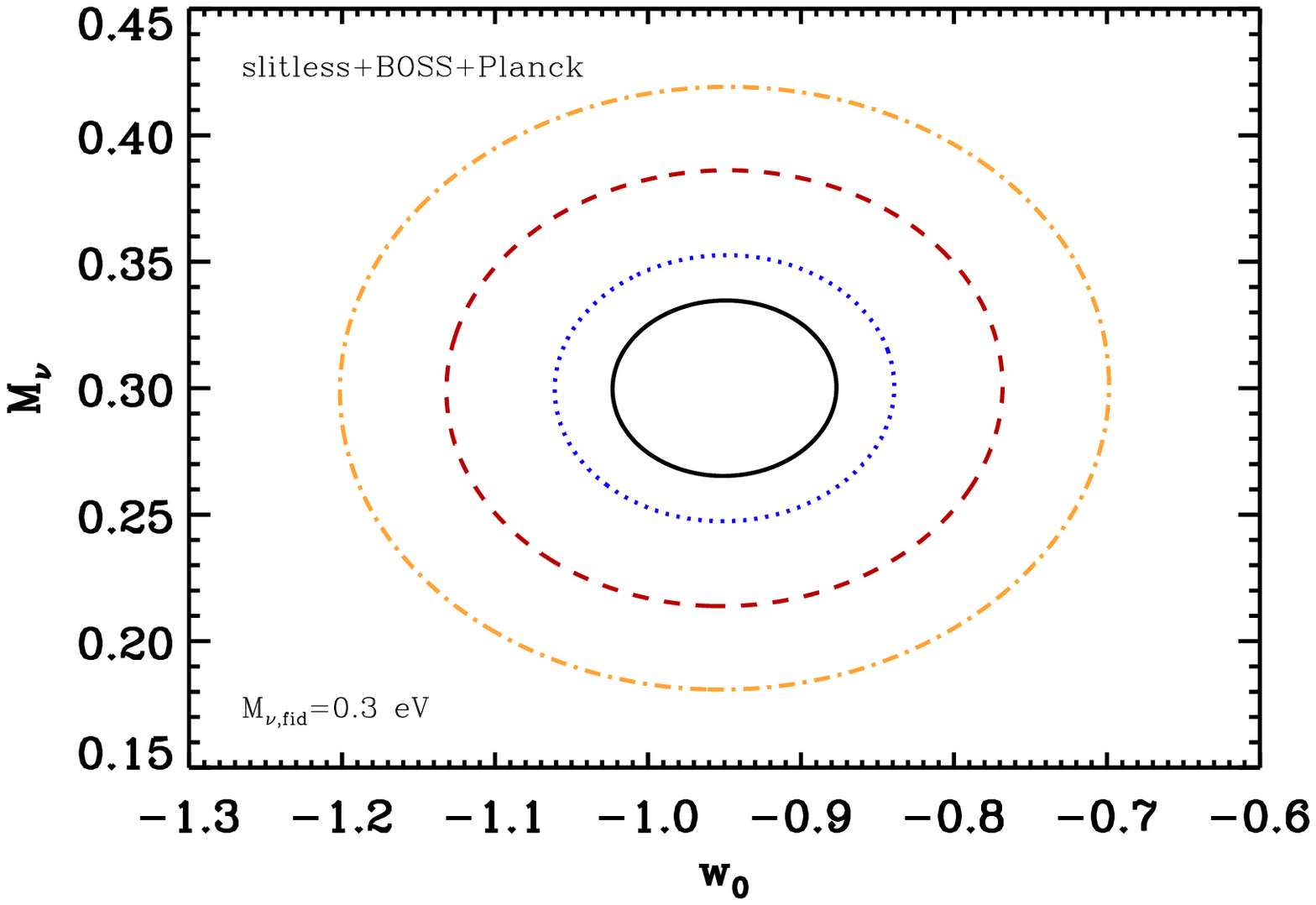}\\
\includegraphics[width=7.3cm]{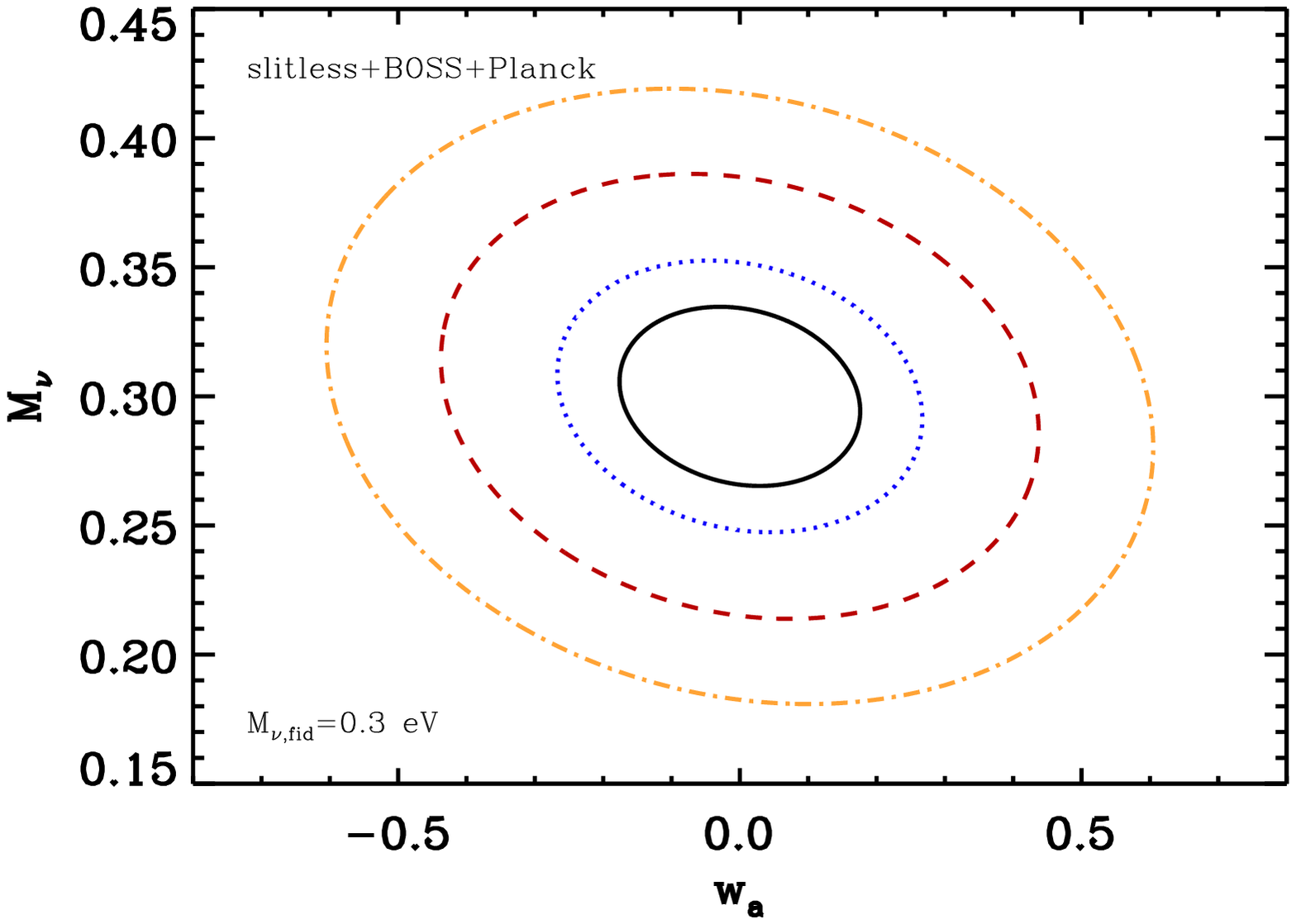}&
\includegraphics[width=7.3cm]{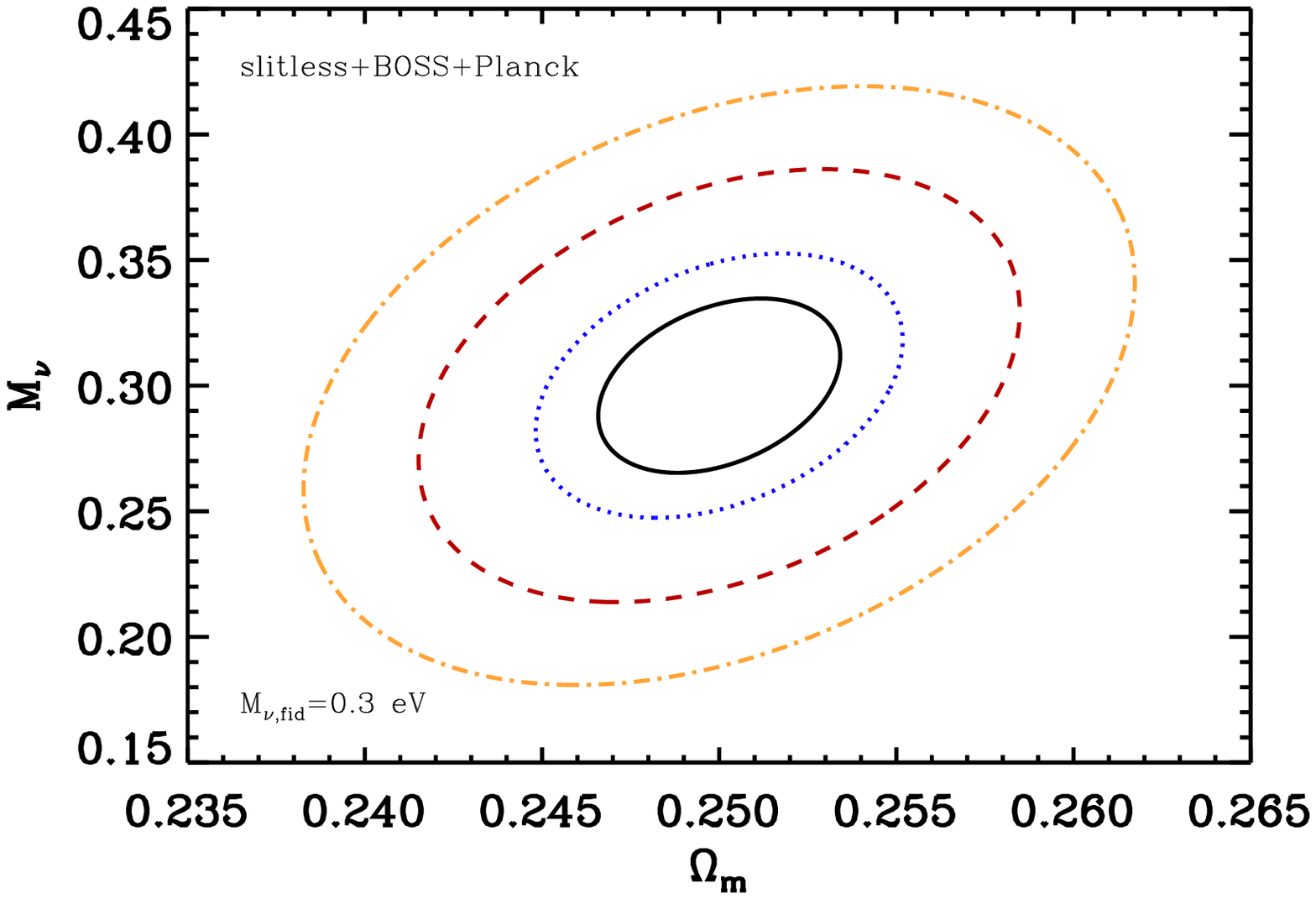}\\
\end{tabular}
\caption{2-parameter $M_\nu$-$q_\alpha$ joint contours with $q_\alpha=
  \Omega_{de}, w_0,w_a,\Omega_m$
    for the fiducial model with $M_\nu=0.3$ eV and a degenerate
    neutrino mass spectrum, obtained after combining the slitless survey data with BOSS
    data and Planck priors. The blue dotted line, the red dashed line and the orange
dot-dashed line represent the 68$\%$ C.L., 95.4$\%$ C.L. and
99.73$\%$ C.L., respectively. The black solid line shows the
1-parameter confidence level at 1--$\sigma$.}
\label{fig_Mnu03_sigmas}
\end{figure*}
\begin{figure*}
\begin{tabular}{l l}
\includegraphics[width=7.3cm]{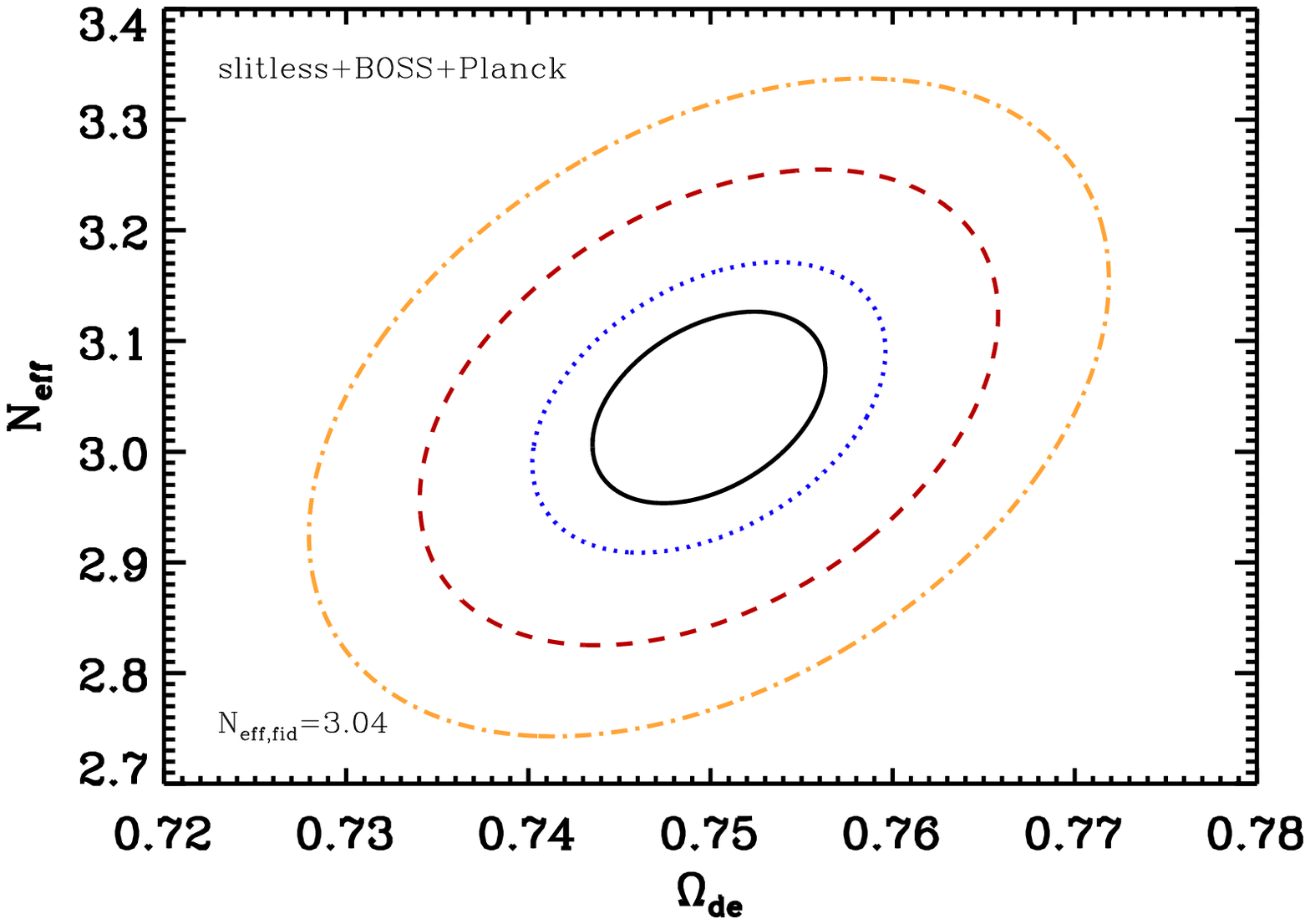}&
\includegraphics[width=7.3cm]{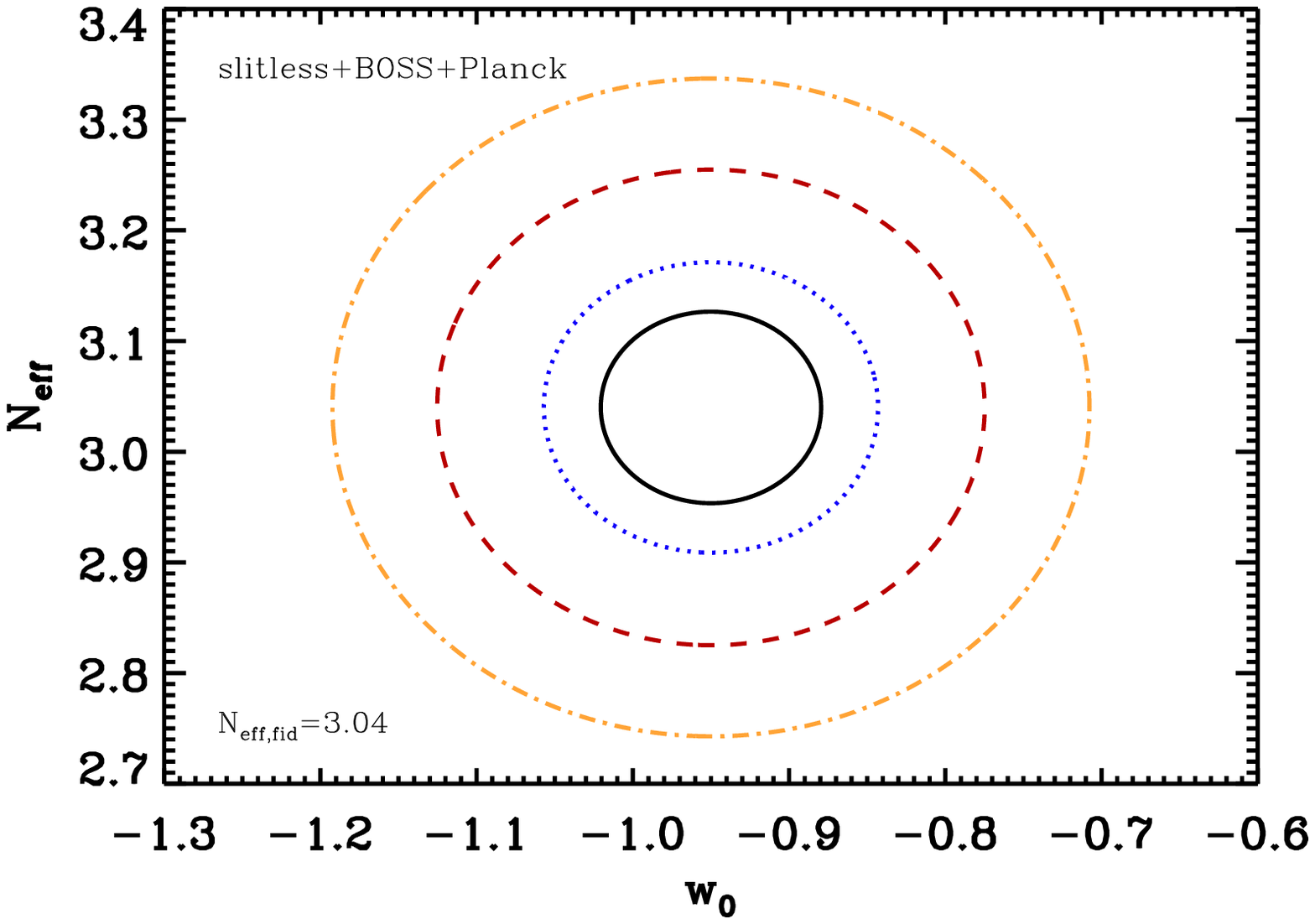}\\
\includegraphics[width=7.3cm]{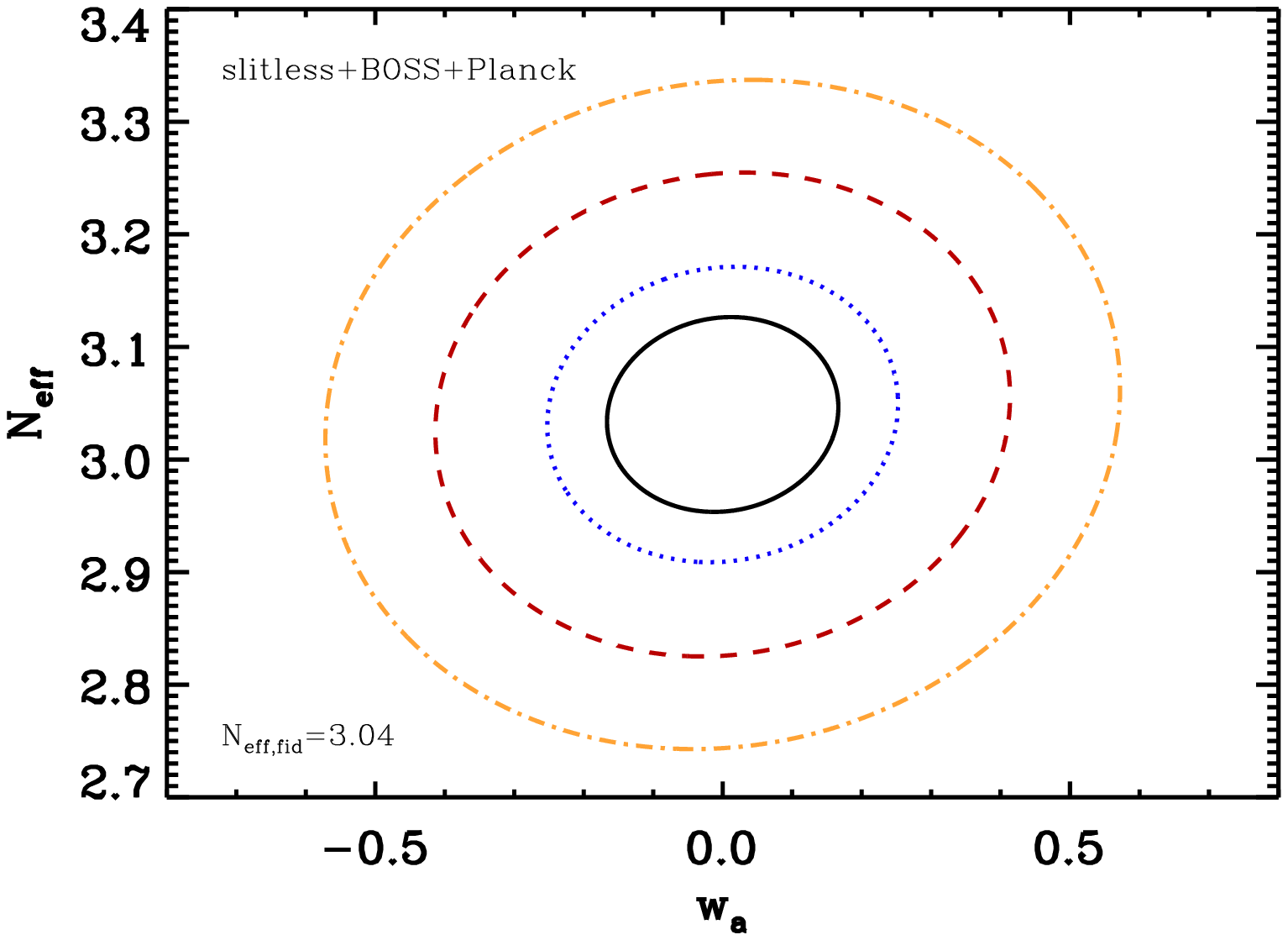}&
\includegraphics[width=7.3cm]{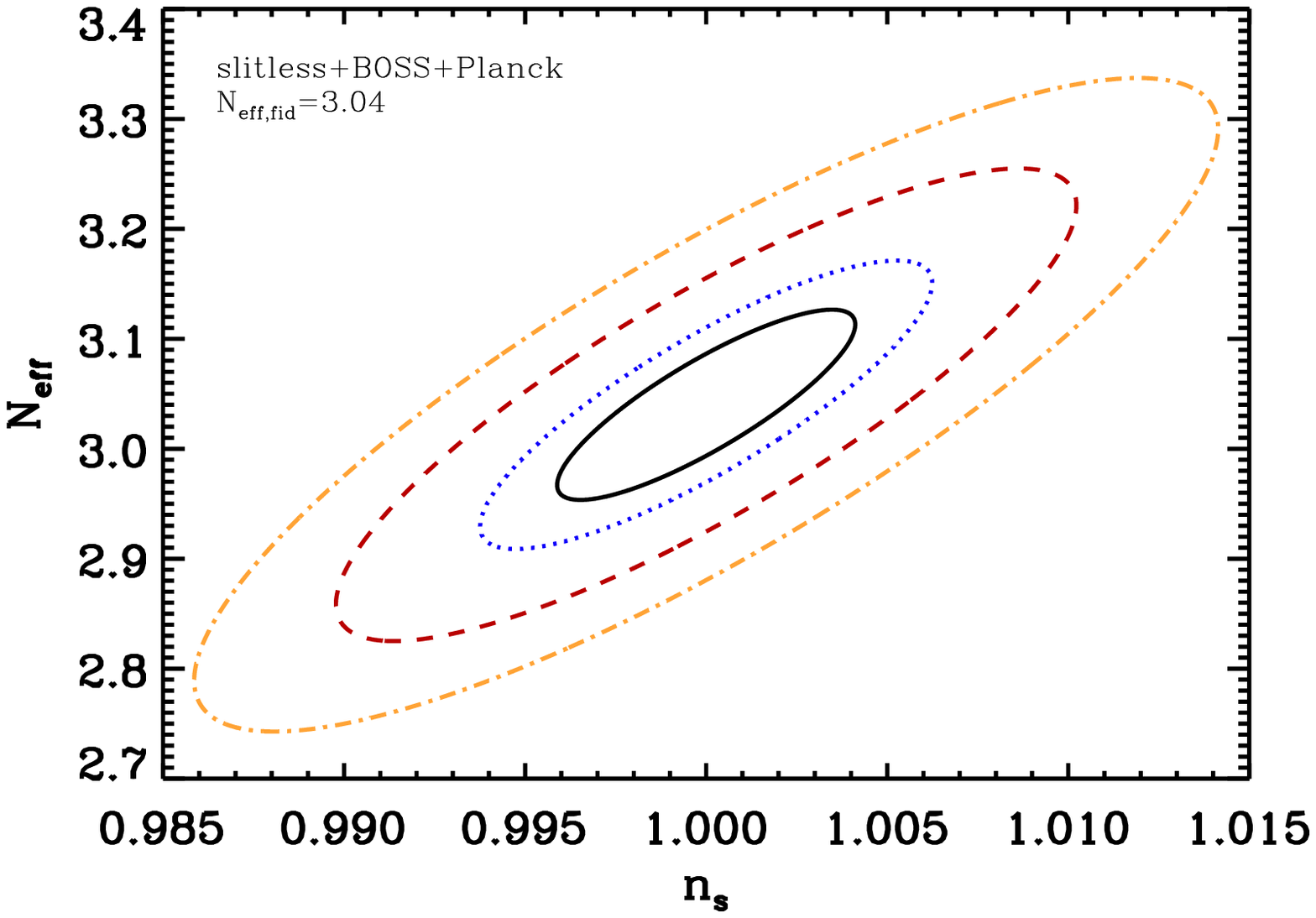}\\
\end{tabular}
\caption{2-parameter $N_{\rm eff}$-$q_\alpha$ joint contours with
  $q_\alpha=\Omega_{de}, w_0,w_a,n_s$ for                                        
  the fiducial model with extra relativistic degrees of freedom 
$N_{\rm eff}=3.04$, 
obtained after combining the slitless survey data with BOSS data and  
    Planck priors. The blue dotted line, the red dashed line and the
    orange dot-dashed line represent the 68$\%$ C.L., 95.4$\%$
    C.L. and 99.73$\%$ C.L., respectively. The black solid line shows
    the 1-parameter confidence level at 1--$\sigma$.}
\label{fig_Neff_sigmas}
\end{figure*} 
\begin{figure*}
\begin{tabular}{l l}
\includegraphics[width=7.3cm]{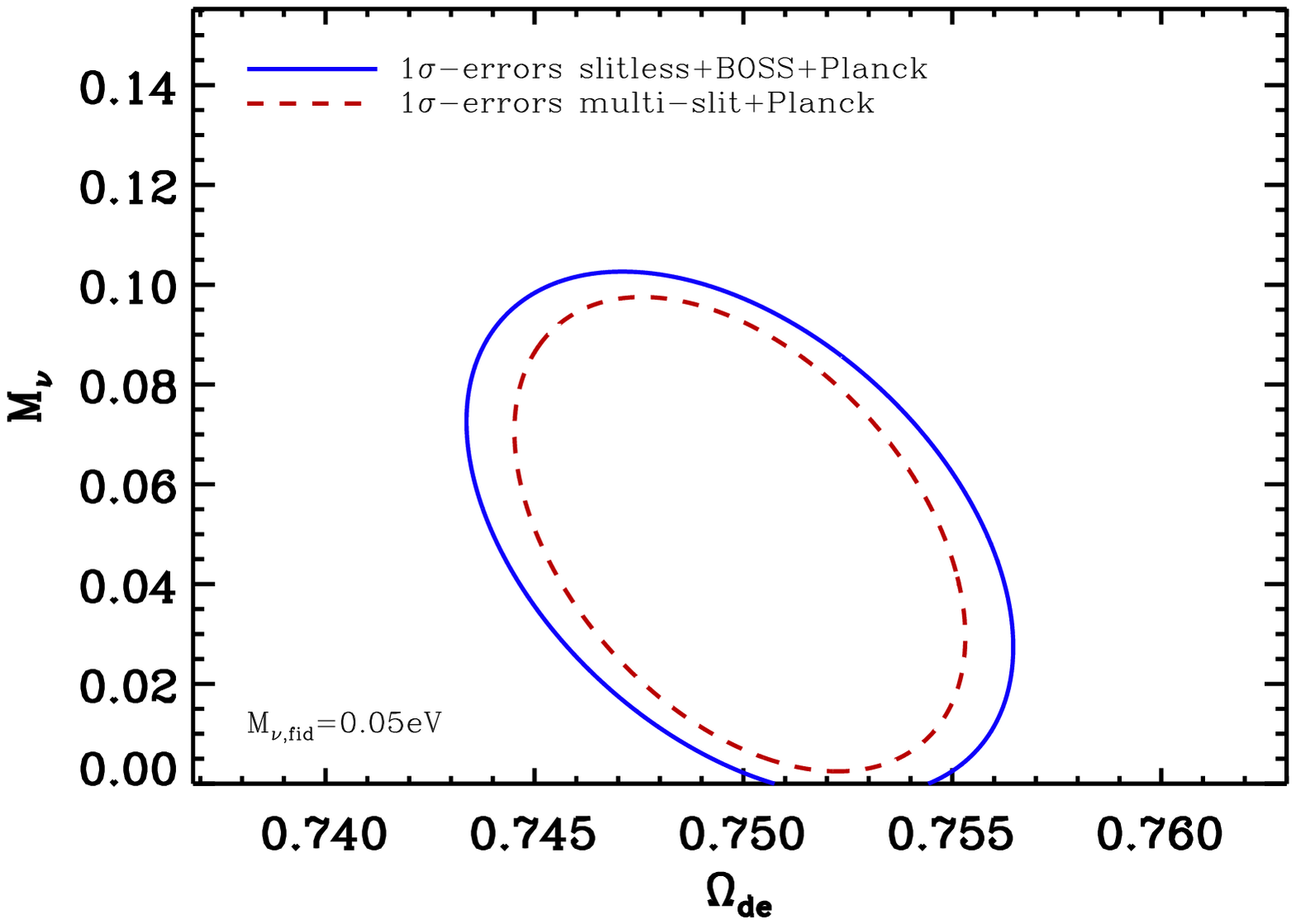}&
\includegraphics[width=7.3cm]{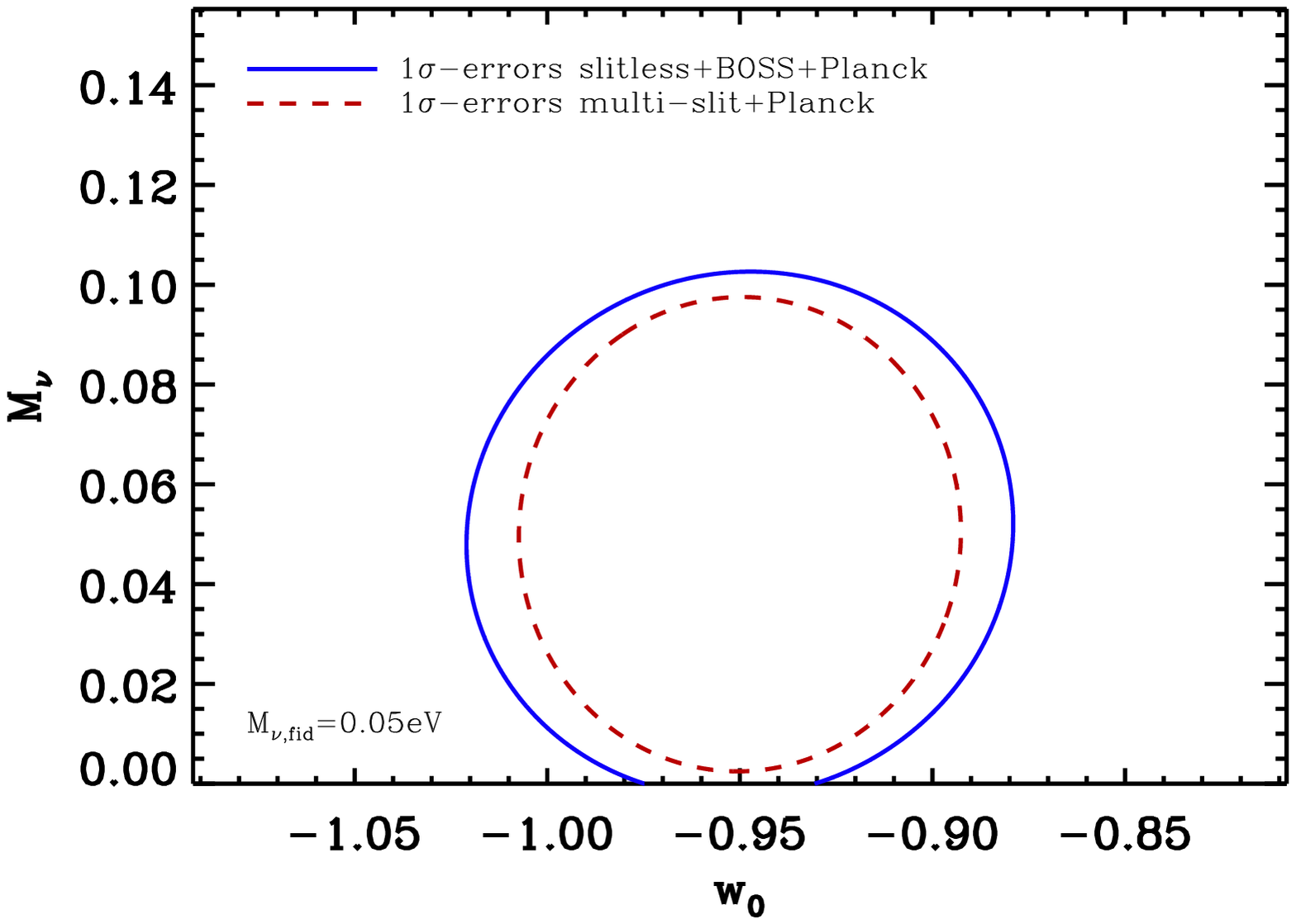}\\
\includegraphics[width=7.3cm]{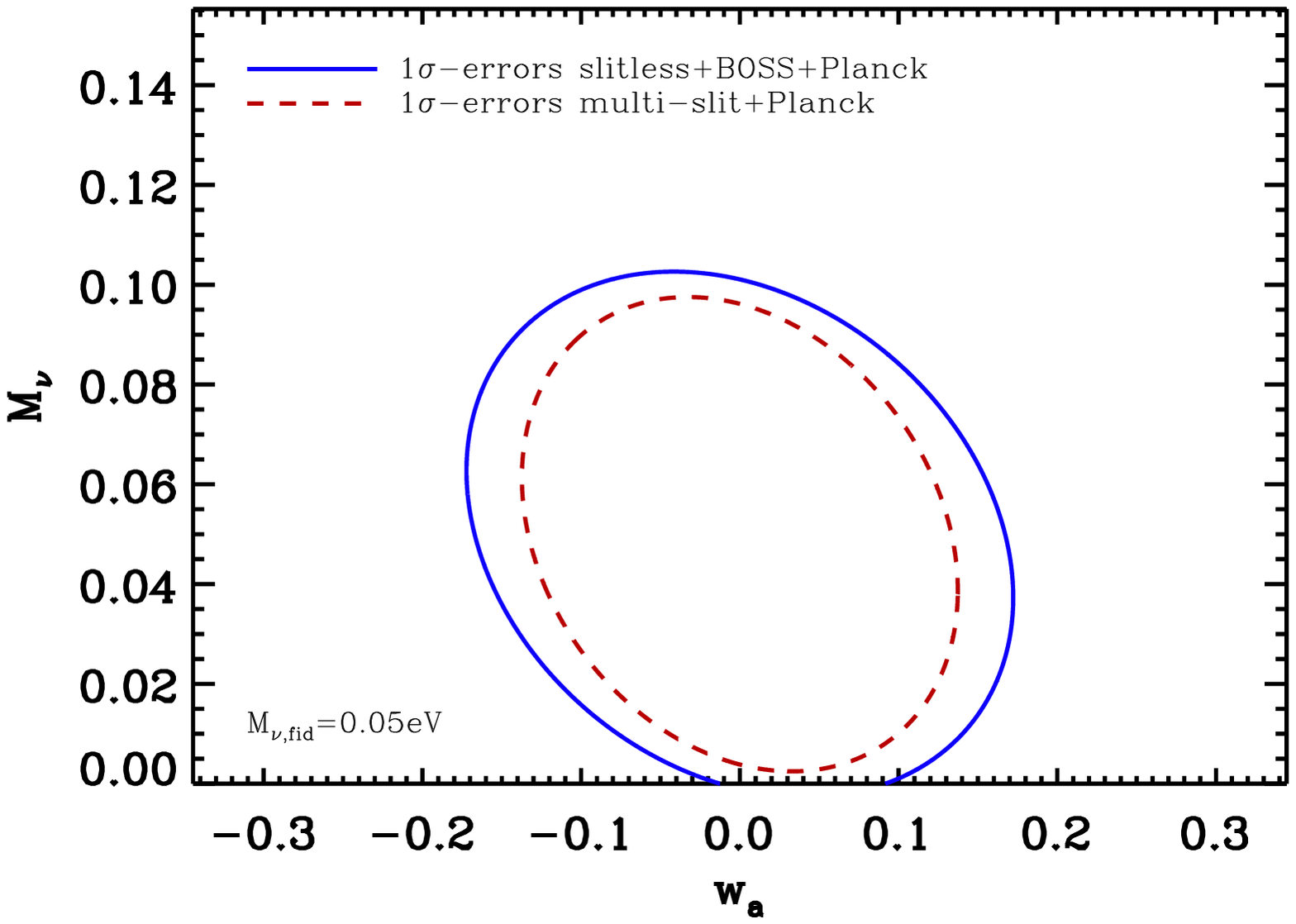}&
\includegraphics[width=7.3cm]{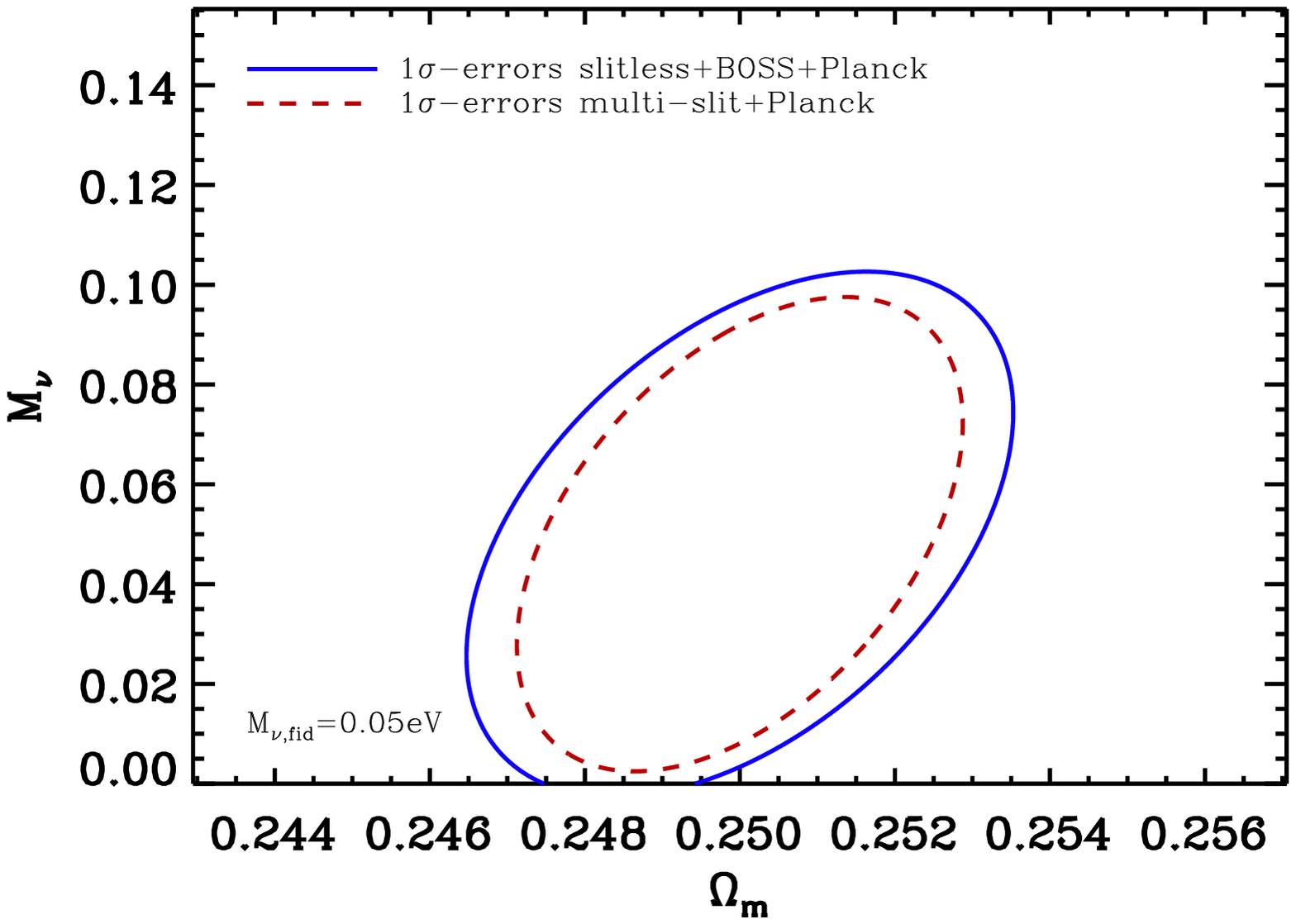}\\
\end{tabular}
\caption{1-parameter confidence levels at 1-$\sigma$ for $M_\nu$ and $q_\alpha$ with
  $q_\alpha=\Omega_{de}, w_0,w_a,\Omega_m$ for the fiducial model with
  $M_\nu=0.05$ eV and a neutrino mass spectrum with normal hierarchy,
  obtained after combining the survey data with Planck priors. The
  blue solid line and the red dashed one represent the
  slitles+BOSS-- and multi-slit--surveys cases respectively, as described in Sec.~3.}
\label{fig_Mnu005_comp}
\end{figure*}
\begin{figure*}
\begin{tabular}{l l}
\includegraphics[width=7.3cm]{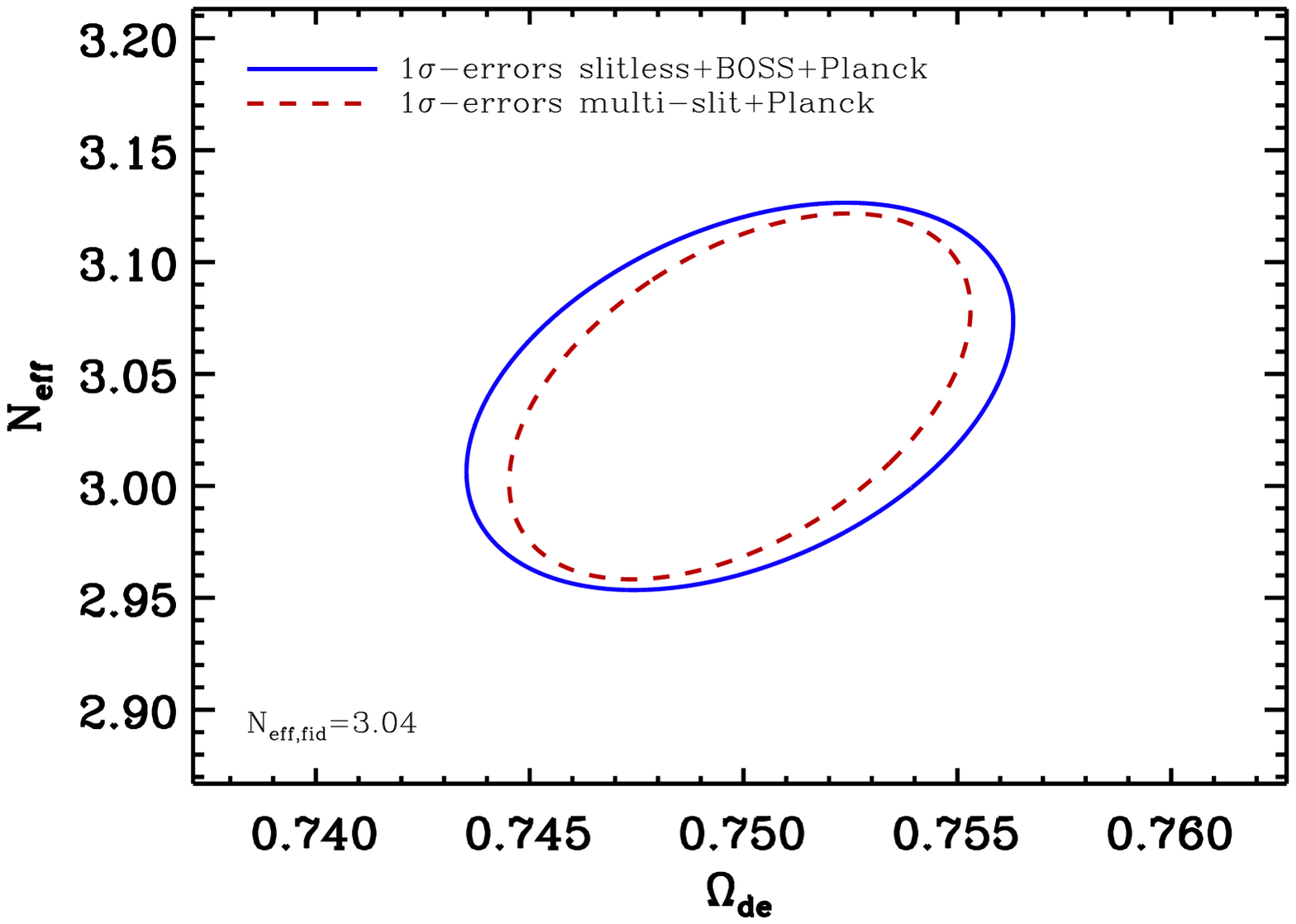}&
\includegraphics[width=7.3cm]{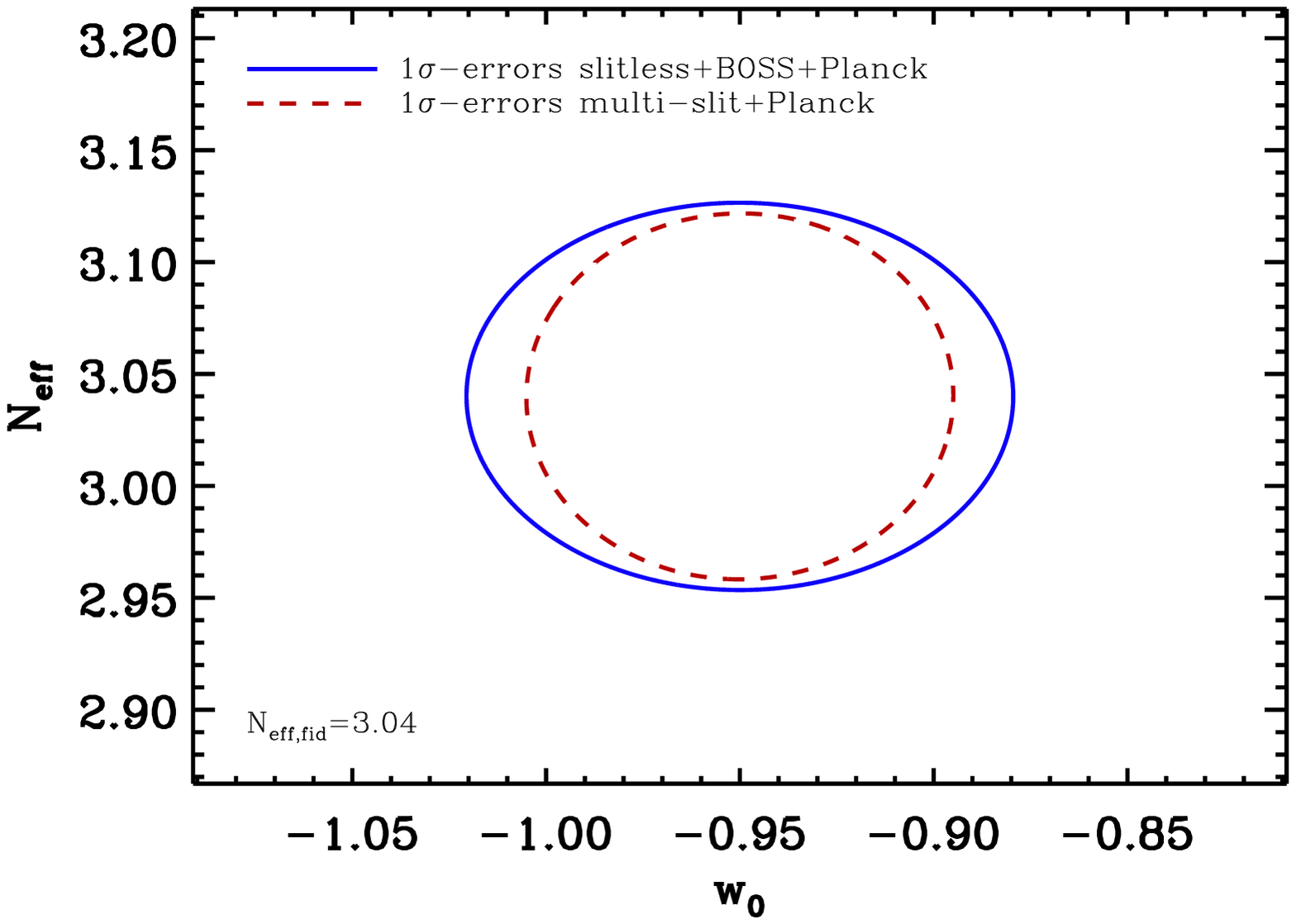}\\
\includegraphics[width=7.3cm]{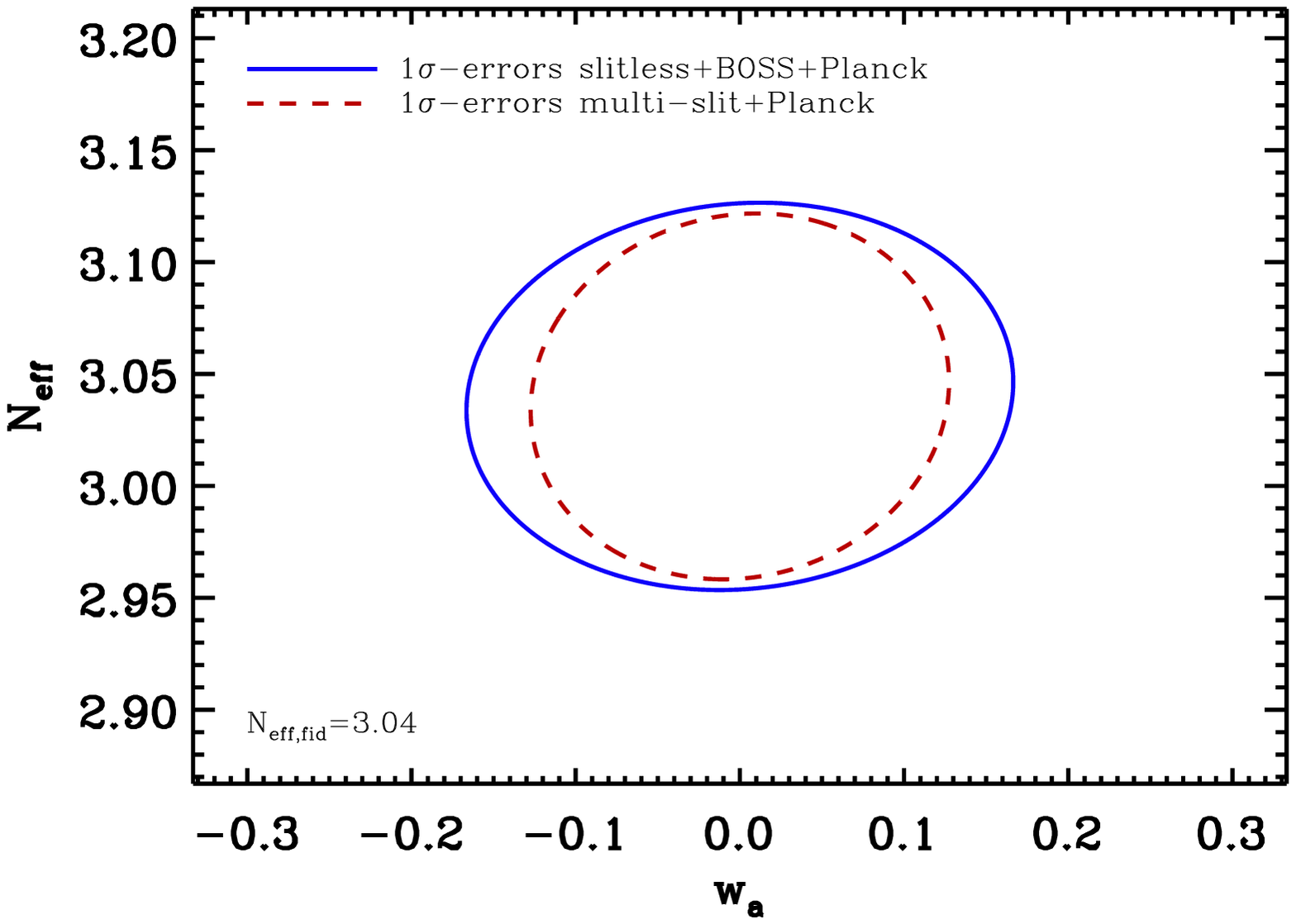}&
\includegraphics[width=7.3cm]{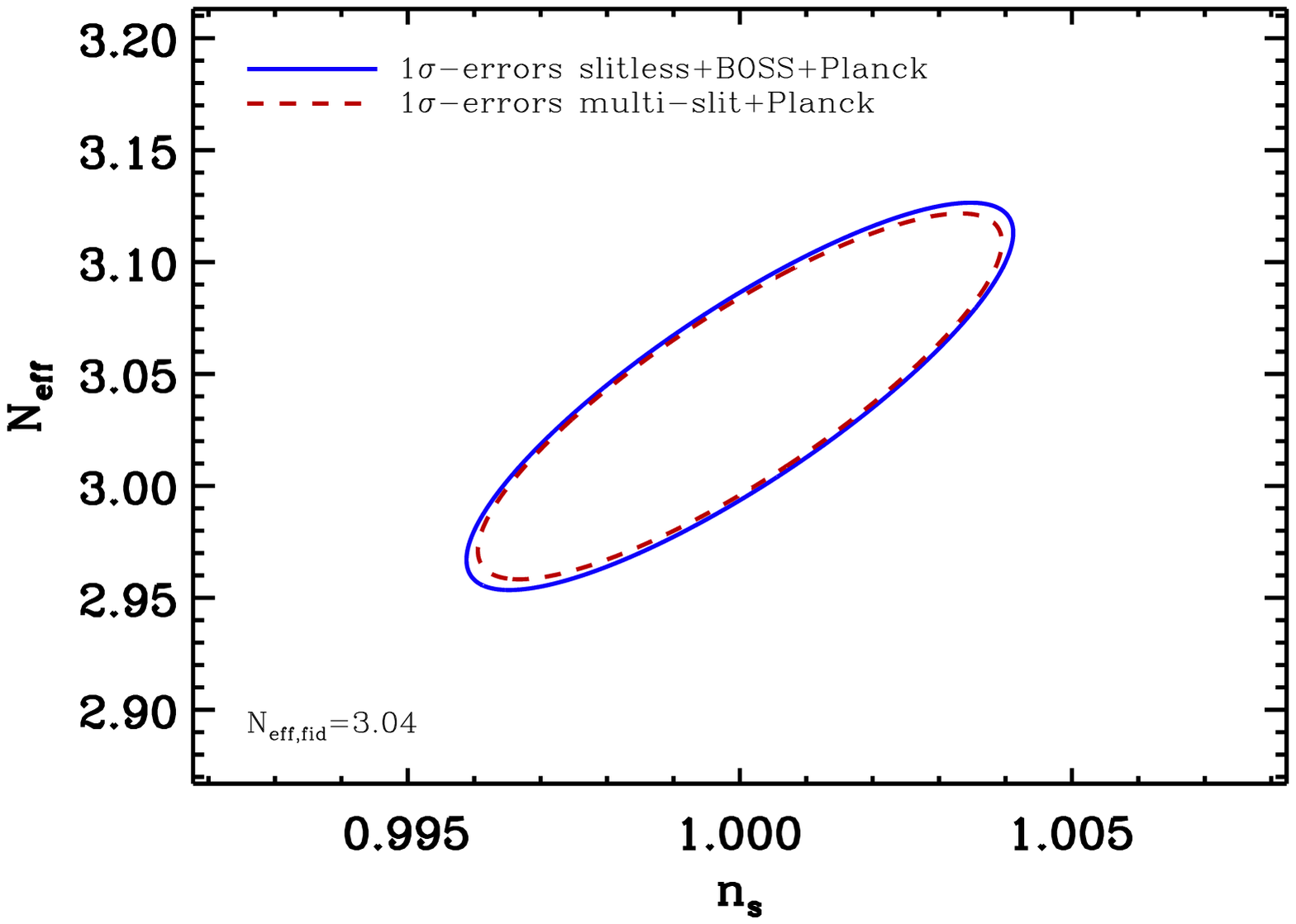}\\
\end{tabular}
\caption{1-parameter confidence levels for $N_{\rm eff}$ and $q_\alpha$ with
  $q_\alpha=\Omega_{de}, w_0,w_a,n_s$ for the fiducial model with
  $N_{\rm eff}=3.04$, obtained after combining the survey data with
  Planck priors. The blue solid line and the red dashed one represent 
the slitles+BOSS-- and multi-slit--surveys cases respectively, as described in Sec.~3.}
\label{fig_Neff_comp}
\end{figure*}
\begin{figure*}
\begin{tabular}{l l}
\includegraphics[width=7.3cm]{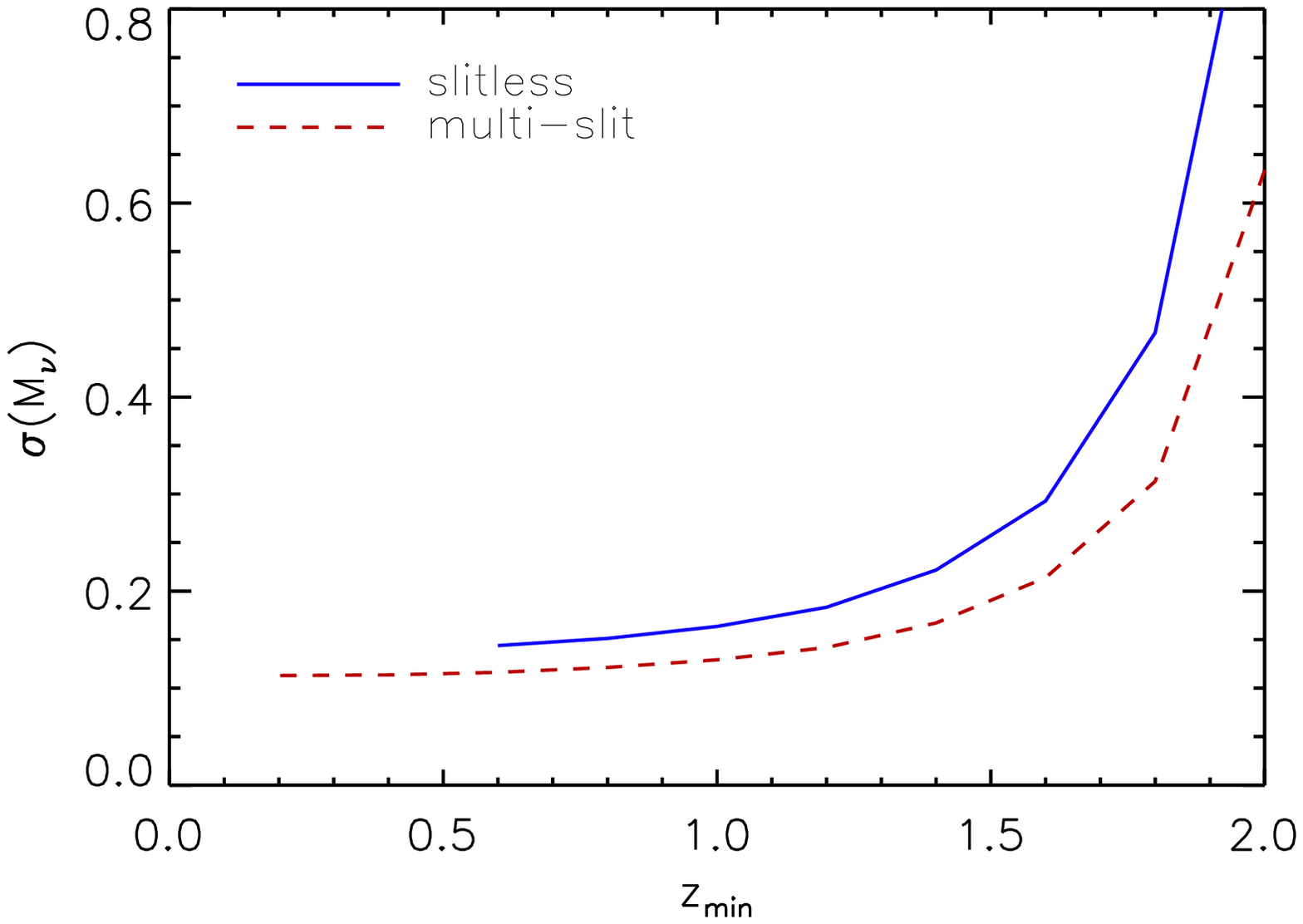}&
\includegraphics[width=7.3cm]{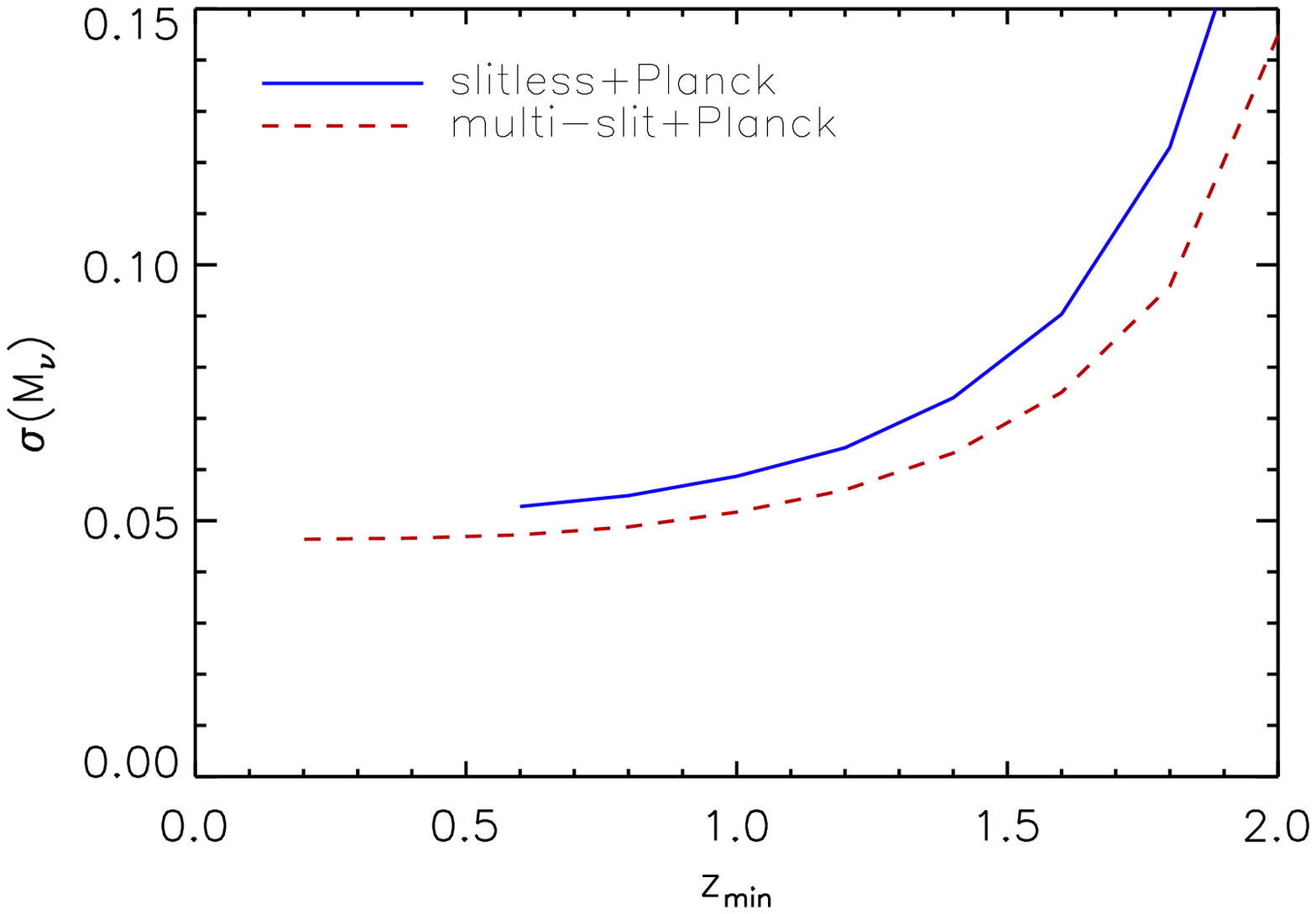}\\
\includegraphics[width=7.3cm]{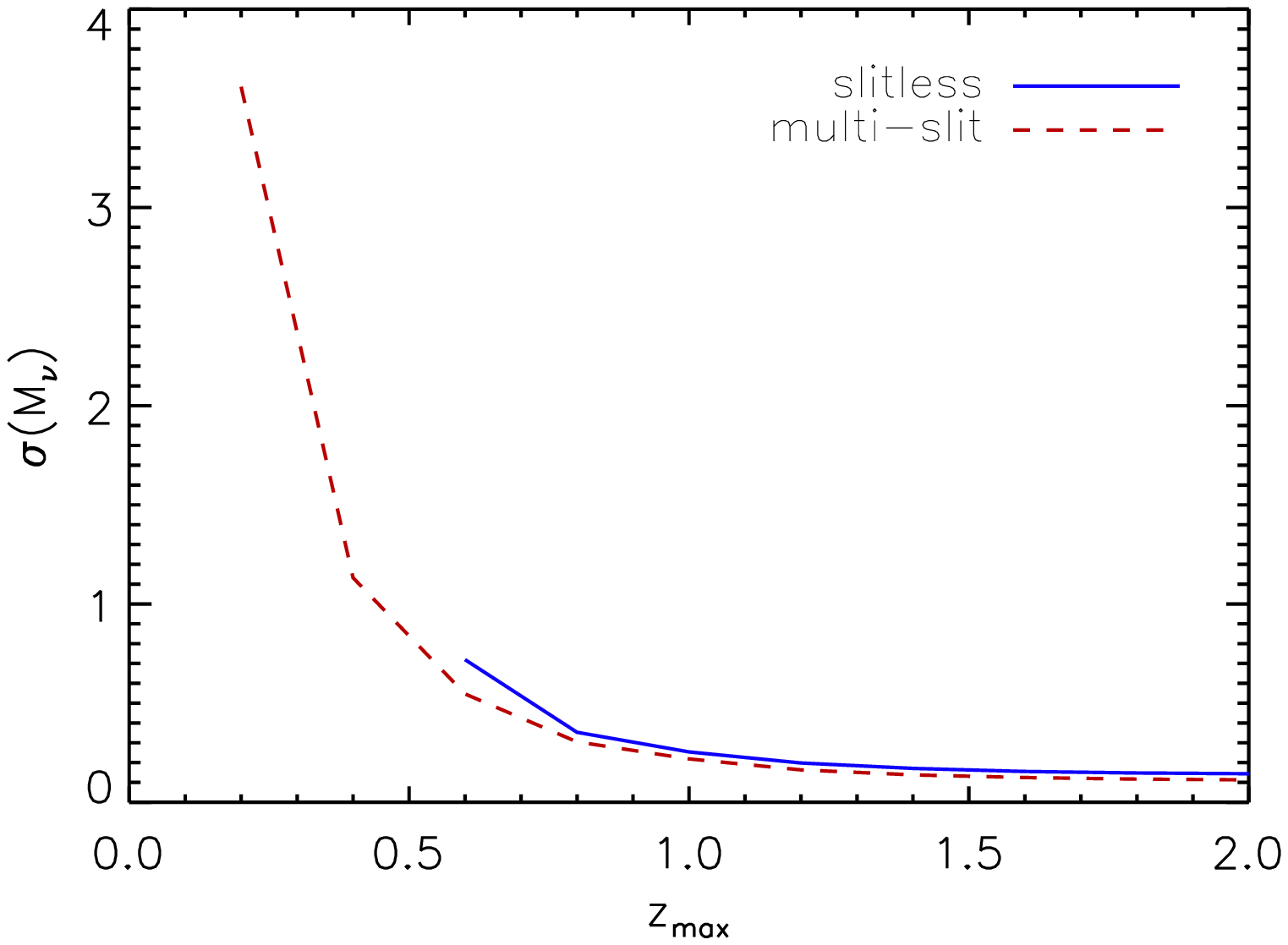}&
\includegraphics[width=7.3cm]{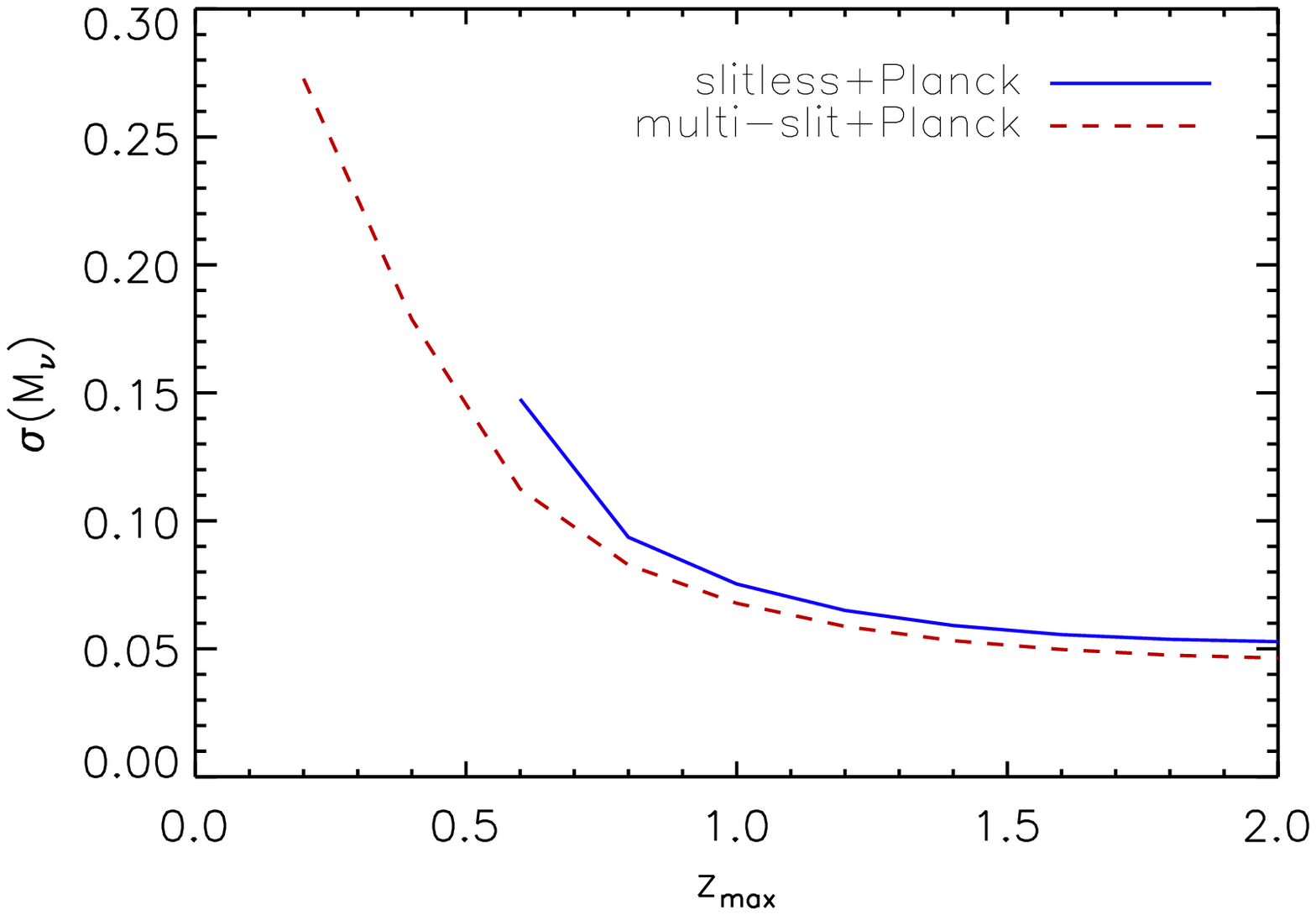}\\
\end{tabular}
\caption{Top: $M_\nu$--errors, for $M_\nu|_{\rm fid}=0.05$ eV, as functions of the minimum redshift
  $z_{min}$ of the surveys, where we have fixed the maximum redshift
  $z_{max}=2.1$ for both the spectroscopic strategies. The lowest
  minimum redshifts considered are $z_{min}=0.5$ and $z_{min}=0.1$ for
  the slitless and multi-slit spectroscopies, respectively.  
  Bottom: $M_\nu$--errors, for $M_\nu|_{\rm fid}=0.05$ eV, as functions of the maximum
  redshift $z_{max}$ of the surveys, where we have fixed the minimum
  redshifts $z_{min}=0.5$ and $z_{min}=0.1$ for the slitless and
  multi-slit spectroscopies, respectively.}
\label{mnu_z_depend}
\end{figure*}
\begin{figure*}
\begin{tabular}{l l}
\includegraphics[width=7.3cm]{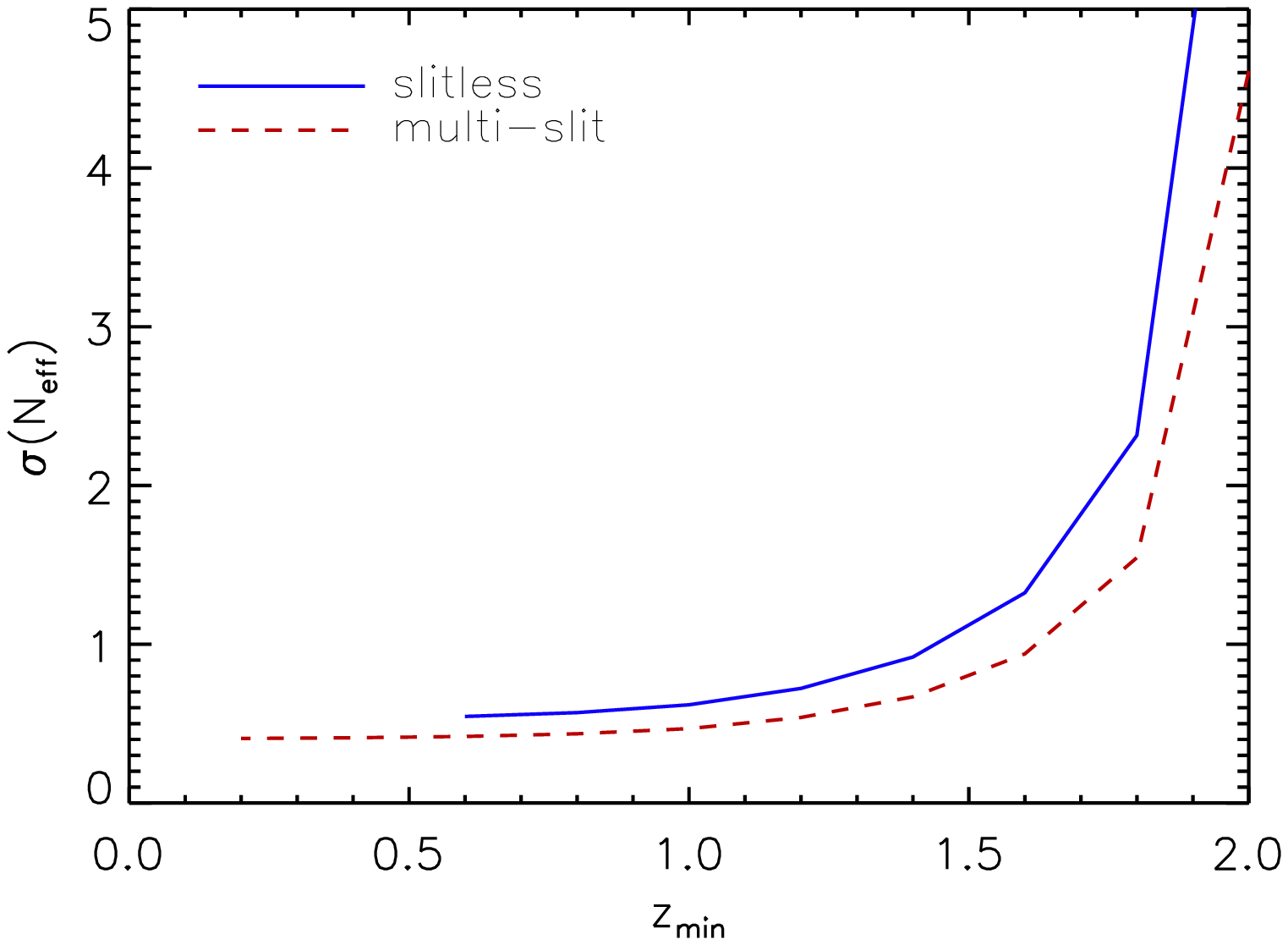}&
\includegraphics[width=7.3cm]{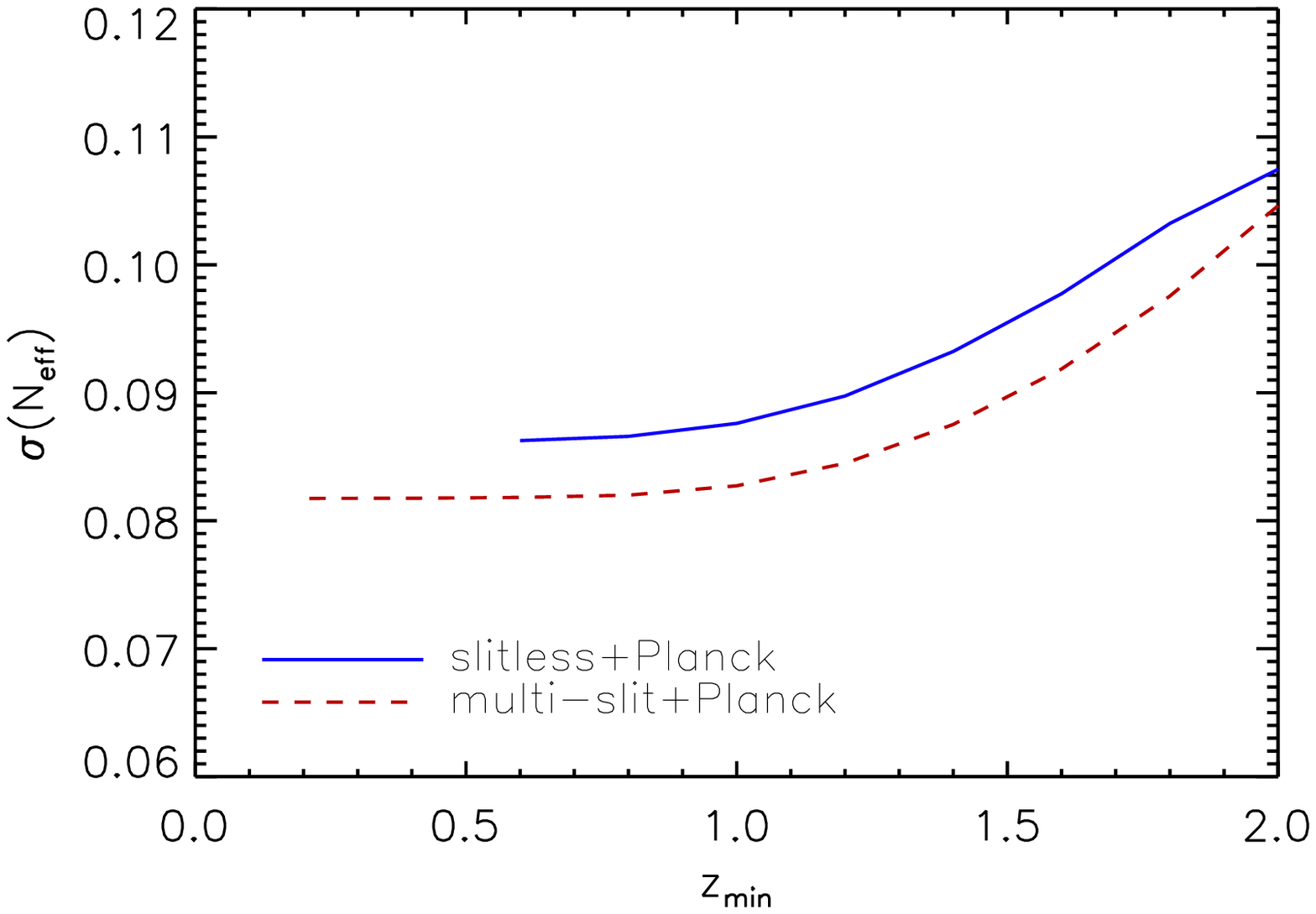}\\
\includegraphics[width=7.3cm]{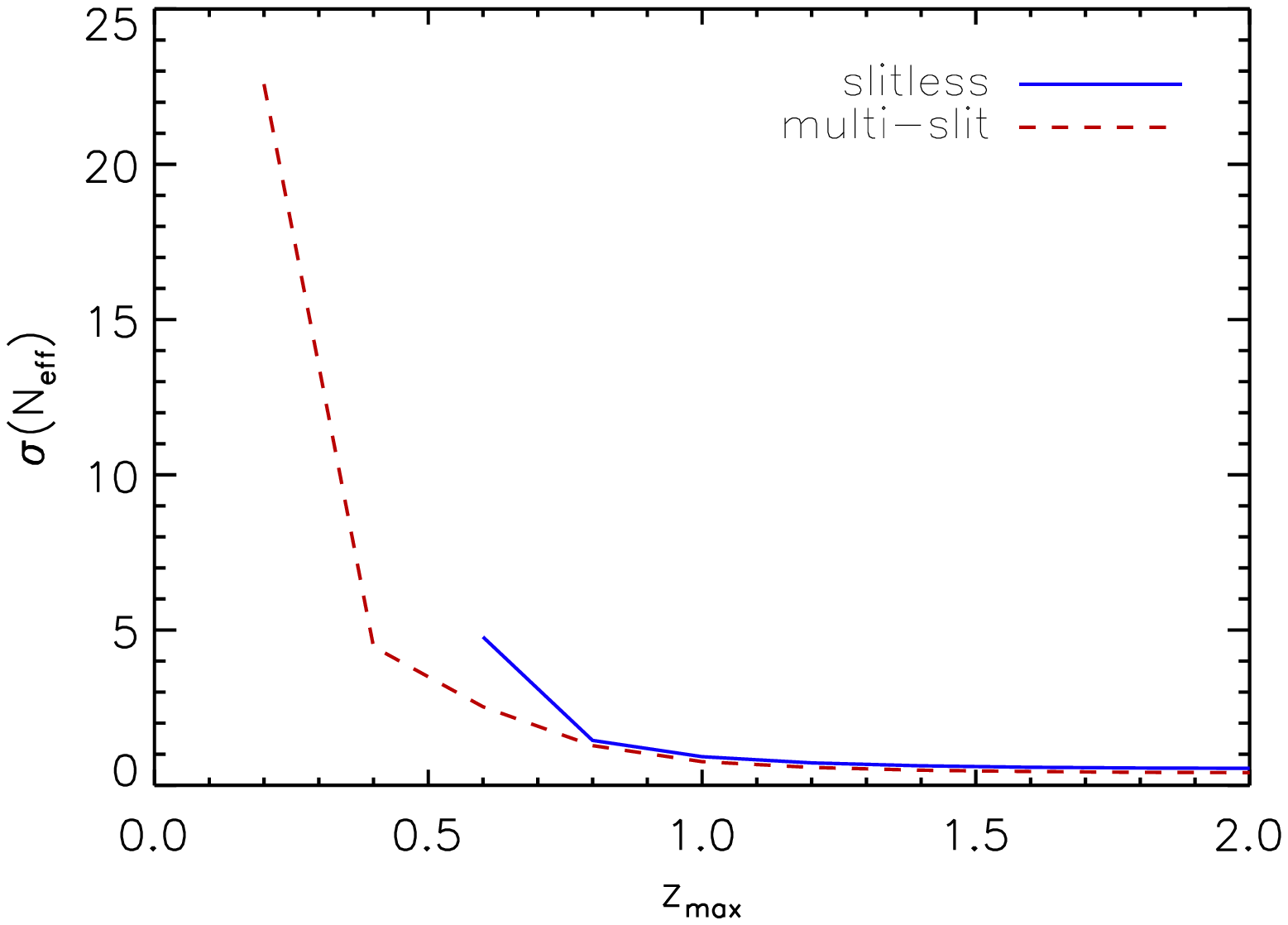}&
\includegraphics[width=7.3cm]{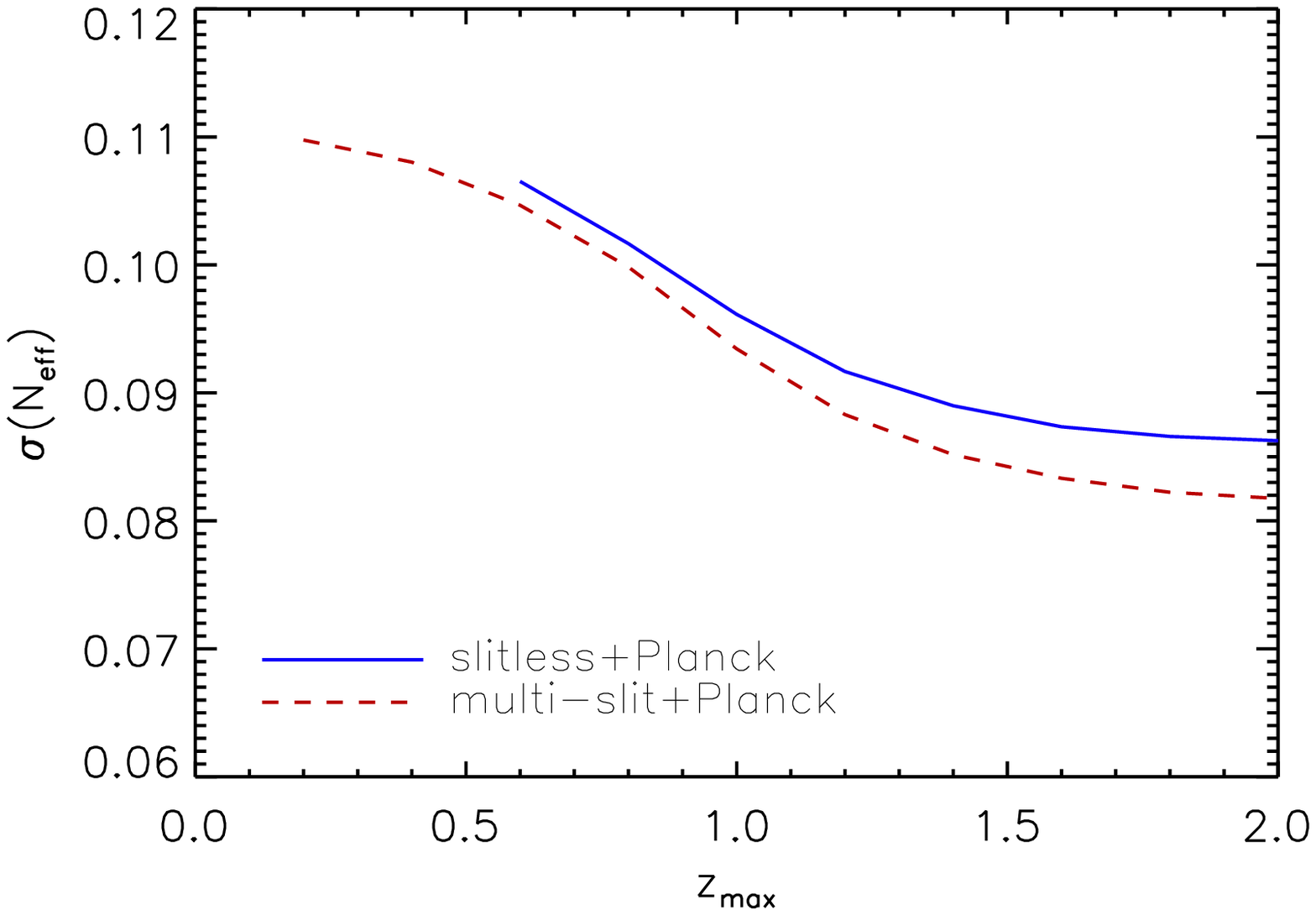}\\
\end{tabular}
\caption{Top: $N_{\rm eff}$--errors as functions of the minimum
  redshift  $z_{min}$ of the surveys, where we have fixed the maximum
  redshift  $z_{max}=2.1$ for both the spectroscopic strategies. The
  lowest  minimum redshifts considered are $z_{min}=0.5$ and
  $z_{min}=0.1$ for the slitless and multi-slit spectroscopies,
  respectively. Bottom: $N_{\rm eff}$--errors as functions of the
  maximum  redshift $z_{max}$ of the surveys, where we have fixed the
  minimum  redshifts $z_{min}=0.5$ and $z_{min}=0.1$ for the slitless
  and  multi-slit spectroscopies, respectively.}
\label{Nnu_z_depend}
\end{figure*}
Let us discuss the findings for the $M_\nu$-- and $N_{\rm eff}$--cosmologies separately.\\

\subsection{$M_\nu$--cosmology: Correlations}

When considering forecasts from LSS alone, we find that $M_\nu$ is correlated
with all the cosmological parameters affecting the galaxy power
spectrum shape and BAO positions at scales $k\leq \kmax$ (see columns
2-6 in the upper panels of Tables~\ref{slitless_corr}-\ref{dmd_corr}).
In particular, there is a quite strong positive correlation 
$r\sim0.55$ between the total neutrino mass
$M_\nu$ and the matter density $\Omega_m$. In fact, since neutrino free-streaming
suppresses the total matter transfer function on scales smaller than the
free-streaming scale $k_{\rm fs}$, increasing the neutrino mass
produces on $T(k,z)$ the opposite effect than
increasing the total matter content of the Universe. For the same
reason, massive neutrinos mimic the effect of a red tilt on the galaxy
power spectrum, resulting in a positive correlation between $M_\nu$ and the scalar
spectral index $n_s$. Moreover, $M_\nu$ is also positively correlated to the baryon
density $\Omega_b$ if the total matter content of the Universe is held
fixed (a larger $\Omega_b$ enhances the BAO and the information content
of $T(k,z)$).
The correlation of $M_\nu$ with the remaining parameters $h$,
$\Omega_{de}$, $w_0$, and $w_a$ can be explained looking at the Alcock-Paczynski
prefactor in Eq.~(\ref{eq:Pobs}), and recalling that for the moment we are
not including information from the amplitude of the galaxy power
spectrum, since we marginalise over the bias, the growth factor, the
redshift space distortions, and
the power spectrum normalisation. In this case we have to consider the
expression of $\hz$ 
in presence of a non-vanishing spatial curvature of the Universe
$\Omega_{\rm K}=1-\Omega_m-\Omega_{de}-\Omega_r$
\begin{eqnarray}
\label{Hz}
H(z)&=&H_0\big\{\Omega_{r}[(1+z)^4-(1+z)^2]+\Omega_m[(1+z)^3-(1+z)^2]+\nonumber \\
&&\Omega_{de}[(1+z)^{3(1+w_0+w_a)}
e^{3w_a(a-1)}-(1+z)^2]+(1+z)^2\big\}^{1/2},
\end{eqnarray}
and remember that $\DA$ is related to the inverse of
$\hz$\footnote{Let us specify that, at the redshifts covered by 
the LSS surveys considered in this work, we can safely consider that
neutrinos are already non-realtivistic, so that they contribute to the
$\Omega_m$ entering in $H(z)$. In fact, since we consider a minimum neutrino mass
$m_{\nu,i}=0.05$ (normal hierarchy), the minimum redshift at which the
massive neutrino component becomes
non-relativistic is $z_{nr,i}\sim 93.5 >> z_{\max}=2.1$ (see Eq.~(A7) of
Ref~\cite{0512374}).}
Given the fiducial values of $h$,
$\Omega_{de}$, $w_0$, and $w_a$, and varying each parameter at a time, 
we see from Eq.~(\ref{Hz}) that increasing $h$ or $w_0$ enhances
the observed power spectrum $P_{\rm obs}$ in contrast to the
suppression induced by the increase of the total neutrino mass,
so that these parameters are positively correlated with $M_\nu$.
On the contrary, from Eq.~(\ref{Hz}) and looking at the sign of 
the term $[(1+z)^{3(1+w_0+w_a)} e^{3w_a(a-1)}-(1+z)^2]$, we also deduce that increasing
$\Omega_{de}$ or $w_a$ produces on $P_{\rm obs}$ the same effect as a
larger neutrino mass, so these parameters are negatively correlated or
``anti-correlated'' to $M_\nu$.

Except for the $n_s$ case, we find that the level of correlation is on
the average stable against the value of the fiducial total neutrino mass
and the spectroscopic strategy adopted. For what concerns the mass
hierarchy,  at a given $M_\nu|_{\rm fid}$, the $n_s$-$M_\nu$ correlation
looks to be larger by $\sim 54\%-72\%$
for normal hierarchy in comparison to the
inverted one, at least when information from LSS alone are used, and
the effect is more evident for slitless spectroscopy. 
In contrast, the $h$-$M_\nu$ and $\Omega_b$-$M_\nu$ correlations slightly decrease 
by $\sim
34\%-24\%$ respectively, for normal hierarchy compared to the inverted 
hierarchy. This can be understood if we consider that, for the same $M_\nu|_{\rm fid}$, in the case 
of normal hierarchy, the transfer function is slightly less suppressed on the scales
of interest with respect to the inverted one (see Fig.~\ref{transfer},
and \S~\ref{mnu-errors} for further comments on the relation between
forecasted $M_\nu$--errors and the neutrino mass hierarchy).

When Planck priors are added to the survey constraints, all degeneracies are either resolved or 
reduced, except for the covariance $M_\nu$-$\Omega_{de}$.
In particular, the correlation between $M_\nu$ and $n_s$ is completely 
resolved, being reduced by $\sim$ one order of magnitude.
In some cases, the
correlation coefficient $r$ can even change  sign (see columns 2-6 in
the lower panels of Tables~\ref{slitless_corr}-\ref{dmd_corr}).
This change in the behaviour of $r$ arises either
due to the presence of dominant parameter degeneracies affecting
the CMB spectrum, or  because of marginalisation of  a high-dimension parameter space
down to two variables. To summarise, after the inclusion of Planck priors, 
the remaining dominant correlations
among $M_\nu$ and the other cosmological parameters are
$M_\nu$-$\Omega_{de}$, $M_\nu$-$\Omega_m$, and $M_\nu$-$w_a$.

\subsection{$M_\nu$--cosmology: Forecasted error-bars}
\label{mnu-errors}
The 1--$\sigma$ errors of the parameters are shown in columns 2-6 of
Tables~\ref{slitless_errors}-\ref{dmd_errors}. 
We see that, with respect to the slitless spectroscopy, 
the multi-slit spectroscopy is able to reduce the
neutrino mass errors of about 20\%-30\%, depending on the fiducial
neutrino mass, if LSS data alone are used. 
In addition, for the same $M_\nu|_{\rm fid}$, the 1--$\sigma$ error on total
neutrino
mass for normal hierarchy is $\sim 17\%-20\%$ larger than
for the inverted one. It looks like that the matter power
spectrum is less able to give information on the total neutrino mass
when the normal hierarchy is assumed as fiducial neutrino mass
spectrum. This is similar to what found in Ref.~\cite{1003.5918} for
the constraints on the neutrino mass hierarchy itself,
when a normal hierarchy is assumed as the fiducial one. On
the other hand, when CMB information are included, the $M_\nu$-errors
decrease by $\sim$35\% in favour of the normal hierarchy, at a given
$M_\nu|_{\rm fid}$. This difference arises from the changes in the
free-streaming effect due to the assumed mass hierarchy, and is in agreement with the results in
Ref.~\cite{0403296}, which confirms that the expected errors
on the neutrino masses depend not only on the sum of neutrino masses,
but also on the order of the mass
splitting between the neutrino mass states.

When Planck priors are added,
we find that the 1--$\sigma$ errors on $M_\nu$ are in the range
$0.03-0.05$ eV, depending on the fiducial total neutrino mass, with an
average difference of 15\% between the two spectroscopic
strategies, favouring again the multi-slit spectroscopy. 
This means that fixing some of the model free parameters,
e.g. 
assuming a $\Lambda$CDM Universe, 
future spectroscopic galaxy surveys, combined with CMB probes,
will be able to measure the minimum total neutrino mass $M_\nu=0.05$ eV
required by oscillation experiments; we will further comment on this
in \S \ref{growth&FoG}.
Finally, depending on $M_\nu|_{\rm fid}$, the total CMB+LSS
dark-energy FoM decreases
only by $\sim 15\%-25\%$ with respect to the value obtained if neutrinos are
supposed to be massless, meaning that the ``$P(k)$--method marginalised
over growth--information'' is quite robust in constraining the 
dark-energy equation of state.\\

\subsection{$N_{\rm eff}$--cosmology: Correlations}
Likewise to the $M_\nu$ case, we compute the 1--$\sigma$ errors and
the correlation coefficients among $N_{\rm eff}$ and the 
cosmological parameters considered in Eq.~(\ref{q_set}), for LSS alone
and in combination with Planck errors.

Interpreting the sign of the correlations is not so straightforward 
for the $N_{\rm eff}$--cosmology, since the number of relativistic
species gives two opposite contributions to $P_{\rm obs}$, and the total
sign of the correlation depends on the dominant one, for each single
cosmological parameter. In fact, from the bottom-left panel of 
Fig.~\ref{transfer}, it is clear
that a larger $N_{\rm eff}$ value 
suppresses the transfer
function $T(k)$ on scales $k\leq \kmax$. On the other hand, from
Eq.~(\ref{Or}) and Eq.~(\ref{Hz}), we see that a larger $N_{\rm eff}$ value
also increases the Alcock-Paczynski prefactor in $P_{\rm obs}$.
For what concerns the dark-energy
parameters $\Omega_{de}$, $w_0$, $w_a$, 
and the dark-matter density
$\Omega_m$, we find that 
the Alcock-Paczynski prefactor dominates, so that $N_{\rm eff}$
is positively correlated to $\Omega_{de}$ and $w_a$, and
anti-correlated to $\Omega_m$ and $w_0$. In contrast, for the
other parameters, the $T(k)$ suppression produces the larger effect and
$N_{\rm eff}$ results to be anti-correlated to $\Omega_b$, and
positively correlated to $h$ and $n_s$. See the last column in the
upper panels of Tables~\ref{slitless_corr}-\ref{dmd_corr} for the LSS
alone case. The degree of the correlation $r$ is stable against the
spectroscopic strategy adopted and is very large in the $n_s$-$N_{\rm
eff}$ case, being $r\sim 0.8$ with and without Planck priors.
For the remaining cosmological parameters, all the correlations are
reduced when CMB information are added, except for the covariance
$N_{\rm eff}$-$\Omega_{de}$, as happens also for the $M_\nu$--cosmology.
To summarise, after the inclusion of Planck priors, the remaining 
dominant correlations among $N_{\rm eff}$ and the other cosmological 
parameters are $N_{\rm eff}$-$n_s$, $N_{\rm eff}$-$\Omega_{de}$, and 
$N_{\rm eff}$-$h$.

\subsection{$N_{\rm eff}$--cosmology: Forecasted errors}

The 1--$\sigma$ errors of the parameters are shown in the last column of
Tables~\ref{slitless_errors}-\ref{dmd_errors}.
Also in this case, compared to the slitless spectroscopy,
the multi-slit spectroscopy is able to reduce the
$N_{\rm eff}$ errors by $\sim$30\% when LSS alone are used. When Planck priors are added,
we find a 1--$\sigma$ error on $N_{\rm eff}$ of $\sim 0.08$, with a
difference of only $6\%$ between the two spectroscopy strategies,
again in favour of the multi-slit one.

\subsection{Constraints on neutrino properties in the context of dark energy surveys}
The main science  goal of the galaxy surveys considered in this work
is to constrain dark energy. Considering neutrinos properties might
degrade the dark energy constraints (because  of the introduction of 
extra parameters in the model), or it may be that such surveys are 
not optimised (e.g., in their redshift coverage) to measure neutrino 
properties and thus perform sub-optimally for these parameters.

We find that, depending on the fiducial $M_\nu$ value, the 
total CMB+LSS dark-energy FoM decreases only by $\sim 15\%-25\%$ with 
respect to the
FoM obtained if neutrinos are assumed to be massless, and that, 
for the $N_{\rm eff}$--cosmology, the total CMB+LSS dark-energy FoM 
decreases only by $\sim 5\%$ with respect to the FoM obtained
by holding $N_{\rm eff}$ fixed. This means that the ``$P(k)$--method
marginalised over growth--information'' is quite robust in constraining the
dark-energy equation of state.

In Figs.~\ref{fig_Mnu03_sigmas}-\ref{fig_Neff_sigmas} we show the
jointly 2-parameter
projected 68$\%$ C.L., 95.4$\%$ C.L. and 99.73$\%$ C.L. contours
in the $\zeta$-$q_\alpha$ sub-space, where $\zeta=M_\nu,N_{\rm eff}$
and $q_\alpha=\Omega_{de},w_0,w_a,\Omega_m,n_s$, 
in the case of the slitless survey in combination with BOSS and Planck
priors, for the $N_{\rm eff}$-- and $M_\nu(=0.3$ eV)--cosmologies.
The black solid lines show the 1-parameter confidence levels at 1--$\sigma$.
The orientation of the ellipses reflects the correlations among the
parameters shown in the lower panels of
Tables~\ref{slitless_corr}-\ref{dmd_corr}.

Moreover, for a visual comparison of the constraints obtained with the
two spectroscopic strategies, in Figs.~\ref{fig_Mnu005_comp}-\ref{fig_Neff_comp}
we show the 1-parameter confidence levels at 1-$\sigma$
in the $\zeta$-$q_\alpha$ sub-space for the combinations
slitless+BOSS+Planck and multi-slit+Planck, respectively. 

Finally, we consider the fiducial
cosmology with $M_\nu|_{\rm fid}=0.05$ eV and, in the top panels of
Fig.~\ref{mnu_z_depend}, we show the $M_\nu$--errors as
functions of the minimum redshift $z_{min}$ of the surveys, with and
without Planck priors. We fix the maximum redshift $z_{max}=2.1$ for both the
spectroscopic strategies, while the lowest minimum redshifts are given
by $z_{min}=0.5$ and $z_{min}=0.1$ for 
the slitless and multi-slit spectroscopies, respectively. 
In the bottom panels of Fig.~\ref{mnu_z_depend}, we show the
$M_\nu$--errors as functions of the maximum redshift $z_{max}$ of the surveys, 
where we have fixed the minimum redshifts $z_{min}=0.5$ and
$z_{min}=0.1$ for the slitless and  
multi-slit spectroscopies, respectively. For the
slitless case, we have
verified that extending the minimum redshift to $z_{min}=0.1$ changes
the neutrino constraints by $\sim$0.8\% only.
We make the same analysis for the $N_{\rm eff}$--errors and the
corresponding trends are shown in Fig.~\ref{Nnu_z_depend}.

Note that concerning the possible degeneracies with the dark-energy 
parameters, we find that the major covariance is with 
$\Omega_{de}$, rather than with the dark energy equation 
of state parameters $w_0$ and $w_a$, and that extending the redshift range of 
the surveys considered would not reduce drastically the forecasted
errors\footnote{Concerning the dependence of $M_\nu$-- and $N_{\rm
  eff}$--errors on the survey area, we note that $F^{\rm LSS}$ is linearly dependent on the
effective survey volume $V_{\rm eff}$ (see Eqs.~(2.8)-(2.9)), 
therefore $\sigma(M_\nu)$ and $\sigma(N_{\rm eff})$, extracted from
LSS data alone, are inversely proportional to the square root of the survey area, for a fixed
redshift range.}.

\subsection{The effects of growth inclusion and random peculiar
  velocities on neutrino mass constraints}
\label{growth&FoG}
It is well known that galaxy peculiar velocities produce
redshift-space distortions (RDS), which can be exploited by a large deep redshift survey 
to measure the growth rate of
density fluctuations $f_g$, within the same redshift bins in which
$H(z)$ is estimated via BAO. In particular, RDS allow to
constrain $f_g$ times the normalisation of the power spectrum,
i.e. $f_g\sigma_8$. In order to include growth information in our Fisher
matrix analysis, following Ref.~\cite{1006.3517}, we rewrite Eq.~(\ref{eq:Pobs})
as
\begin{align}
P_{obs}(k_{{\rm ref} \perp},k_{{\rm ref} \parallel},z) =
\frac{D_A(z)_{\rm ref}^2  H(z)}{D_A(z)^2
  H(z)_{\rm ref}} 
\left[ \sigma_{8g}(z)+ f_g(z,k)\sigma_{8}(z)\frac{k_{{\rm
        ref}\parallel}^2}{k_{{\rm ref}\perp}^2+k_{{\rm
        ref}\parallel}^2} \right]^2\times C(k,z) + P_{shot},
\label{Pobs_s8}
\end{align}
where
$C(k,z)\equiv P_{\rm matter}(k,z)/ \sigma^2_{8}(z)$, and 
$\sigma_{8g}(z)=b(z) \sigma_{8}(z)$. We refer to this as the ``full
$P(k)$--method, with growth--information included''. Let us
  stress that we model RDS as an additive component, using as free
parameters $(b\sigma_8,f_g\sigma_8)$. $f_g\sigma_8$ can be measured without knowing the bias
  $b$ or the amplitude of the matter fluctuations $\sigma_8$;
  therefore this parameter choice reduces possible systematic errors
  due to estimates of the bias \cite{0807.0810}.

While, for large separations, galaxy peculiar
velocities give information on the growth of structures, on small
scales random peculiar velocities cause the so-called Fingers of
God (FoG), stretching compact structures along the line-of-sight \cite{0807.0810,scoccimarro}.
Although, on the scales of interest in this work, this effect is
expected to be moderate \cite{0808.0003,1011.2842}, we now include it in the
Fisher analysis by introducing a Gaussian distribution for the
pairwise velocity dispersion in configuration space, which produces
a Gaussian damping $e^{-k^2\mu^2\sigma_v^2}$ of the observed
galaxy power spectrum $P_{obs}$. This effect is degenerate with possible
inaccuracies $\sigma_z$ in the observed redshifts due to a line-of-sight smearing
of the structures, so we absorb it in the Gaussian damping factor of Eq.~(\ref{eq:Pm}). 

For the sake of simplicity, we consider only the fiducial cosmology
with $M_\nu=0.05$ eV.
In fact, in this case, the scale dependence of the growth rate $f_g$
due to free-streaming massive neutrinos can be assumed to be
negligible. This is evident from the bottom-right panel of Fig.~\ref{transfer}, where we show the
scale dependence correction to $f_g$ for $M_\nu=0.05$ eV, given by
the function $\mu(k,f_\nu,\Omega_{de}) \equiv f_g(M_\nu\neq 0)/f_g(M_\nu= 0)$, 
where $f_\nu=\Omega_\nu/\Omega_m$, introduced by
Ref.~\cite{0709.0253}
in their Eqs.~(16)-(17). In this case the growth suppression is only of the order of
$0.2\%$, affecting scales $k>0.1 \,h$/Mpc.

Under these assumptions, when we include the effect of FoG in the
Fisher analysis, we consider also $\sigma_v(z_i)$ as a scale independent
variable, which is treated as
a nuisance parameter to be marginalised over in each redshift bin, together with
\{$\sigma_{8g}(z_i)$, $P_{shot}^i$\}, 
and we project the errors on \{$H(z_i)$, $D_A(z_i)$, $f_g(z_i)\sigma_{8}(z_i)$,
$\omega_m$, $\omega_b$, $\zeta$, $n_s$, $h$\} into the final set of cosmological
parameters \{$\Omega_m$, $\Omega_{de}$, $\Omega_b$, $h$, $\zeta$,
$w_0$, $w_a$, $n_s$, $\sigma_8$\}. This parameter set differs from Eq.~(\ref{q_set})
only for the substitution $\Delta^2_{\cal R}(k_0) \rightarrow \sigma_8$,
whose fiducial value is fixed by $\Delta^2_{\cal R}(k_0)=2.45\times
10^{-9}$ and the other parameters of the $M_\nu|_{\rm
  fid}(=0.05 {\rm eV})$--cosmology described in \S \ref{Fiducial cosmologies} .
We refer to this as the ``full
$P(k)$--method, with FoG and growth--information included''.

In what follows we compare the $M_\nu$ constraints obtained by the
three $P(k)$--methods considered in this work, for the fiducial
cosmology with $M_\nu|_{\rm fid}=0.05$ eV.

\subsubsection{Growth inclusion effects}
\label{growth-effects}
We find that neutrino mass errors are quite stable at
$\sigma(M_\nu)=0.05$ eV, against the
adopted method (whether growth--information are included or
marginalised over), and decrease only by $10\%$--$20\%$ when
$f_g\sigma_8$ measurements are included. 
We can understand this result as follows.
$\Omega_\nu$ affects the shape of the power spectrum, i.e.
enters the transfer function $T(k,z)$, which is sampled
on a very large range of scales, including the $P(k)$ turnover
scale, by the nearly full-sky surveys under consideration.  
On the other hand, $f_g$ is only slightly dependent on
$M_\nu$ for $M_\nu|_{\rm fid}=0.05$ eV (see the bottom-right panel of
Fig.~\ref{transfer}). Consequently, the effect on the observed power
spectrum shape dominates over the information extracted by measurements
of $f_g\sigma_8$. This quantity, in turn, generates new correlations
with $M_\nu$ via the $\sigma_8$-term, which now we constrain
simultaneously with the other cosmological parameters, and which
actually is anti-correlated with $M_\nu$\footnote{In particular, using
  the ``full $P(k)$--method, with growth--information included'', we
  find $\sigma(\sigma_8)\sim 0.011,0.0085$ for BOSS+slitless and
  multi-slit, respectively, and  
$\sigma(\sigma_8)\sim 0.0014,0.0013$ for BOSS+slitless+Planck and
multi-slit+Planck, respectively. Using the ``full $P(k)$--method, with
FoG and growth--information included'', we find
$\sigma(\sigma_8)\sim 0.012,0.0096$ for BOSS+slitless and multi-slit,
respectively, and again
$\sigma(\sigma_8)\sim 0.0014,0.0013$ for BOSS+slitless+Planck and                                
multi-slit+Planck, respectively.} \cite{Marulli2010_inprep}.
 
On the other hand, if we suppose that early dark-energy is negligible,
the dark-energy parameters $\Omega_{de}$, $w_0$ and $w_a$ do not enter
the transfer function, and consequently growth information have 
relatively more weight when added to constraints from $H(z)$ and
$D_A(z)$ alone. 

As a result, the $M_\nu$--errors are quite insensitive to growth
inclusion, hence almost independent of the adopted $P(k)$--method.  
This is in contrast to the dark-energy parameter constraints 
(see e.g. Ref.~\cite{0709.0253}).
In fact, we find that, with respect to the ``full $P(k)$--method, 
marginalised over growth--information'',
the ``full $P(k)$--method, with growth--information included'' is able
to increase the dark-energy FoM by $\sim 50$\% and $\sim 60$\%
for the slitless and multi-slit strategies, respectively, from both
LSS data alone and in combination with Planck priors, when
massive neutrinos are assumed in the fiducial cosmology.

To summarise, we find that, due to the slight dependence of $f_g$
on $M_\nu$ when $M_\nu|_{\rm fid}=0.05$ eV, and due to the further
degeneracy with $\sigma_8$ (correlated also with the dark-energy
parameters), we do
not find a total effective gain on the accuracy of
$M_\nu$ measurements, when growth-information are added. On the other 
hand, the value of the dark-energy FoM does increase when 
growth-information are included, even                                  
if it decreases by a factor $\sim 2-3$ with respect to cosmologies
where neutrinos are assumed to be massless, due to the correlation
among $M_\nu$ and the dark-energy parameters. As confirmation of this
degeneracy, we find that, when growth-information are added and if the dark-energy
parameters $\Omega_{de}$, $w_0$, $w_a$ are held fixed to their
fiducial values, the errors 
$\sigma({M_\nu})$ decrease to $0.027$ eV and $0.025$ eV, for the 
slitless and multi-slit spectroscopies combined with Planck,
respectively.

\subsubsection{Incoherent velocity inclusion effects}
\label{fog-effects}
We expect that dark-energy parameter errors
are somewhat sensitive to FoG effects. This can be understood
in terms of correlation functions in the redshift-space;
the stretching effect due to random peculiar
velocities contrasts the flattening effect due to large-scale bulk
velocities.
Consequently, these two competing effects act along opposite 
directions on the dark-energy parameter constraints.

We find that the dark-energy FoM obtained with the ``full
$P(k)$--method, 
with FoG and growth--information included'' result
to be FoM$^{\rm LSS}=51,\,95$ and FoM$^{\rm LSS+CMB}=268,\,391$,
for the slitless and multi-slit spectroscopies respectively,
i.e. very similar to the FoMs obtained from the ``full $P(k)$--method,
marginalised over growth--information'', as shown in
Tables~\ref{slitless_errors}-\ref{dmd_errors}, 
which therefore can be considered a more stable approach against
galaxy peculiar velocity uncertainties.

On the other hand, the neutrino mass errors are expected to
be almost stable at $\sigma({M_\nu})=0.05$             
when FoGs effects are taken into account by marginalising over
$\sigma_v(z)$, increasing only by $10\%$--$14\%$ with respect to the
``full $P(k)$--method, marginalised over growth--information''.
Moreover, in this case, if the dark-energy parameters 
$\Omega_{de}$, $w_0$, $w_a$ are held fixed to their fiducial values, the errors
$\sigma({M_\nu})$ become $0.029$--$0.028$ eV, for the
slitless and multi-slit spectroscopies combined with Planck,
respectively. In other words, we get errors which
are only $\sim$11\% larger than the ones obtained without marginalising
over $\sigma_v(z)$ under the same 
assumptions on $\Omega_{de}$, $w_0$, and $w_a$, as described in \S \ref{growth-effects}. 

\subsection{Systematic effects, non-linearities and bias}
\label{systematics}
Fisher-matrix based forecasts are not particularly well suited to quantify 
systematic effects. The errors reported so far are statistical errors, 
which are meaningful only as long as they dominate over  systematic
errors. 
It is therefore important to consider  sources of systematics and 
their possible effects on the recovered parameters.
Possible sources of systematic errors of major concern 
are the effect of non-linearities and the effects of galaxy bias.
In our analysis so far we have used the linear theory 
matter power spectrum and applied scale-independent bias to it.

The description of non-linearities in the matter power spectrum 
in the presence of massive neutrinos is a relatively new subject. 
It has been addressed in several different ways: Refs.~\cite{Wong:2008ws,Saito1,Saito2,Saito3}  
use perturbation theory, Ref.~\cite{Pietroni} used the time-RG flow
approach 
and Refs.~\cite{Brandbyge1,Brandbyge2,Brandbyge3,1003.2422} used
different 
schemes of N-body simulations.  From the above references 
it is clear that the effect of massive neutrinos on the matter 
power spectrum  in the  non-linear regime must be explored via 
N-body simulations to encompass all the relevant effects.  
Different simulations schemes and approximations agree 
already at or below the \% level (for neutrino masses  
allowed by current observations and $k<1$ Mpc/$h$) 
indicating that non-linear effects on the matter power 
spectrum can in the future be modelled to the required accuracy.  

On the scales considered in this work the effects of 
non-linearities are small and statistical--errors could 
be further reduced considering smaller scales. As shown in 
\cite{Saito1}, the  effect of non linearities does not erase  
or reduce the effect of massive neutrinos.  In fact,  by comparing 
Fig.~4  of \cite{Saito1} and e.g.,  Fig.~4 of \cite{Brandbyge2},  
it is apparent that the difference between the massless and massive 
neutrino case  is enhanced by non-linearities. 
Thus  while  it will be mandatory to include non-linearities  
in the actual  data analysis, the forecasted errors  
are not made artificially smaller by using  the linear 
matter power spectrum to compute our Fisher matrices. 

Pushing to smaller scales however would worsen the 
systematic  effect of  scale-dependent and/or  
non-linear bias.  Here we have made the simplifying 
assumption that bias is scale-independent up to $k_{max}$ 
but the redshift dependence is not known and is 
marginalised over.  Bias is known to be scale-independent 
on large, linear  scales but to become non-linear and 
scale-dependent for small scales and/or for very 
massive halos. A scale-dependence of bias may mimic 
in part the effect of massive neutrinos. The 
scale-dependence of bias however is expected not 
to have the same redshift dependence as massive 
neutrino effects, thus offering the possibility 
to break a possible degeneracy. 
A  scale-dependence of bias may cancel in part the effect of
massive neutrinos.  It arises because halos, especially massive or
rare  ones, are non-linearly biased with respect to the dark
matter. Scale-dependent bias (at least for dark matter halos hosting
galaxies) is  found first to increase with increasing $|k|$ then
decrease, but the bias of the galaxies hosted in the dark halos may be
more complicated.  This will be an important limitation in any
practical application  especially  if we want to  include mildly
non-linear scales.  There are,  however,  several possibilities to
control or quantify systematics introduced by bias.  In fact,
the bias  behaviour  varies  for differently
selected objects (different colour or different brightness): splitting
the sample in differently-biased tracers will thus help disentangle
the  systematic effect from the cosmological signal (approach similar
to that of e.g.,\cite{1006.2825}).
Finally on large, linear scales, the neutrino  mass splitting leaves a
specific signature on the shape of the power spectrum that can also be
used as a cross check of the $M_{\nu}$ signal as illustrated in
Ref.~\cite{DePutter_etal}. A more quantitative investigation of scale-dependent bias is beyond the scope of
this paper, but we plan to study this important issue as more
data and simulations become available.

\begin{table*}
\caption{$\sigma(M_\nu)$ and $\sigma(N_{\rm eff})$ marginalised errors from LSS+CMB}
\setlength{\tabcolsep}{0.3pt}
\begin{tabular}{|l c c c c c c|}
\hline
\hline
{}&{}&{}&\footnotesize{{\small General cosmology}}&{}&{}&{}
\\
\hline
\footnotesize{{\small {\rm fiducial}$\to$ }}&
\footnotesize{{\small $M_\nu$=0.3 eV}}${}^a$&
\footnotesize{{\small $M_\nu$=0.2 eV}}${}^a$ &
\footnotesize{{\small $M_\nu$=0.125 eV}}${}^b$ &
\footnotesize{{\small $M_\nu$=0.125 eV}}${}^c$ &
\footnotesize{{\small $M_\nu$=0.05 eV}}${}^b$ &
\footnotesize{{\small $N_{\rm eff}$=3.04}}${}^d$
\\
\hline
\footnotesize{{\small slitless+BOSS+Planck}} & $0.035$ & $0.043$ & $0.031$ & $0.044$ & $0.053$ & $0.086$\\
\footnotesize{{\small multi-slit+Planck}} & $0.030$ & $0.038$ & $0.027$ & $0.039$ & $0.046$ & $0.082$\\
\hline
\hline
{}&{}&{}&\footnotesize{{\small $\Lambda$CDM cosmology}}&{}&{}&{}
\\
\hline
\footnotesize{{\small slitless+BOSS+Planck}} & $0.017$ & $0.019$ &$0.017$ 
& $0.021$ & $0.021$ & $0.023$\\
\footnotesize{{\small multi-slit+Planck}} & $0.015$ & $0.016$ &$0.014$
& $0.018$ & $0.018$ & $0.019$\\
\hline
\end{tabular}
\begin{flushleft}
${}^a$\footnotesize{for degenerate spectrum: $m_1\approx m_2\approx
  m_3$};
${}^b$\footnotesize{for normal hierarchy: $m_3\neq 0$, $m_1\approx
  m_2\approx 0$}\\
${}^c$\footnotesize{for inverted hierarchy: $m_1\approx m_2$,
  $m_3\approx 0$};
${}^d$\footnotesize{fiducial cosmology with massless neutrinos}
\end{flushleft}
\label{summary}
\end{table*}

\section{Conclusions}
\label{Conclusions}
In this work we have forecasted errors on the total neutrino
mass $M_\nu$ and the effective
number of relativistic species $N_{\rm eff}$, by combining Planck
priors with future data from space-based spectroscopic galaxy redshift surveys in the
near-IR. We have considered two survey strategies based on
slitless and multi-slit spectroscopies. The assumed set of cosmological
parameters is very general and takes into account a time-varying
dark-energy equation of state, as well as a non-vanishing spatial curvature of
the Universe. We exploited information from  the galaxy power spectrum shape
and BAO positions, marginalising over galaxy bias;  thus our findings do not depend on bias
measurement accuracy (as long as, on the large scales  
considered, bias is scale independent or its scale 
dependence is known), or modelling  of the redshift dependence of bias \cite{LahavDES}. 

The 1--$\sigma$ errors are shown in
Tables~\ref{slitless_errors}-\ref{dmd_errors}, 
and the correlation coefficients in Tables~\ref{slitless_corr}-\ref{dmd_corr}. In
Figs.~\ref{fig_Mnu03_sigmas}-\ref{fig_Mnu005_comp} we show the joint
2-parameter confidence levels.

Regarding $M_\nu$--errors, we find that
the multi-slit spectroscopy is able to reduce the
neutrino mass errors of about 20\%-30\% compared to the slitless 
spectroscopy, depending on the fiducial total
neutrino mass, if LSS data alone are used. When Planck priors are
added, the 1--$\sigma$ errors on $M_\nu$ are in the range
$0.03-0.05$ eV, depending on the fiducial neutrino mass, with an
average difference of 15\% between the two spectroscopic
strategies, favouring the multi-slit spectroscopy. 

Moreover, depending on the fiducial $M_\nu$--value, the total CMB+LSS
dark-energy FoM, with growth--information marginalised over, decreases
only by $\sim 15\%-25\%$ with respect to the value obtained if
neutrinos are assumed to be massless (or their mass is assumed to be
perfectly known), meaning that the ``$P(k)$--method
marginalised                               
over growth--information'' is quite robust to assumptions 
about model cosmology when constraining the
dark-energy equation of state. The situation is different when we include
growth-information, since in this case the value of
the dark-energy FoM decreases by a factor $\sim 2-3$ with respect to
cosmologies that assume massless neutrinos.

Considering the fiducial cosmology with $M_\nu|_{\rm fid}=0.05$ eV,
in \S \ref{growth&FoG} we checked the stability of
$M_\nu$--errors to the inclusion of growth--information and peculiar
velocity uncertainties.
We compared the following approaches:
the ``full $P(k)$--method, marginalised over growth--information'',
the ``full $P(k)$--method, with growth--information included'', and
``full $P(k)$--method, with FoG and growth--information included''.
We found that $M_\nu$--errors are quite stable at
$\sigma(M_\nu)=0.05$ eV, against the
adopted method. This result is as expected, if we consider that,
unlike dark energy parameters, $M_\nu$ affects the shape of the power
spectrum via a redshift-dependent transfer function $T(k,z)$, which is
sampled
on a very large range of scales including the $P(k)$ turnover scale,
therefore this effect
dominates over the information extracted from measurements of
$f_g\sigma_8$.

Regarding $N_{\rm eff}$--errors, again we find that, 
compared to the slitless spectroscopy,
the multi-slit spectroscopy is able to reduce the
$N_{\rm eff}$--errors by $\sim$30\% when LSS alone are used. When Planck
priors are added,
we find $\sigma(N_{\rm eff})\sim 0.08$, with only a $6\%$ difference
between the two spectroscopy strategies,
again in favour of the multi-slit one. The total CMB+LSS dark-energy FoM
decreases only by $\sim 5\%$ with respect to the value obtained
holding $N_{\rm eff}$ fixed at its fiducial value,
meaning that also in this case the ``$P(k)$--method
marginalised over growth--information'' is not too 
sensitive to assumptions about  model cosmology when constraining the dark-energy
equation of state. 

Finally, in Table~\ref{summary} we summarise the dependence
of the $M_\nu$-- and $N_{\rm eff}$--errors on the model cosmology,
for the two spectroscopic strategies combined with
Planck. We conclude that, if $M_\nu$ is $>0.1$ eV, 
these surveys will be able
to determine the neutrino mass scale independently
of the model cosmology assumed. If  $M_\nu$ is $<0.1$ eV,
the sum of neutrino masses, and in particular the minimum neutrino
mass required by neutrino oscillations, can be measured
in the context of a $\Lambda$CDM model.

This means that future spectroscopic galaxy surveys, such as Euclid or 
SPACE, JEDI, and possibly WFIRST in the future, 
will be able to cover the entire  parameter space  
for neutrino mass allowed by oscillations experiments

Moreover, as summarised in Fig.~\ref{hierarchy}, 
they will be competitive with future 3D cosmic shear photometric surveys,
which, in combination with Planck priors, will give similar constraints on
$M_\nu$ and $N_{\rm eff}$ \cite{Kitching_nu}. Since, these two kinds of LSS
probe are affected by different systematics, their constraints 
on neutrino masses and relativistic degrees of freedom
will provide a consistency check of the two independent measurement methods.

We conclude that future nearly all-sky spectroscopic galaxy surveys
will detect the cosmic neutrino background at high statistical significance, 
and provide a measurement of the neutrino mass
scale. This will provide an important confirmation of  
our model for the early Universe, and crucial 
insights into  neutrino properties, highly 
complementary to future particle physics experiments. 

\acknowledgments
CC acknowledges L. Moscardini for useful discussions.
AC and CC acknowledge the support from the Agenzia
Spaziale Italiana (ASI-Uni Bologna-Astronomy Dept.
``Euclid-NIS'' I/039/10/0), and MIUR PRIN 2008 ``Dark energy and
cosmology with large galaxy surveys''.
LV is supported by FP7-PEOPLE-2007-4-3-IRG n. 202182,
FP7-IDEAS-Phys.LSS 240117 and MICINN grant AYA2008-03531.

\appendix

\section{Planck priors}
\label{sec:Planck}
In this work we use the Planck mission parameter
constraints as CMB priors, by estimating the cosmological parameter errors
via measurements of the temperature and polarisation power
spectra. 
As CMB anisotropies, with the exception of the integrated
Sachs-Wolfe effect, are not able to constrain the equation of state of
dark-energy $(w_0,w_a)$\footnote{On the contrary, using $(w_0,w_a)$ as
model parameters to compute the CMB Fisher matrix could artificially
break exiting degeneracies.}, we follow the prescription laid out by
DETF \cite{albrecht09}.

We do not include any B-mode in our
forecasts and assume no tensor mode contribution
to the power spectra. 
We use the 100 GHz, 143 GHz,
and 217 GHz channels as science channels. These channels
have a beam of $\theta_{\rm fwhm}=9.5'$, $\theta_{\rm fwhm}=7.1'$, and
$\theta_{\rm fwhm}=5'$, respectively, 
and sensitivities of $\sigma_T= 2.5 \mu K/K$, $\sigma_T= 2.2 \mu K/K$,
$\sigma_T= 4.8 \mu K/K$ for temperature, and $\sigma_P = 4\mu K/K$,
$\sigma_P = 4.2\mu K/K$, $\sigma_P = 9.8\mu K/K$ for polarisation,
respectively.  
We take $f_{\rm sky} = 0.80$ as the sky fraction in order to account for
galactic obstruction, and use a minimum $\ell$-mode $\ell_{\rm
  min}=30$ in order to avoid problems with polarisation foregrounds
and not to include information from the late
Integrated Sachs-Wolfe effect, which depends on the specific
dark-energy model. 
We discard temperature and polarisation data at
$\ell > 2000$  to reduce sensitivity to contributions from patchy
reionisation and point source contamination (see \cite{albrecht09} and
references therein).

We assume a $\Lambda$CDM fiducial cosmology, and choose the
following set of parameters to describe the temperature and
polarisation power spectra 
$\vec{\theta}= (\omega_m, \omega_b, \zeta,
100\times \theta_S,\ln (10^{10}\Delta^2_{\cal R}(k_0)), n_S, \tau)$, where $\theta_S$ is the
angular size of the sound horizon at last scattering, and
$\tau$ is the optical depth due to reionisation. Note that a different
parameter set is assumed in \cite{Mukherjee:2008}. 

The Fisher matrix for CMB power spectrum is given by \cite{Zalda:1997,Zalda:1997b}:
\begin{equation}
  F_{ij}^{CMB}=\sum_{l}\sum_{X,Y}\frac{\partial
    C_{X,l}}{\partial\theta_{i}}\mathrm{COV^{-1}_{XY}}\frac{\partial
    C_{Y,l}}{\partial\theta_{j}},
  \label{eqn:cmbfisher}
\end{equation} where $\theta_i$ are the parameters to constrain, 
$C_{X,l}$ is the harmonic power spectrum for the
temperature-temperature ($X\equiv TT$), 
temperature-E-polarisation ($X\equiv TE$) and the 
E-polarisation-E-polarisation ($X\equiv EE$) power spectrum. The covariance
$\rm{COV}^{-1}_{XY}$ of the errors for the various power spectra is
given by the fourth moment of the distribution, 
which under Gaussian assumptions is entirely given in terms of the $C_{X,l}$ with 
\begin{eqnarray}
{\rm COV}_{T,T} & = & f_\ell\left(C_{T,l}+W_T^{-1}B_l^{-2}\right)^2 \\
{\rm COV}_{E,E} & = & f_\ell\left(C_{E,l}+W_P^{-1}B_l^{-2}\right)^2  \\
{\rm COV}_{TE,TE} & = & f_\ell\Big[C_{TE,l}^2+\\
&&\left(C_{T,l}+W_T^{-1}B_l^{-2}\right)\left(C_{E,l}+W_P^{-1}B_l^{-2}\right)\Big] \nonumber \\
{\rm COV}_{T,E} & = & f_\ell C_{TE,l}^2  \\
{\rm COV}_{T,TE} & = & f_\ell C_{TE,l}\left(C_{T,l}+W_T^{-1}B_l^{-2}\right) \\
{\rm COV}_{E,TE} & = & f_\ell C_{TE,l}\left(C_{E,l}+W_P^{-1}B_l^{-2}\right)\; ,
\end{eqnarray}
where $f_\ell = \frac{2}{(2\ell+1)f_{\rm sky}}$,
$W_{T,P}=\sum_c W^c_{T,P}$, $W^c_{T,P}=(\sigma^c_{T,P}\theta^c_{\rm fwhm})^{-2}$  
being the weight per solid angle for temperature and polarisation respectively, 
with a 1--$\sigma$ sensitivity per pixel of $\sigma^c_{T,P}$ and a beam
of $\theta^c_{\rm fwhm}$ extent, for each frequency channel $c$. 
The beam window function is given in terms of the full width half
maximum (fwhm) beam width by 
$B_{\ell}^2 =\sum_c (B^c_{\ell})^2 W^c_{T,P}/W_{T,P}$, where 
$(B^c_\ell)^2= \exp\left(-\ell(\ell+1)/(l^c_s)^2\right)$, 
$l^c_s=(\theta^c_{\rm fwhm})^{-1}\sqrt(8\ln2)$ 
and $f_{\rm sky}$ is the sky fraction \cite{9702100}. 

We then calculate the Planck CMB Fisher matrix with the help of the
publicly available CAMB code \cite{Lewis:1999bs}. Finally, we transform 
the Planck Fisher matrix for the DETF parameter set to the
final parameter sets $\bf q$ considered in this work (see \S
\ref{Fiducial cosmologies} and \S \ref{growth&FoG}),
using the transformation
\begin{equation}
F_{\alpha \beta}^{\rm CMB}= \sum_{ij} \frac{\partial
  \theta_i}{\partial q_{\alpha}}\,                       
F_{ij}^{\rm CMB}\, \frac{\partial \theta_j}{\partial q_{\beta}}.
\end{equation}

\addcontentsline{toc}{section}{References}
\bibliographystyle{JHEP}
\bibliography{CVWC2010_new_figs_arxiv}

\end{document}